\documentclass[longauth]{aa}
\usepackage[varg]{txfonts}
\usepackage{graphicx}
\usepackage[colorlinks=true,linkcolor=blue,citecolor=blue]{hyperref}
\usepackage{subfig}
\usepackage{booktabs}
\def\degb{^{\circ}}
\usepackage{soul}
\newcommand{\ra}[1]{\renewcommand{\arraystretch}{#1}}

\begin{document}

\authorrunning{G. Singh et al.}
\titlerunning{Asymmetries in the inner regions of HD\,141569 } 

\title{Revealing asymmetrical dust distribution in the inner regions of HD\,141569 }

 \author{
 G. Singh\inst{\ref{lesia}} 
 \and T. Bhowmik\inst{\ref{lesia}} 
 \and A. Boccaletti\inst{\ref{lesia}} 
 \and P. Th\'ebault\inst{\ref{lesia}} 
 \and Q. Kral\inst{\ref{lesia}} 
 \and J. Milli\inst{\ref{ipag}}
 \and J. Mazoyer\inst{\ref{lesia}} 
 \and E. Pantin\inst{\ref{cea}} 
 \and J. Olofsson \inst{\ref{valpa},\ref{mpia},\ref{nucleo}} 
 \and R. Boukrouche\inst{\ref{oxford}} 
 \and E. DiFolco\inst{\ref{lab}} 
 \and M. Janson\inst{\ref{stockholm}} 
 \and M. Langlois \inst{\ref{cral},\ref{lam}}
 \and A.-L. Maire\inst{\ref{mpia}, \ref{liege}} 
 \and A. Vigan \inst{\ref{lam}}
 \and M. Benisty\inst{\ref{ipag}} 
 \and J.-C. Augereau\inst{\ref{ipag}} 
 \and C. Perrot\inst{\ref{valpa},\ref{nucleo}} 
 \and R. Gratton\inst{\ref{padova}}
 \and T. Henning\inst{\ref{mpia}} 
  \and F. M{\'e}nard\inst{\ref{ipag}}
 \and E. Rickman\inst{\ref{geneva},\ref{esa}} 
 \and Z. Wahhaj\inst{\ref{eso}} 
 \and A. Zurlo\inst{\ref{lam},\ref{santiago},\ref{escuela}} 
 \and B. Biller\inst{\ref{roe}} 
 \and M. Bonnefoy\inst{\ref{ipag}} 
 \and G. Chauvin\inst{\ref{ipag}} 
 \and P. Delorme\inst{\ref{ipag}} 
 \and S. Desidera\inst{\ref{padova}}
 \and V. D'Orazi\inst{\ref{padova}}
 \and M. Feldt\inst{\ref{mpia}}
 \and J. Hagelberg\inst{\ref{geneva}}
 \and M. Keppler\inst{\ref{mpia}}
 \and T. Kopytova\inst{\ref{mpia}}
 \and E. Lagadec\inst{\ref{oca}}
 \and A.-M. Lagrange\inst{\ref{ipag}\ref{lesia}}
 \and D. Mesa\inst{\ref{padova}}
 \and M. Meyer\inst{\ref{michigan},\ref{eth}}
 \and D. Rouan\inst{\ref{lesia}}
 \and E. Sissa\inst{\ref{padova}}
 \and T.O.B. Schmidt\inst{\ref{lesia},\ref{hamburg}}
 \and M. Jaquet \inst{\ref{lam}} 
 \and T. Fusco \inst{\ref{lam},\ref{onera}}
 \and A. Pavlov \inst{\ref{ipag}} 
 \and P. Rabou \inst{\ref{ipag}} 
}
 \institute{
 LESIA, Observatoire de Paris, Universit\'e PSL, CNRS, Universit\'e de Paris, Sorbonne Universit\'e, 5 place Jules Janssen, 92195 Meudon, France\label{lesia}\\
 \and CNRS, IPAG, Universit\'e Grenoble Alpes, 38000 Grenoble, France\label{ipag}\\
 \and CEA, IRFU, DAp, AIM, Universit\'e Paris-Saclay, Universit\'e Paris Diderot, Sorbonne Paris Cit\'e, CNRS, F-91191 Gif-sur-Yvette, France\label{cea}\\
 \and Instituto de F\'isica y Astronom\'ia, Facultad de Ciencias, Universidad de Valpara\'iso, Av. Gran Breta\~na 1111, Playa Ancha, Valpara\'iso, Chile\label{valpa}\\
 \and Max Planck Institut f\"ur Astronomie, K\"onigstuhl 17, 69117 Heidelberg, Germany\label{mpia} \\
 \and N\'ucleo Milenio Formaci\'on Planetaria - NPF, Universidad de Valpara\'iso, Av. Gran Breta\~na 1111, Valpara\'iso, Chile\label{nucleo} \\
 \and University of Oxford, Department of Physics, Sub-department of Atmospheric, Oceanic \&
Planetary Physics (AOPP)\label{oxford}\\
 \and Laboratoire d’Astrophysique de Bordeaux, Univ. Bordeaux, CNRS, B18N, all\'ee Geoffroy Saint-Hilaire, 33615 Pessac, France\label{lab}\\
 \and Department of Astronomy, Stockholm University, AlbaNova University Center, SE-109 91 Stockholm, Sweden\label{stockholm} \\
 \and Univ Lyon, Univ Lyon1, Ens de Lyon, CNRS, Centre de Recherche
Astrophysique de Lyon, UMR5574, F-69230 Saint-Genis-Laval, France\label{cral}\\
 \and Aix Marseille Univ, CNRS, CNES, LAM, Marseille, France \label{lam}\\
 \and STAR Institute, Universit\'e de Li\`ege, All\'ee du Six Ao\^ut 19c, B-4000 Li\`ege, Belgium\label{liege}\\
 \and Geneva Observatory, University of Geneva, Chemin Pegasi 51, 1290 Versoix, Switzerland\label{geneva}\\
 \and European Space Agency (ESA), ESA Office, Space Telescope Science Institute, 3700 San Martin Drive, Baltimore, MD 21218, USA\label{esa}\\
 \and European Southern Observatory, Alonso de Cordova 3107, Vitacura, Casilla 19001, Santiago, Chile\label{eso}\\
 \and N\'ucleo de Astronom\'ia, Facultad de Ingenier\'ia y Ciencias, Universidad Diego Portales, Av. Ejercito 441, Santiago, Chile\label{santiago}\\
 \and Escuela de Ingenier\'ia Industrial, Facultad de Ingenier\'ia y Ciencias, Universidad Diego Portales, Av. Ejercito 441, Santiago, Chile\label{escuela}\\
 \and SUPA, Institute for Astronomy, The University of Edinburgh, Royal Observatory, Blackford Hill, Edinburgh, EH9 3HJ, UK\label{roe}\\
 \and INAF - Osservatorio Astronomico di Padova, Vicolo dell’ Osservatorio 5, 35122, Padova, Italy\label{padova}\\
 \and Université Côte d'Azur, Observatoire de la Côte d'Azur, CNRS, Laboratoire Lagrange, France\label{oca}\\
 \and Department of Astronomy, University of Michigan, Ann Arbor, MI 48109, USA\label{michigan}\\
\and Institute for Particle Physics and Astrophysics, ETH Zurich, Wolfgang-Pauli-Strasse 27, 8093 Zurich, Switzerland\label{eth}\\
\and Hamburger Sternwarte, Gojenbergsweg 112, D-21029 Hamburg,
Germany\label{hamburg}\\
 \and DOTA, ONERA, Universit\'e Paris Saclay, F-91123, Palaiseau France\label{onera}\\
\email{garima.singh@obspm.fr}}

  \abstract
  {The combination of high-contrast imaging with spectroscopy and polarimetry offers a pathway to studying the grain distribution and properties of debris disks in exquisite detail. Here, we focus on the case of a gas-rich debris disk around HD\,141569A, which features a multiple-ring morphology first identified with SPHERE in the near-infrared.}
  { We obtained polarimetric differential imaging with SPHERE in the $H$-band to compare the scattering properties of the innermost ring at 44\,au with former observations in total intensity with the same instrument. In polarimetric imaging, we observed that the intensity of the ring peaks in the south-east, mostly in the forward direction, whereas in total intensity imaging, the ring is detected only at the south. This noticeable characteristic suggests a non-uniform dust density in the ring. With these two sets of images, we aim to study the distribution of the dust to solve for the actual dust distribution.} 
  {We implemented a density function varying azimuthally along the ring and generated synthetic images both in polarimetry and in total intensity, which are then compared to the actual data. The search for the best-fit model was performed both with a grid-based and an MCMC approach. Using the outcome of this modelization, we further measured the polarized scattering phase function for the observed scattering angle between 33$\degb$ and 147$\degb$ as well as the spectral reflectance of the southern part of the ring between 0.98\,$\muup$m and 2.1\,$\muup$m. We tentatively derived the grain properties by comparing these quantities with MCFOST models and assuming Mie scattering.}           
  {We find that the dust density peaks in the south-west at an azimuthal angle of $220\degb\sim 238\degb$ with a rather broad width of $61\degb\sim 127\degb$. The difference in the intensity distributions observed in polarimetry and total intensity is the result of this particular morphology. Although there are still uncertainties that remain in the determination of the anisotropic scattering factor, the implementation of an azimuthal density variation to fit the data proved to be robust. Upon elaborating on the origin of this dust density distribution, we conclude that it could be the result of a massive collision when we account for the effect of the high gas mass that is present in the system on the dynamics of grains. In terms of grain composition, our preliminary interpretation indicates a mixture of porous sub-micron sized astro-silicate and carbonaceous grains.} 
  {The SPHERE observations have allowed, for the first time, for meaningful constraints to be placed on the dust distribution beyond the standard picture of a uniform ring-like debris disk. However, future studies with a multiwavelength approach and additional detailed modeling would be required to better characterize the grain properties in the HD\,141569 system. }


\maketitle

%
%
%
%
\section{Introduction}
\label{s:intro}

The formation of planetary systems is a complex process involving several stages that are not yet fully understood. When the gas from the primordial disk is dissipated and planets are formed, typically after 5-10 Myrs, the remaining mass of the system is distributed in planetesimals, which, due to the gravitational interplay with planets, are usually arranged in the form of one or several rings orbiting as far as tens of astronomical units (au) from the star (depending on the stellar mass), akin to the asteroid and Kuiper belts in our own Solar System. The current paradigm implies that planetesimals in these rings undergo mutual destructive collisions, initializing a collisional cascade that produces fragments of various sizes, all the way down to small dust grains \citep{wyatt2008}. These small dust grains at the tail end of the collisional cascade can be detected in the form of a photometric excess in the infrared, but often in imaging as well. Systems in which such second generation fragment production is underway are referred to as debris disks.

A significant fraction of these debris disks have been imaged, from the visible up to the millimetric domains, revealing a great variety of spatial structures, such as arms, spirals, radial gaps, warps, clumps, and two-sided asymmetries \citep{Hughes2018}. Several scenarios have been explored for the possible origin of these structures, such as dynamical interactions with planets \citep[e.g.,][]{reche2009, thebault2012b, nesvold2015}, interactions with leftover or second-generation gas \citep{takeuchi2001, lyra2013}, or dynamical perturbations of companion stars \citep{thebault2012a}. One scenario that has received a lot of attention in the past decade involves giant impacts between Ceres-to-planet-sized embryos, which could create transient bright spots or arms \citep{thebault2018}, but also more long-lived structures. The typical long-lived outcome of a large-scale collision is an asymmetric structure with a bright clump at the location of the initial breakup, through which all debris must pass, and a more extended and diffuse structure on the opposite side \citep{jackson2014, kral2015}. The survival time of these asymmetries can be of the order of $10^5$ up to several $10^6$\,years.

Observations in direct imaging at near-infrared (NIR) or optical wavelengths are sensitive to the stellar light scattered by the dust. The way the light is scattered depends on the disk morphology (size and inclination, in particular), grains properties (composition, shape, and size), and surface density. As a result, direct imaging has the ability to provide constraints on these characteristics. The latest facility instruments, Spectro-Polarimetric High-contrast Exoplanet REsearch
 \citep[SPHERE,][]{Beuzit2019}, Gemini Planet Imager \citep[GPI,][]{Macintosh2014}, and Subaru Coronagraphic Extreme Adaptive Optics project \citep[SCExAO,][]{scexao} have now achieved an impressive record of disk studies which have been efficiently complemented by Atacama Large
Millimeter/submillimeter Array (ALMA) results \cite[e.g.,][]{mg2018, olofsson2019}.

In the evolution of planetary systems, the disk around the Herbig Ae/Be star HD\,141569A
\citep[$\sim5$\,Myr,][]{Weinberger2000,merin2004} occupies a unique place as it shares the characteristics of both protoplanetary and debris disks, and is regarded as a turning point object between these two categories. Initially classified as a transitional disk, it is now considered a gas-rich debris disk. Studying this object provides the opportunity to shed light on a very specific moment in the evolution of planetary systems. With favorable characteristics in proximity \citep[$110.63^{+0.91}_{-0.88}$\,pc,][]{gaia2018} and size (400\,au), this unique system has been the subject of intensive studies, the objective being to reach the innermost regions where planets are expected to form. 

The disk of HD\,141569A was first resolved in the NIR coronagraphic observations with HST/NICMOS2 at 1.6\,$\muup$m \citep{Augereau1999a} and at 1.1\,$\muup$m \citep{Weinberger1999}, allowing for two ring-like components to be resolved, located at roughly 250\,au and 410\,au from the star. Follow-up observations at visible wavelengths laid out a detailed view of the structure of these rings composed of arcs and spirals \citep{Mouillet2001, Clampin2003}, later confirmed in the NIR with an 8-m class ground-based telescope \citep{Biller2015, Mazoyer2016}.  

Based on the spectral energy distribution (SED), the presence of material nested within the two outer rings, inwards of 100\,au, was suggested by \citet{Augereau1999a} and then established owing to mid-IR thermal imaging \citep{Fisher2000, Marsh2002, Thi2014}. More recently, \citet{Konishi2016} and \citet{Perrot2016}, with HST/STIS and SPHERE, respectively, identified substructures in the central area in scattered light. In particular, SPHERE provided a unique view of the inner system by revealing, for the first time, several rings the most prominent being located at about 45 au. This finding was confirmed with \emph{L} band imaging by \citet{Mawet2017} and in polarimetry with GPI \citep{Bruzzone2020}. The innermost ring is of great interest because it features a strong north-south brightness asymmetry along the major axis of the disk, whereas we would rather expect an east-west asymmetry due to scattering according to the disk inclination ($\sim57^{\circ}$). For this reason, \citet{Perrot2016} suggested that the density of the innermost ring is not constant azimuthally, which has not, thus far, been accounted for in modeling works \citep{Bruzzone2020}. As for the gas content, the most detailed observations obtained at a high angular resolution with ALMA {($^{13}$CO $J\,=\,1\rightarrow0$ transition acquired at a spatial resolution of 0.76$''$\,$\times$\,0.56$''$)} allowed us to assess the CO gas distribution at similar distances than in scattered light \citep{Difolco2020}. The corresponding map is asymmetrical with a significant excess of emission in the western part of the disk (from position angles of 220$^{\circ}$ to 290$^{\circ}$).

The purpose of the analysis presented in this paper is to take advantage of both the polarimetric and total intensity data obtained with SPHERE in order to solve for the density and scattering functions. Section\,\ref{s:obs&reduct} presents the observations and data reduction methods. The morphology of the disk and the modeling of the innermost ring both in polarized and total intensities are explained in Sects.\,\ref{s:morpho} and\,\ref{s:model}, respectively. In Sects.\,\ref{s:SPF} and \ref{s:spectro}, we present the methods for extracting the phase function of the ring, followed by the spectroscopy of the southern part of the ring in the total intensity data. In Sect.\,\ref{s:grain_properties} we model the grain properties. In Sect.\,\ref{s:discuss} we discuss the plausible physical origin of the ring.

%
%
%
%
\begin{table*}[th!]
\caption{Log of SPHERE observations.}
\begin{center}
\begin{tabular}{l  l c l c l c l c l c l c l c l c c }
\hline
\hline
Date UT &  Filter & Field rotation & Pol. cycle & DIT &       N$_\mathrm{exp}$  &    T$_\mathrm{exp}$ & seeing  &      $\tau_0$ & TN  \\
 &   &  ($^\circ$) &   &   (s) &   & (s) &  ($''$) & (ms) & ($^\circ$)  \\ \hline 
   2015-05-16 &  IRDIS-H23 & 42.70 & - & 64 & 64 & 4096 & 0.83$\pm$0.07 & 3.6$\pm$0.3 & -1.712 \\
   2015-05-16 &  IFS-YJ & 42.70 & - & 64  & 64 & 4096 & 0.83$\pm$0.07 & 3.6$\pm$0.3 & -1.712 \\
   2016-03-30 & IRDIS-K12 & 42.57 & - & 64 & 64 & 4096 & 0.86$\pm$0.16 & 3.1$\pm$0.6 & -1.756\\
   2016-03-30 & IFS-YH & 42.57 & - & 64 & 64 & 4096 & 0.86$\pm$0.16 & 3.1$\pm$0.6 & -1.756\\
2017-03-21 &  IRDIS - H & stabilized & 10    &   64  &   80 &    5120 &  0.75$\pm$0.12   &   3.4$\pm$0.7 & -1.700\\
   2017-06-15 & IRDIS-K12 & 59.35 & - & 96 & 64 & 6144 & 0.52$\pm$0.10 & 6.1$\pm$2.0 & -1.739\\
   2017-06-15 & IFS-YH    & 59.35 & - & 96 & 64 & 6144 & 0.52$\pm$0.10 & 6.1$\pm$2.0 & -1.739\\
\hline
\end{tabular}
\end{center}
\label{tab:log}
\tablefoot{The columns from left to right indicate the date of observations in UT, the filter, the field rotation, the number of polarimetric cycles, the individual exposure time (DIT) in seconds, the total number of exposures, the total exposure time in seconds, the DIMM seeing measured in arcseconds, the correlation time $\tau_0$ in milliseconds, the variation of the flux during the sequence in \%, and the true north (TN) offset in degrees.}
\end{table*}

%
%
%
%
\section{Observation and data reduction}
\label{s:obs&reduct}

HD\,141569 (V=7.12, H=6.86) was observed with SPHERE \citep{Beuzit2019} on March 21st, 2017, with the Infrared Dual-Band Imager and Spectrograph \citep[IRDIS,][]{Dohlen2008} in dual-beam polarimetric imaging \citep[DPI,][]{Langlois2014, Van2020}. We used FIELD tracking in the BB\_H filter ($\lambda=1.625\,\muup$m, $\Delta\lambda=0.29\,\muup$m). Ten polarimetric cycles were obtained in which the half-wave plate was rotated at four orientations (0$^{\circ}$, 22.5$^{\circ}$, 45$^{\circ}$, 67.5$^{\circ}$) to measure the linear polarization parameters $Q+$, $U+$, $Q-$, $U-$. To preserve the polarimetric efficiency, we adapted the derotator angle for each polarimetric cycle following the procedure recommended in \citet{deBoer2020}. The log of the observations is shown in Table\,\ref{tab:log}. We used an apodized Lyot coronagraph \citep{Carbillet2011} with a diameter of 185\,mas (configuration N\_ALC\_YJH\_S) to suppress the starlight for observations dated $2015-05-16$ and $2017-03-21$, and with a diameter of 240\,mas (configuration N\_ALC\_Ks) for observations on $2016-03-30$ and $2017-06-15$.
 
The total intensity data has been acquired in May 2015 using the dual band imaging \citep[DBI,][]{Vigan2010} mode of IRDIS with filters $H2$ and $H3$, and with the Integral Field Spectrograph \citep[IFS,][]{Claudi2008} in $YJ$ mode ($0.95-1.35\muup$m, $R\sim54$). More details of the observations can be found in \citet{Perrot2016}. 
The second and third epochs were obtained in  March 2016 and June 2017 with a different setup (IRDIFS-ext), which combines IRDIS in the $K1$ and $K2$ filters and IFS in $YJH$ mode ($0.95-1.55\,\muup$m, $R\sim33$). All the data were processed using the data reduction and handling (DRH) pipeline \citep{Pavlov2008,Mesa2015} at the SPHERE Data Centre (DC)\footnote{http://sphere.osug.fr} \citep{Delorme2017}. This preliminary data-reduction process included sky and dark subtractions, flat-field corrections, star-centering using calibration spots, distortion correction \citep{Maire2016}, bad-pixel removal, and wavelength calibration. The point-spread function (PSF) was obtained by moving the star outside of the coronagraphic mask and placing a neutral density (ND2) in the beam. The data are processed with Karhunen-Lo\`eve Image Projection \citep[KLIP,][]{Soummer2012} to perform advanced angular differential imaging \citep[ADI,][]{Marois2006ADI}. 

The polarimetric data, on the other hand, were reduced with IRDAP, the IRDIS  Data reduction for Accurate Polarimetry \citep[][version 1.3.0]{Van2020}. IRDAP addresses the crosstalk and the Instrumental Polarisation (IP) effects of both the telescope and IRDIS. The pipeline first pre-processesed the data by applying dark subtraction, flat fielding, bad-pixel correction, and frame centering. It then computes the double-difference and double-sum images \citep{Tinbergen1996} to obtain the linear Stokes parameters ($Q$ and $U$) and the corresponding total intensities ($I_{\mathrm{Q}}$ and $I_{\mathrm{U}}$), which are later corrected for IP effects and crosstalk. IRDAP provides the linearly polarized intensity, angle, and degree of linearly polarized light of the source, and
computes azimuthal Stokes parameters $Q_{\mathrm{\phi}}$ and $U_{\mathrm{\phi}}$ \citep{deBoer2020}, with or without the star polarization removed. This is to say that IRDAP can identify the polarization signal of the star that may be present in the images possibly due to a spatially unresolved inner disk component. Such a stellar polarization signal creates a halo of polarized light that produces a butterfly pattern in the $Q_{\mathrm{\phi}}$ and
$U_{\mathrm{\phi}}$ images, thereby distorting the images at close separations from the star. Throughout the paper, we  use the star polarization subtracted images. Under the assumptions of single scattering and low inclinations \citep{canovas2015}, $Q_{\mathrm{\phi}}$ thus represents the polarimetric signal, and $U_{\mathrm{\phi}}$ the noise.

%
%
%
%
\begin{figure*}
  \centering
    \includegraphics[width=0.48\textwidth]{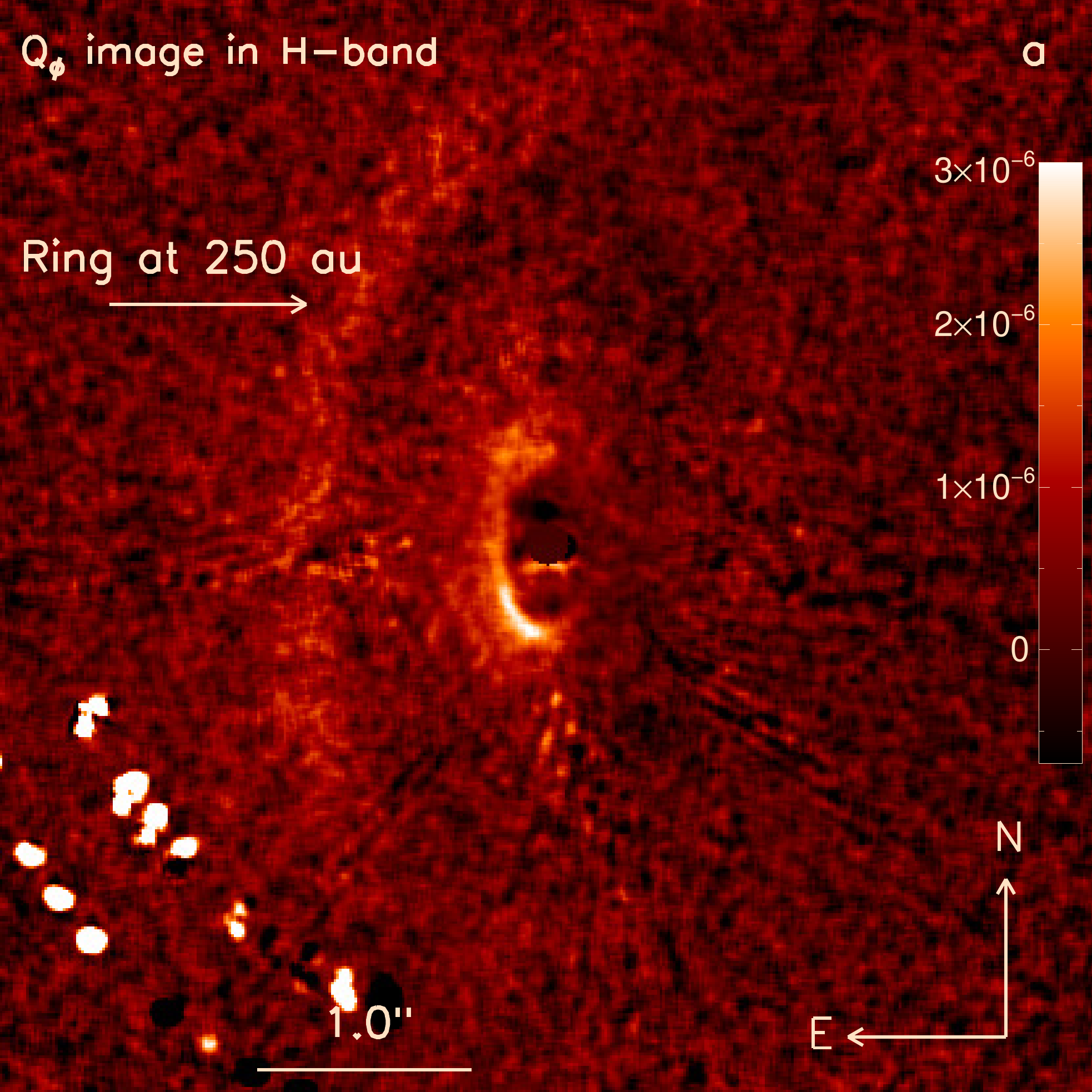}
\includegraphics[width=0.48\textwidth]{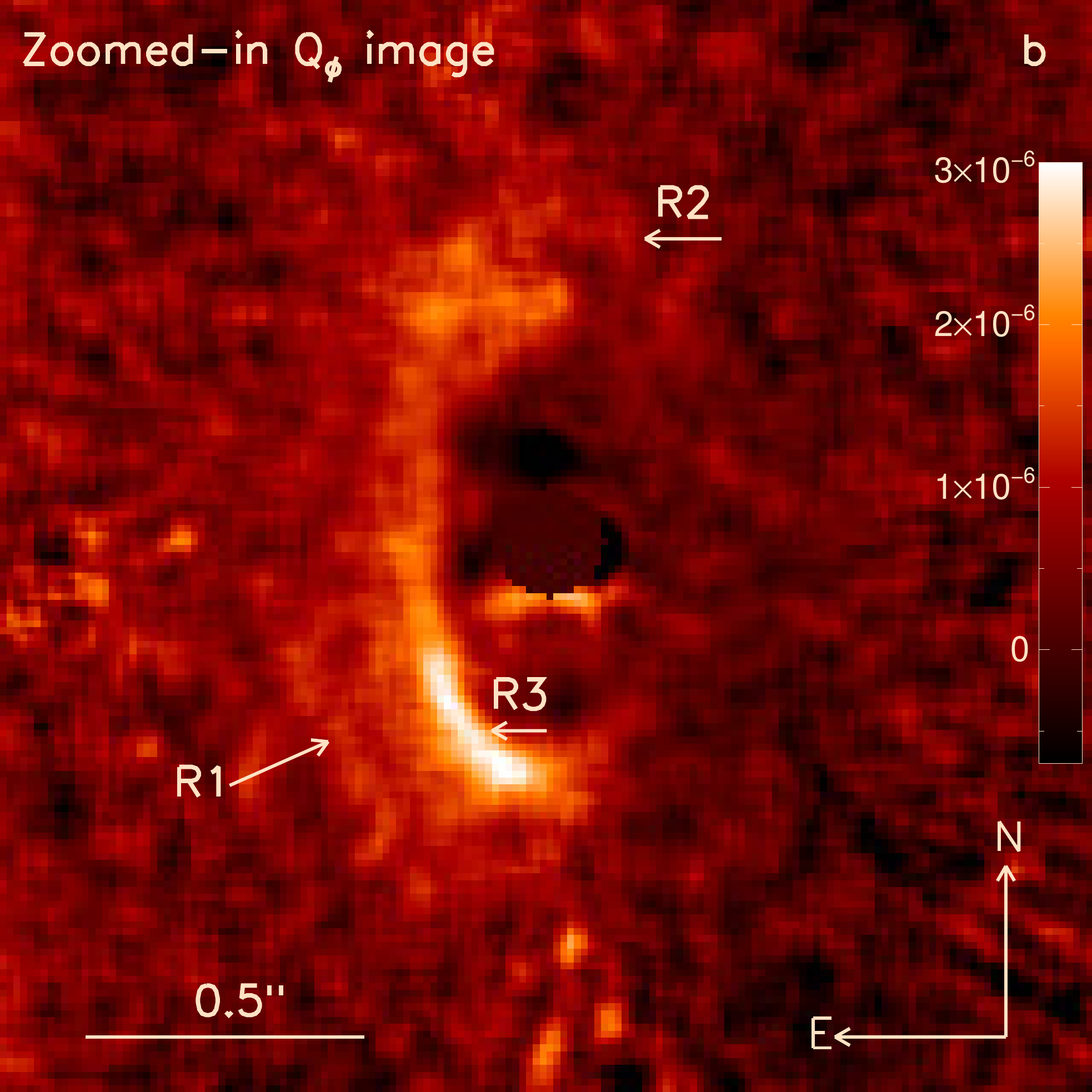}

    \includegraphics[width=0.48\textwidth]{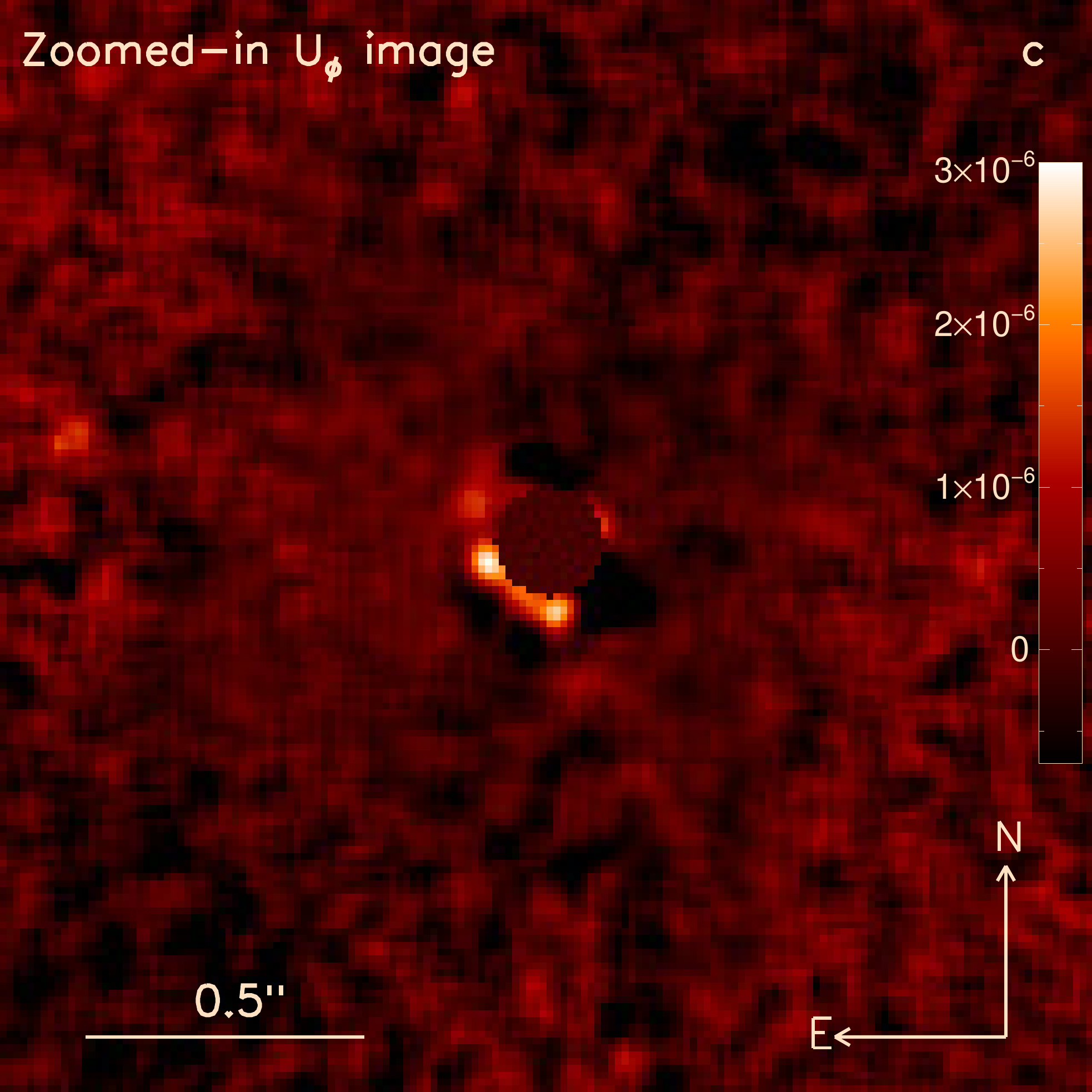}
    \includegraphics[width=0.48\textwidth]{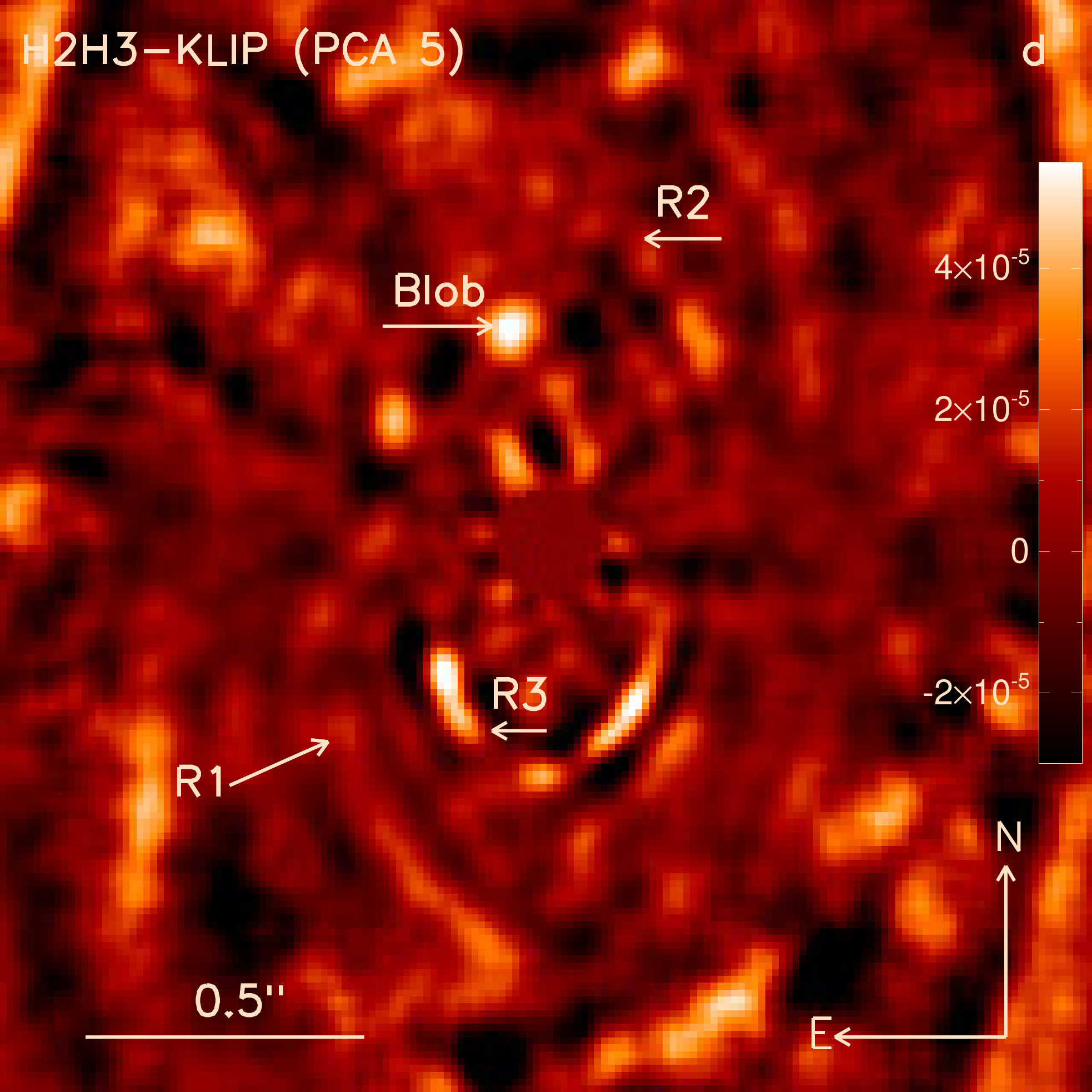}
  \caption{SPHERE observations of HD\,141569A in polarized intensity (2017) and total intensity (2015) displayed in contrast units. (a,\,b) and (c) present the linear polarization intensity $Q_{\mathrm{\phi}}$ and $U_{\mathrm{\phi}}$ images. (d) shows the total intensity image in the same region as in (b). The image (a) presents the wide angle field of view (5$''\times$5$''$) in polarimetry, while the images (b), (c) and (d) are at the same spatial scale and focused on the innermost region (2$''\times$2$''$) of the system. All the labels and arrows in (b) and (d) are at the same position. }
  \label{fig:hd141_fig}
\end{figure*}

%
%
%
%
\section{Morphology of the disk}
\label{s:morpho}

This section describes the $H-$band polarimetric image of the disk around HD\,141569A, and provides a comparison with the total intensity data by \citet{Perrot2016}. Figure\,\ref{fig:hd141_fig}a displays the polarimetric $Q_{\phi}$ image in a 5$''\times$5$''$ field of view, which is smoothed with a five-pixel (61\,mas compared to the diffraction-limited resolution of 42\,mas) Gaussian kernel. While the previously known outer ring at 250\,au is barely detected, we can still appreciate an east-west brightness asymmetry that is in apparent contradiction with the total intensity data, where this ring is more homogeneous. None of the fine structures identified in this ring by \citet{Perrot2016}  are visible in Fig.\,\ref{fig:hd141_fig}a due to the poor signal to noise ratio (S/N). The even fainter outer most ring at 400\,au is not visible in the DPI data for the same reason. In DPI, the transmission of the instrument is reduced as compared to the total intensity observations due to the use of the IRDIS polarizers and also because only a fraction of the light is actually polarized. The black and white circular marks in the lower left corner of Fig.\,\ref{fig:hd141_fig}a are caused by clusters of bad pixels. \footnote{The apparent rotation of the bad pixel clusters is due to the frequent adaptation of the derotator angle during the observing sequence to preserve the polarimetric efficiency \citep{deBoer2020}.} 

The zoom-in $Q_{\mathrm{\phi}}$ and $U_{\mathrm{\phi}}$ images in Fig.\,\ref{fig:hd141_fig}b and c show the inner 2$''\times$2$''$ region of the system, which are smoothed similarly as in Fig.\,\ref{fig:hd141_fig}a. In the $Q_{\mathrm{\phi}}$ image, we can recognize presumably the main components discovered in \citet{Perrot2016}, namely the ringlets R1, R2, R3 as labelled in the zoom-in total intensity image (Fig.\,\ref{fig:hd141_fig}d), which are also visible in its corresponding S/N map in Fig.\,\ref{fig:hd141_fig_SNR}. Starting from the shortest  distances, R3 is located at around 45\,au. In Fig.\,\ref{fig:hd141_fig}b, R3 features an interesting intensity distribution, with a brighter lobe in the southeast, and becomes almost undetected in the western side. R2 is a sort of hook-like structure apparently emanating from the northeast side of the inner ring R3 and extending upwards to the north. It can be confused with the northern tip of R3 but is actually at a larger  distance. A portion of R2 is likely also visible in the southeast. R1 could be an arc, a portion of a ring, or a spiral structure extending in the south-east direction. 

In the IFS images, the western side of R2 and the southern and the north-western part of R3 are clearly visible at all wavelengths. The reduced images with their corresponding S/N maps are presented in Figs.\,\ref{fig:dbi_yjk} and\,\ref{fig:dbi_yjk_SNR}, respectively. To reinforce the identification of the rings, we took advantage of the three epochs and followed the procedure described in \cite{Gratton2019} in which the data are processed with both angular and spectral differential imaging \citep[ASDI-PCA,][]{Mesa2015}, and their corresponding S/N maps are multiplied. The result shown in Fig.\,\ref{fig:asdi} definitely confirms the detection of R2 and gives a stronger evidence of the north-south asymmetry in R3. However, we refrain from deriving any physical parameters from the outcome of this procedure given that the intensity is not well preserved.

In the IRDIS image as shown in Fig.\,\ref{fig:hd141_fig}d, a bright blob is seen in the northeast at a distance of 387\,mas with a diameter of 90\,mas. This blob is not detected in any of the IFS wavelength channels (Fig.\,\ref{fig:dbi_yjk}), nor in the IRDIS K band images \citep{Perrot2016}. If it were a dust feature, it would  also be visible in the polarised intensity map, however, this is not the case. Moreover, it is aligned with the wind direction. Altogether, these arguments lead us to conclude that this source is an artifact superimposed over R3. In addition, the clump-like feature reported at the southern ansa by \cite{Perrot2016} is detected with both IRDIS (Fig.\,\ref{fig:hd141_fig}d) and IFS in total intensity (Fig.\,\ref{fig:dbi_yjk}). On the contrary, the absence of this feature in the $Q_{\mathrm{\phi}}$ image (Fig.\,\ref{fig:hd141_fig}b) could indicate an ADI-induced artifact \citep{Milli2012}.  

While both DPI and ADI images reveal material at the very same location, the polarized and total intensities diverge strongly, owing to a combination of phase function effects, ADI self-subtraction effects, and, possibly, density effects. Focusing on the ring R3, the total intensity is predominantly larger in the southern part. The minor axis of this ring is strongly affected by self-subtraction as a result of a moderate field rotation (42.7$^{\circ}$) and short angular separations (0.41$''$). As for the case of HR\,4796, any forward scattering peak expected toward the east would be damped by the self-subtraction \citep{Milli2017}. On the contrary, the polarized image reveals the disk in a larger range of position angles. The polarized intensity of the ring (see details in Sect.\,\ref{s:Pol_model}) is increasing from the north toward the east, and is reaching an apparent plateau between PA $\sim$115$^{\circ}$ and $\sim$180$^{\circ}$. Then, the intensity drops again toward the west side while the disk becomes undetectable. Therefore, a natural assumption would be that the density of R3 is larger in the southeast, but this is at odds with respect to the total intensity image, which shows a nearly east-west symmetrical intensity in the southern part. 
 
\begin{figure*}
  \centering
  \includegraphics[width=0.48\textwidth]{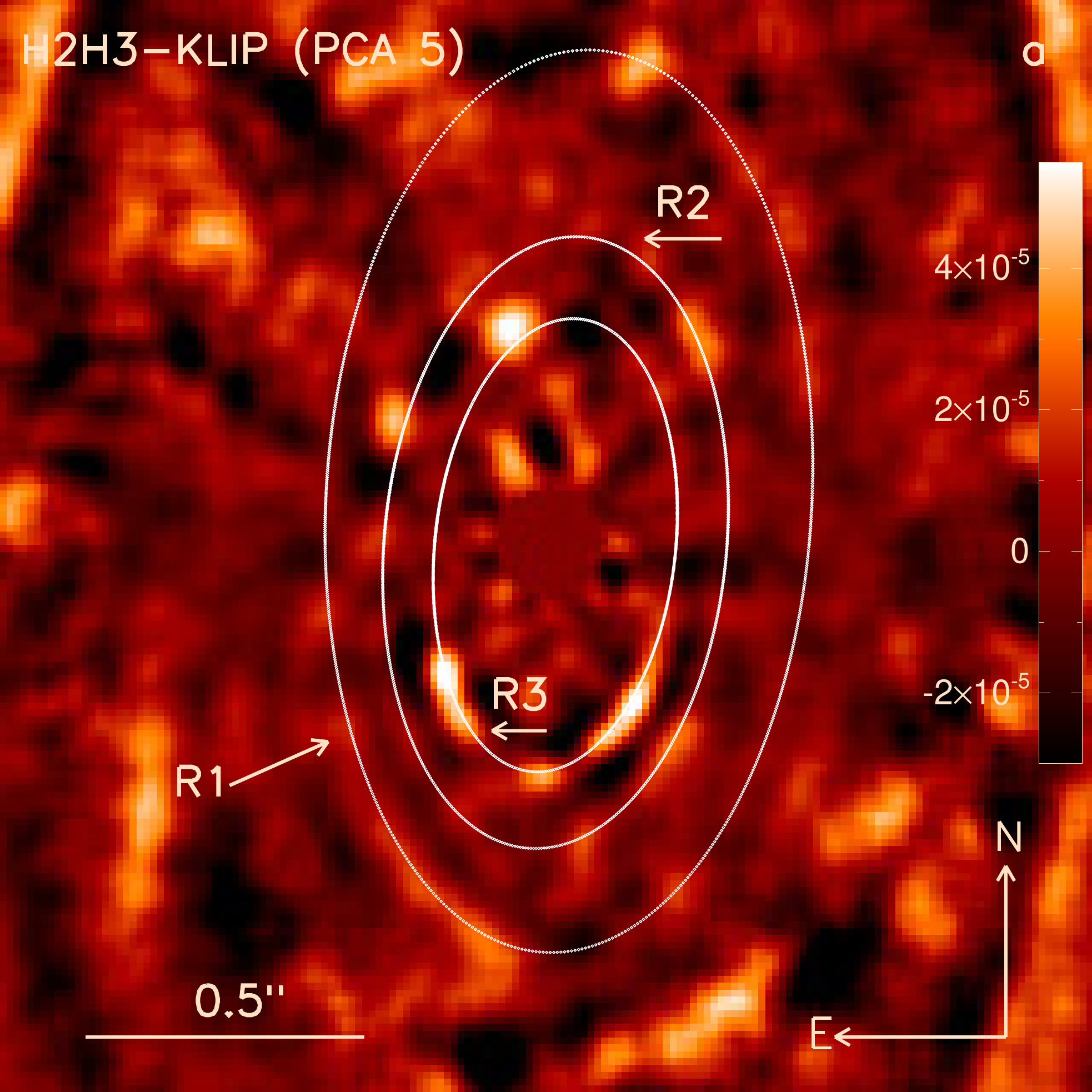}
  \includegraphics[width=0.48\textwidth]{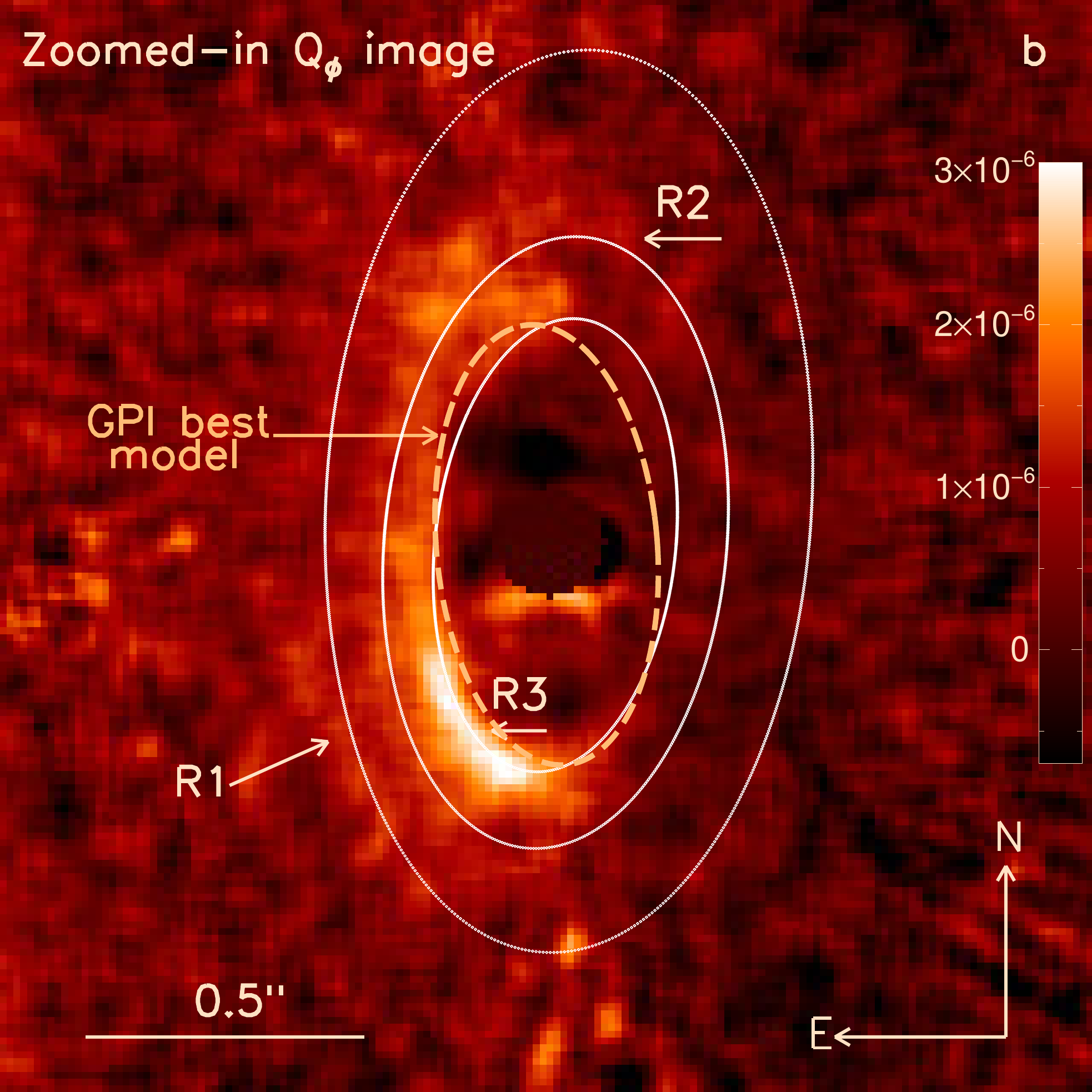}
  \caption{Elliptical fits in the ADI-reduced total intensity image (a) display the geometrical parameters of the inner (R3, 47\,au), middle (R2, 64\,au), and outer (R1, 93\,au) rings constrained in \citet{Perrot2016}. The same ellipses when overlapped in the IRDAP-reduced polarized intensity (b) indicate that the position of these structures were not well constrained in the total intensity data. The labels and arrows indicating R1, R2 and R3 in both images are exactly at the same position. The image (b) also shows the orientation of the best-fit model of GPI at the PA of 5$^{\circ}$. Both images are at the same spatial scale and represent the measured contrast. }
  \label{fig:dbi_elli}
\end{figure*}

The contours displayed in Fig.\,\ref{fig:dbi_elli}a represent the elliptical fits performed on the ADI image by \citet[][Table\,A.2]{Perrot2016} to measure the geometrical parameters of R1, R2, and R3. When overlaid in the polarimetric $Q_{\mathrm{\phi}}$ image in Fig.\,\ref{fig:dbi_elli}b, the elliptical fits differ mostly in the northern part, where they are largely unconstrained in the former total intensity observations (Fig.\,\ref{fig:dbi_elli}a). Also, the ringlets do not look as sharp as in total intensity. This is likely due to the ADI technique itself, which is known to impact the local shape and  photometry of extended objects \citep{Milli2012}. DPI is not affected by such a bias and, in this particular case, it is certainly more reliable in constraining the observed morphology.

\begin{figure}
  \centering
  \includegraphics[width=0.4\textwidth]{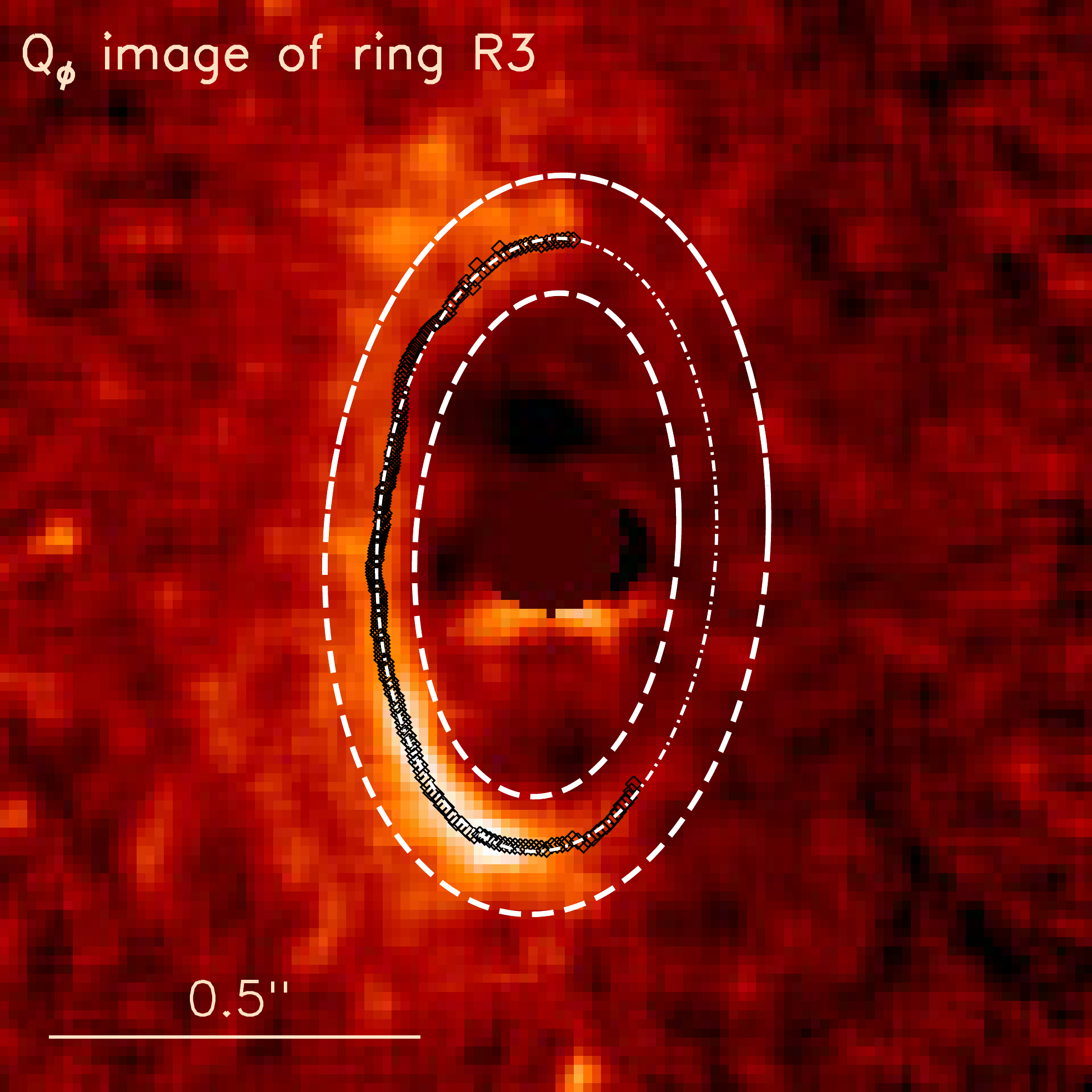}
  \caption{Elliptical fit (dot-dashed white line) to the spine of the ring R3 (black diamonds) as measured within the elliptical mask (dashed white lines).}
  \label{fig:dpi_fit}
\end{figure}

To better constrain the geometry of the ring R3, we isolated the structure by imposing an aperture on the $Q_{\mathrm{\phi}}$ image encompassing only R3. We used a full elliptical mask as an aperture with semi-major and semi-minor axes of 0.35$''$ and 0.18$''$ for the inner ellipse and 0.48$''$ and 0.27$''$ for the outer ellipse as shown in Fig.\,\ref{fig:dpi_fit} in white dashes. The same aperture is used in Sect.\,\ref{s:model} in the context of the modeling. The spine of the ring is measured as for instance in \citet{Boccaletti2013} by applying a Gaussian fit to each  direction (steps of 1\,$^{\circ}$). We used a non-linear least squares algorithm to fit an non-centered ellipse, and obtained the following parameter values: $i=56.5\pm0.7^{\circ}$, $PA= 356.8\pm0.4^{\circ}$, $x_\mathrm{off}=0.2\pm0.2$\,au, $y_\mathrm{off}=0.6\pm0.2$\,au. The inclination and PA are in close agreement with \citet{Perrot2016}. 

These results are broadly consistent with previous observations with GPI. \citet{Bruzzone2020} reported a diffuse ring between 0.4$''$ and 0.9$''$ in the H-band GPI polarized image, featuring a global east-west asymmetry as expected from the forward scattering of sub-micron sized grains. They also report the observed south-eastern brightness difference that is seen in Fig.\,\ref{fig:dbi_elli}b, particularly obvious after the subtraction of a symmetrical model \citep[Fig. 8, ][]{Bruzzone2020}. They interpreted this excess flux in the South as an arc-like over-density along the southern part, in reminiscence of the other arc-like structures observed at larger separation \citep{Clampin2003}. In their modeling approach, they did not consider the ringlets individually. As a result, they derived rather unconstrained values for the inclination ($60\pm 10^{\circ}$) and $PA$ ($5\pm10 ^{\circ}$, dashed light yellow ellipse in Fig.\,\ref{fig:dbi_elli}b), which stand in contrast with the accuracy of our measurements to a certain extent.

%
%
%
%

%
%
%
%

%
%
%
%
\section{Modeling of the innermost ring}
\label{s:model}
\subsection{General assumptions}

It has been suggested in \citet{Perrot2016} that the observed intensity distribution of R3 (Fig.\,\ref{fig:hd141_fig}d) cannot be explained only by the light scattering properties of the dust of a homogeneous ring-like disk given the east-west inclination, thus arguing for an
azimuthal variation of the dust density.

In the polarimetric image (Fig.\,\ref{fig:hd141_fig}b), most of the disk intensity is located to the east, which confirms that the eastern side is the front side of the disk, if the scattering is predominantly forward. The lack of polarized and total intensity to the north further indicates probable depletion of dust. Thus, the DPI observation complements the ADI data in breaking the degeneracy due to the phase function of the dust, and confirms that the dust density of R3 is likely not azimuthally constant. 

We used GRaTeR, a radiative transfer code \citep{Augereau1999b}, to generate synthetic scattered-light images of the ring R3, which we assumed to be optically thin \citep{Thi2014}. The scattering phase function (SPF), which determines the amount of light scattered at a particular scattering angle ($\theta$) for total intensity observation is given by the Henyey Greenstein \citep[HG,][]{Henyey1941} function ($f_{I,\theta}$). We consider Rayleigh approximation to account for the angular dependence of the linear polarization while modeling our DPI data. It has been experimentally demonstrated that the degree of linear polarization (DOLP) as a function of scattering angles for zodiacal and cometary grains has a bell shape \citep[e.g.,][]{Leinert1976, Bertini2017, Frattin2019}, which is well approximated by the Rayleigh scattering function. Since debris disks are expected to be reservoirs of cometary grains, the assumption of representing the polarized SPF with a model following a linear combination of HG and Rayleigh approximation was adopted in the analysis of HIP79977 \citep{Engler2017} and HR\,4796 \citep{Milli2019}. Thus, for our data as well, we assume that the polarized SPF is expressed by: 

\begin{equation}
\label{Eq:PF}
f_{P,\theta} = f_{I,\theta} \cdot \frac{1-cos^2 (\theta)}{1+cos^2(\theta)}. 
\end{equation}
Previous works \citep{Perrot2016, Bruzzone2020} have assumed that the density function varies radially ($R$) and vertically ($Z$). Here, we considered an additional component ($A$) to parameterize the azimuthal variation. The density function $n$ is expressed  as follows: 
\begin{equation}
\label{Eq:DF}
n(r,z,\varphi)= n_0 \cdot R(r)\cdot Z(r,z)\cdot A(\varphi),
\end{equation}
where $n_0$ is the peak of the density function, $r$ the radial dimension in the disk plane, $z$ the vertical dimension perpendicular to the disk plane, and $\varphi$ the azimuthal dimension in the disk plane. 

In addition, we considered two approaches to explore the parameter space:\,grid-search and Monte-Carlo Markov Chain \citep[MCMC, emcee package][]{emcee2013} to derive the best model that could explain the observed data. This is to ensure that we obtain the best solution with well-constrained posterior probability distributions and also to compare the reliability of these widely used methods.

%
%
%
%
\subsection{Azimuthal density distribution}
\label{s:density}

In the effort to account for the origin of  azimuthally variable dust density, the two leading scenarios, namely, the collisional breakup of a large planetesimal as well as the interaction with gas (further discussed in Sect.\,\ref{s:discuss}), make relatively similar predictions regarding the azimuthal geometry of the ring, at least from a qualitative point of view. \citet{jackson2014} have indeed shown that in the aftermath of a planetesimal breakup, the dusty ring that forms peaks at the location of the breakup, remains relatively localized in an angular domain (whose width depends on the size of the parent body) around this peak, and is minimum on the opposite side. As for the gas-related scenario, we would expect the dust to follow, to a certain extent, the azimuthal profile of the gas, which for the inner regions of the HD\,141569 system, has been shown to be concentrated in a particular range of angular separations and position angles around this peak \citep{Difolco2020}.

In principle, the density distribution associated to the afore mentioned scenarios should be modelled in 2D, both in the radial and azimuthal directions, but the S/N of the polarimetric image does not allow us to reliably explore the radial density distribution beyond the region where the R3 ring peaks. Thus we took a conservative approach and considered a 1D prescription. As a consequence, and for the sake of simplicity, we established a hypothesis  that assumes the dust density peaks at a given angular position in the ring and that the region of high-density has a given azimuthal width. We note that, in the simulations of \citet{jackson2014}, the dust density conveniently follows a Lorentzian function \citep[see Fig.9a of][]{jackson2014}. We thus implemented an azimuthal density gradient in the GRaTeR code such that the typical dust density function is multiplied by a Lorentzian expressed as:
\begin{equation}
A(\varphi)=\dfrac{d}{1+(\dfrac{\varphi-\varphi_0}{w})^2},
\label{eq:density}
\end{equation}
where $d$ is the maximum density, $\varphi_0$ is the position angle of the collision site, and $w$ is the width of the Lorentzian function. 

Throughout the paper, we fixed the position angle of the disk $PA$ to $356^{\circ}$ and the disk aspect ratio to $0.02$. While the later is chosen as a guess, \citet{Olofsson2020} showed that it can affect the measurement of phase function depending on the inclination of the disk. Therefore, to model the ring R3 we considered the following seven free parameters, which control the quantities $R$, $Z$ and $A$:

\begin{enumerate}
\item The anisotropic scattering factor, $g$ in the HG function. Forward scattering by the dust particles favours the value of $g$ between 0 and 1 while the backward scattering renders $g$ between 0 and -1. Here, $g$ refers to $g_\mathrm{pol}$ for polarimetric intensity and $g_\mathrm{total\_int}$ for total intensity,
\item The radius $r_0$ (au) of the disk where the dust density peaks,
\item The inclination $i$ ($^\circ$) of the disk,
\item The radial density profiles defined by two power law indexes $\alpha_\mathrm{in}$ and $\alpha_\mathrm{out}$,
\item The azimuth of the peak in dust particle density, $\varphi_0$ ($^\circ$),
\item The angular width of the Lorentzian function $w$ ($^\circ$).
\end{enumerate}

%
%
%
%
\subsection{Modeling of the polarized intensity}
\label{s:Pol_model}

To justify the need of introducing an azimuthal density variation in the modeling, we first performed a test by choosing $g_\mathrm{pol}$ as the only free parameter and considering that the density is constant along the ring. Here, our objective is to show that whatever phase function we assume, it does not emulate the north-south asymmetry. 

\begin{figure}
\includegraphics[width=0.24\textwidth]{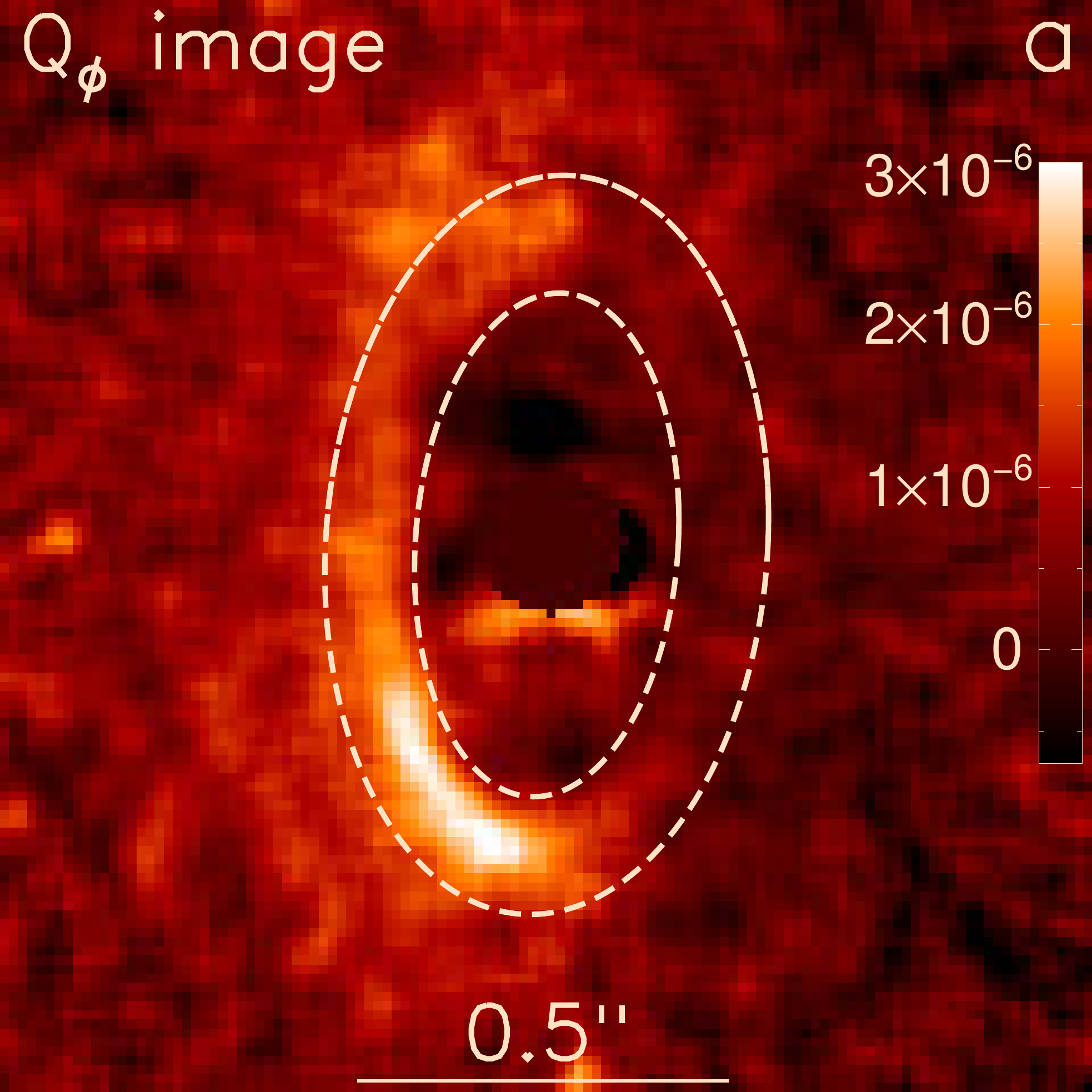} 
 
\includegraphics[width=0.24\textwidth]{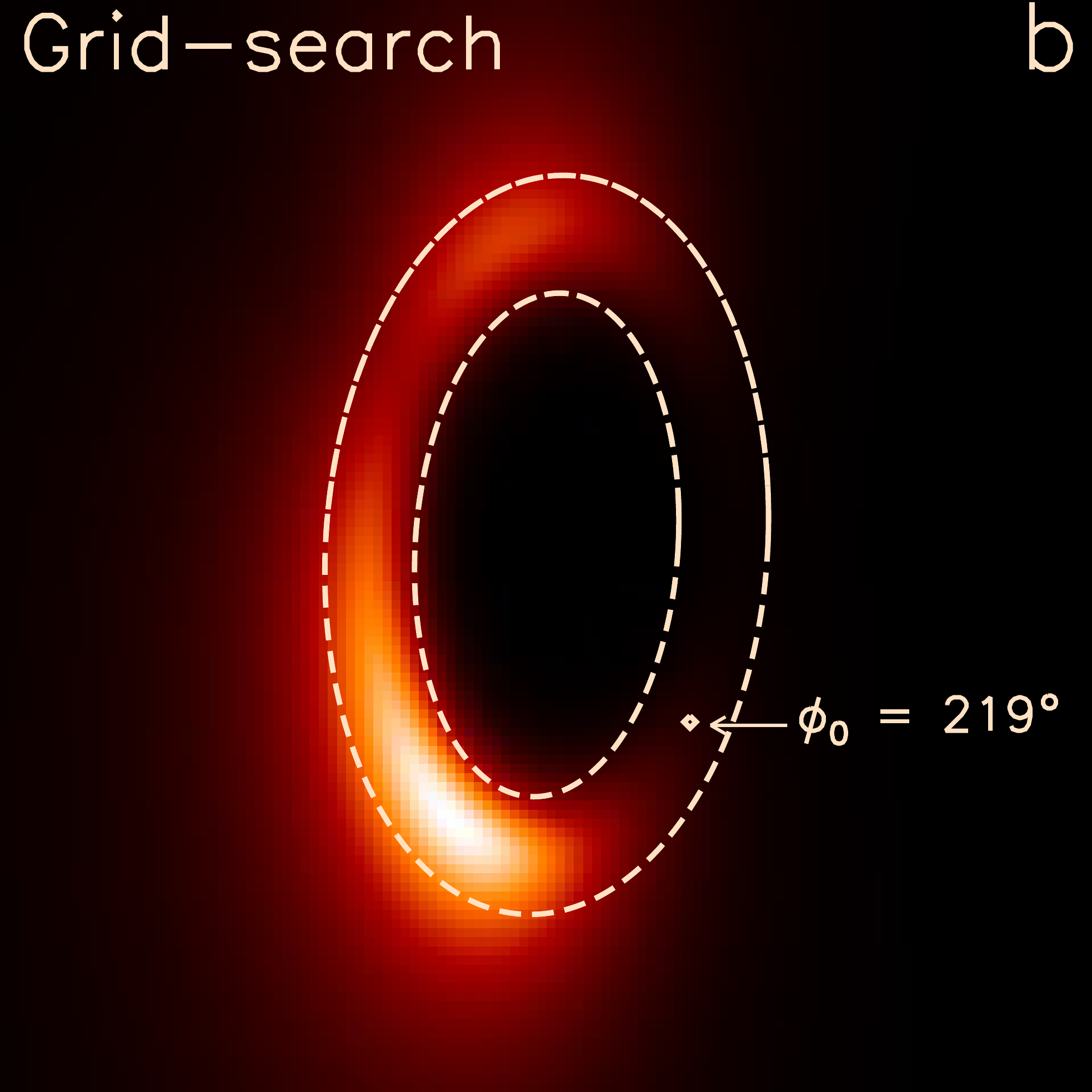}
\includegraphics[width=0.24\textwidth]{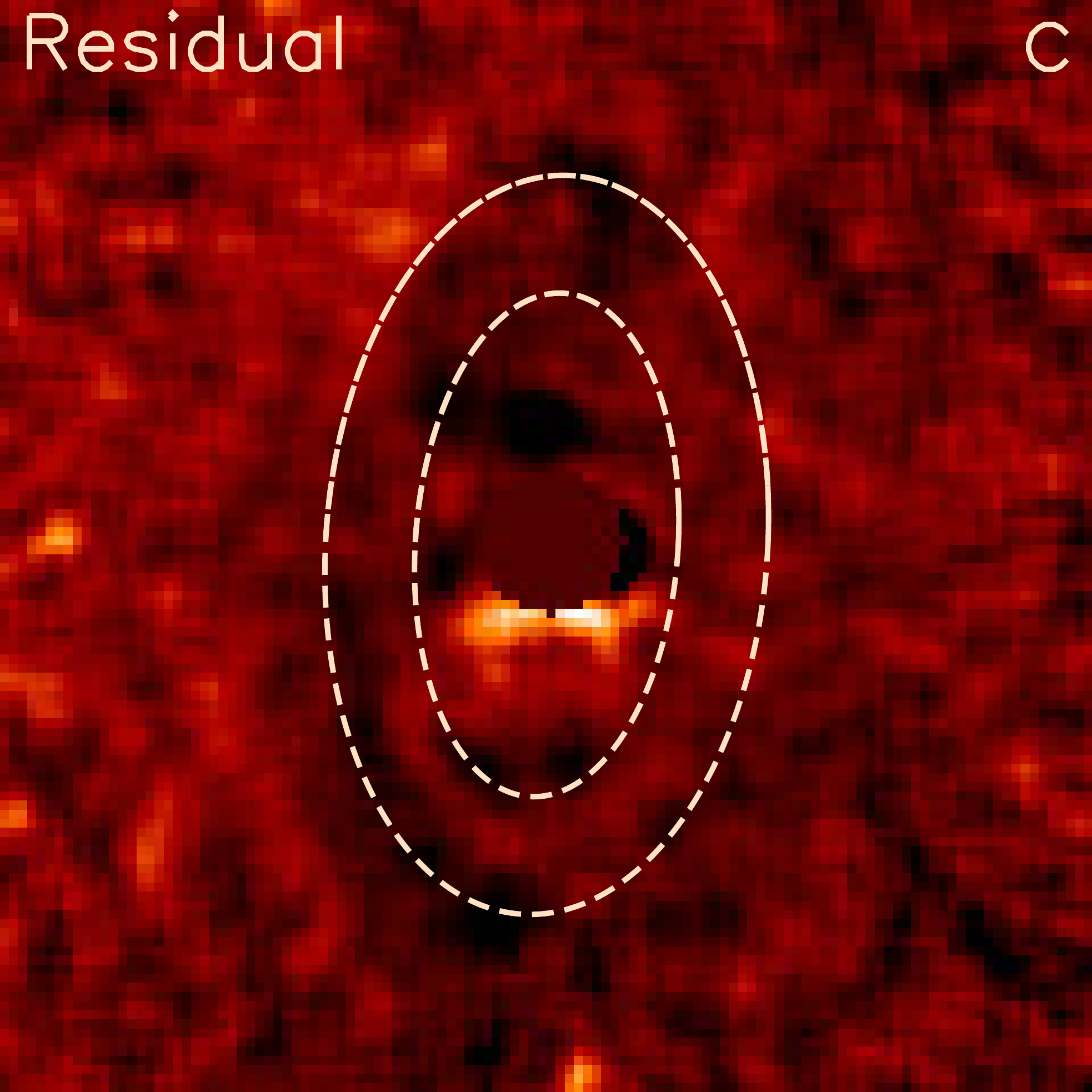}
 
\includegraphics[width=0.24\textwidth]{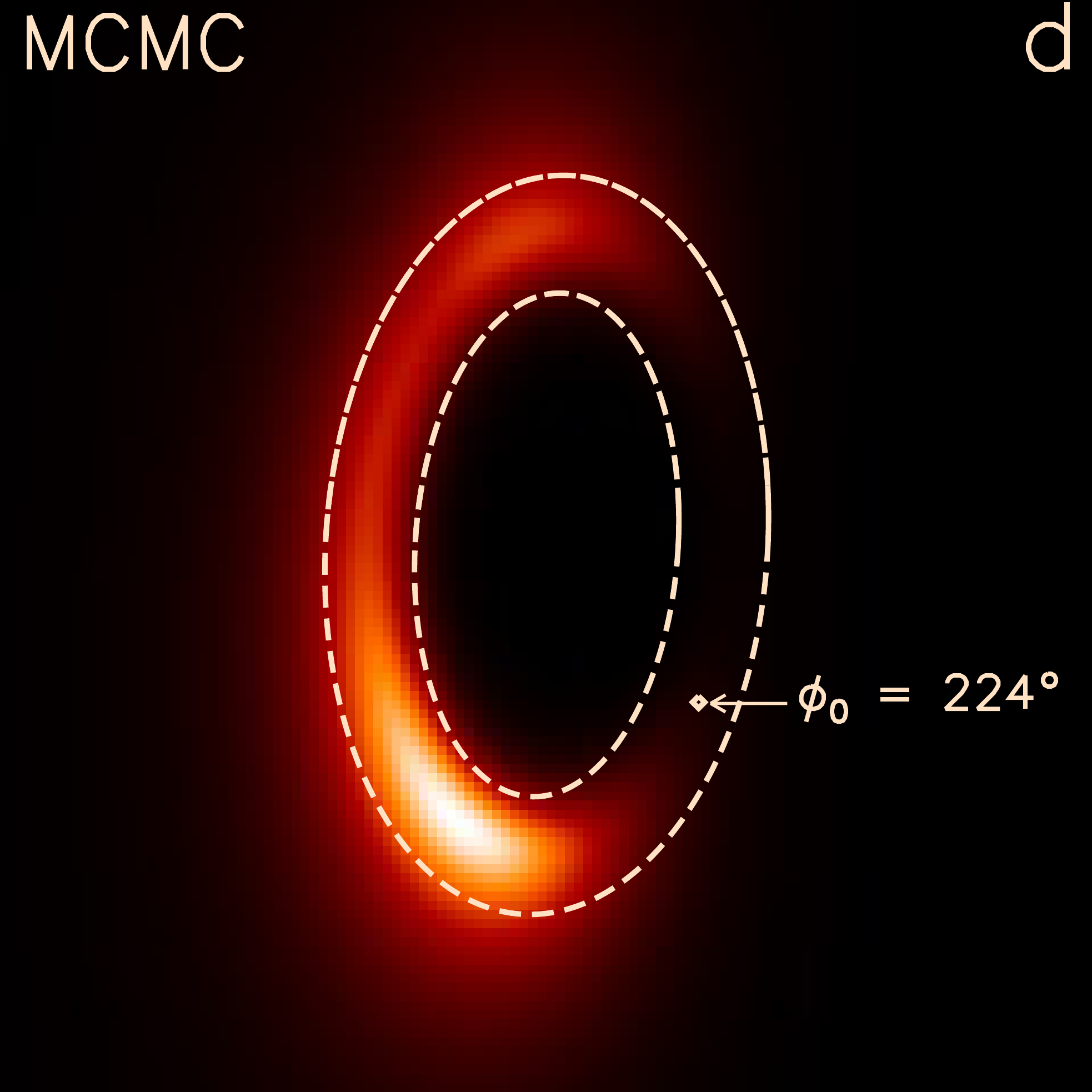}
\includegraphics[width=0.24\textwidth]{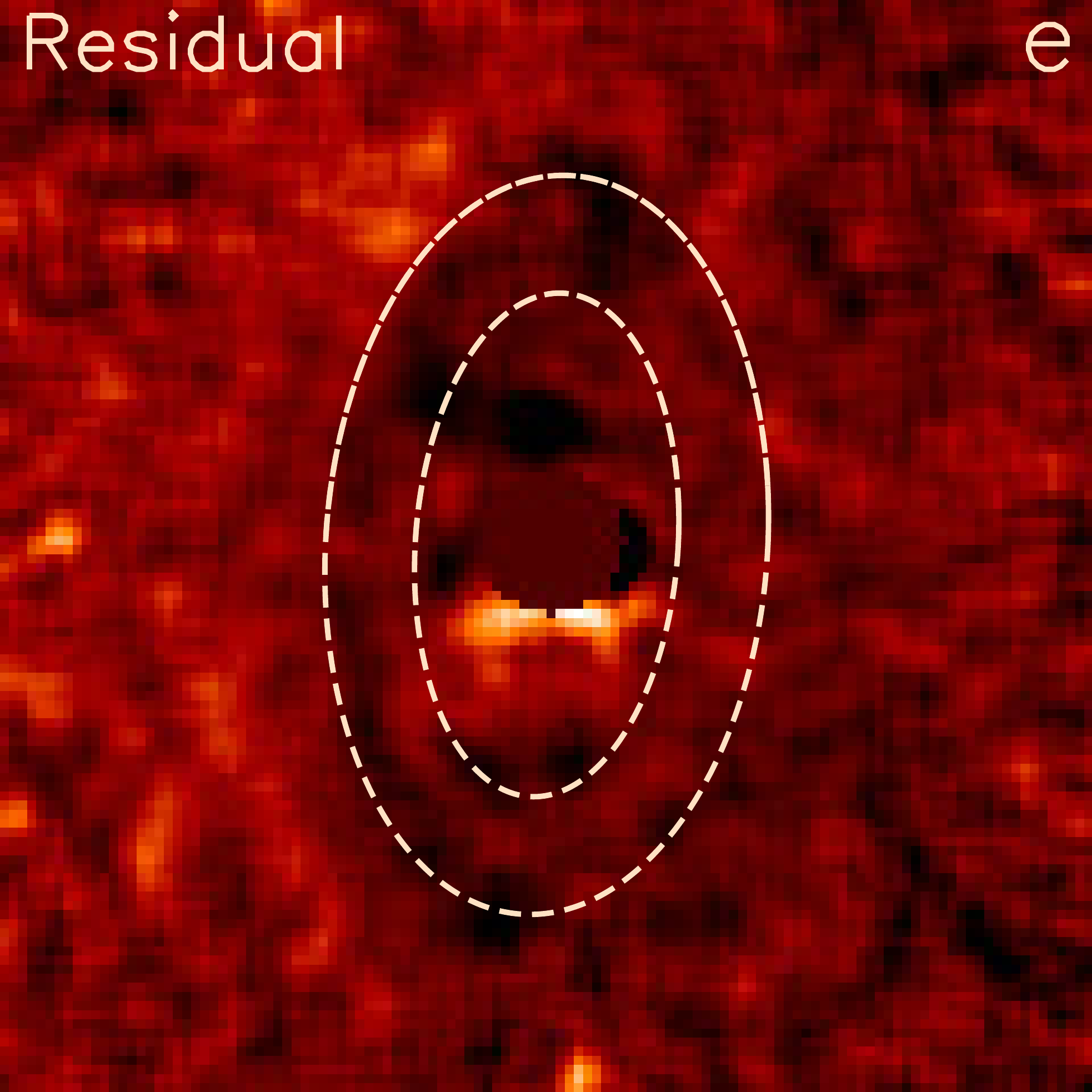} 

\includegraphics[width=0.24\textwidth]{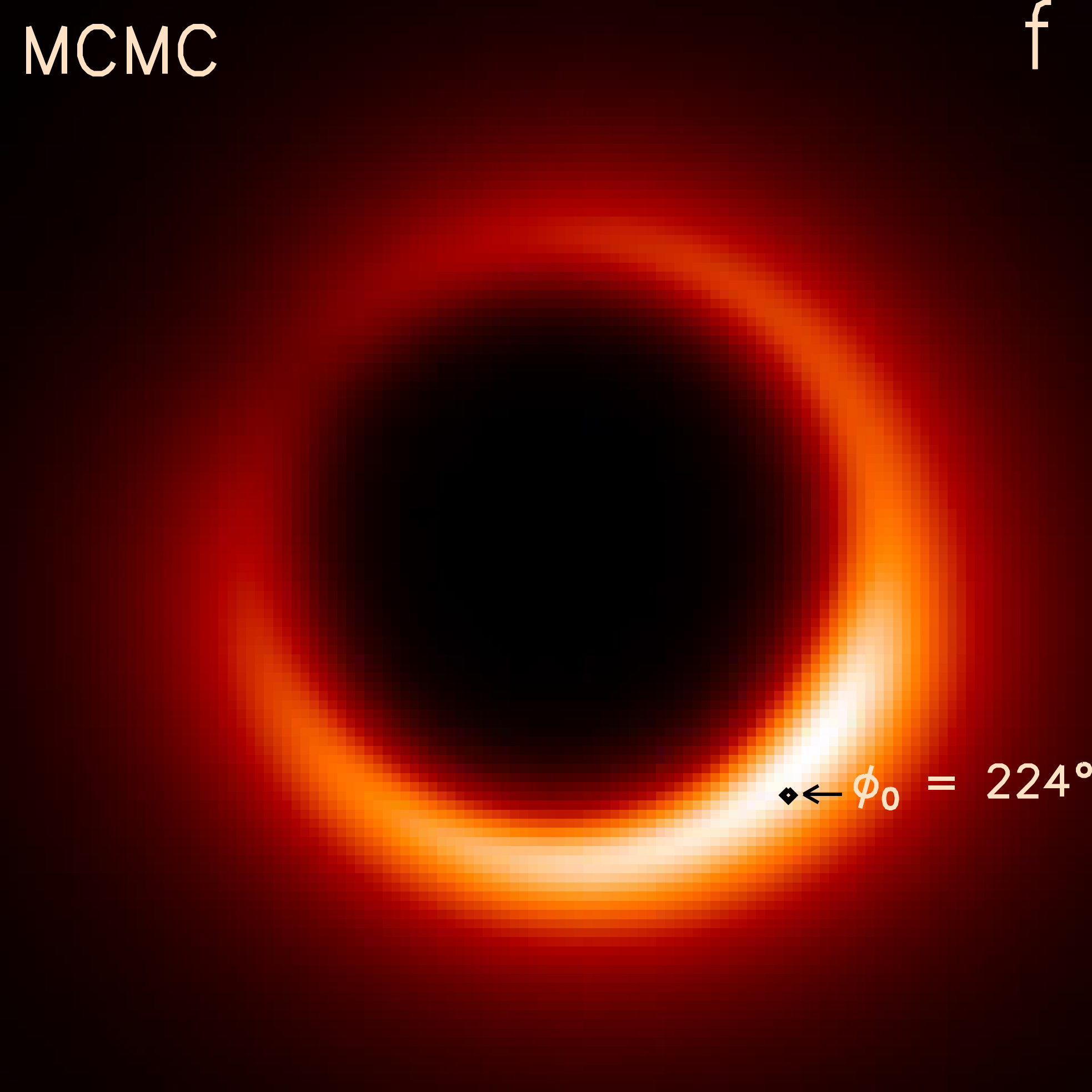} 
  \caption{DPI data is displayed as a reference in the top row, together with the best-fit models (left) and the corresponding residuals (right) resulting from the grid-search (second row), and the MCMC (third row) analysis methods. Table\,\ref{tab:bestParam_DPI} provides the parameters of the model for these two cases. The ellipses in dashed white lines show the aperture within which the reduced $\chi^{2}$ minimization has been performed. The bottom row shows the deprojected density function corresponding to the MCMC best-fit model.}
   \label{fig:dpi_modeling}
\end{figure}

We simulated a series of 50 GRaTeR models from $g_\mathrm{pol}$=0.1 to $g_\mathrm{pol}$=0.99, and set the other four parameters to: $r_0= 45$\,au, $i=57^{\circ}$, $\alpha_\mathrm{in}=20,$ and $\alpha_\mathrm{out}=-20$, as defined in \citet{Perrot2016}. To make a comparison with the DPI image, the models were convolved with the PSF and scaled globally in intensity to the data, which were binned to $2\times2$-pixels. The best-fit model is determined with the reduced $\chi^{2}$ minimization, which is measured within an aperture encompassing the ring R3 (Fig.\,\ref{fig:dpi_modeling}a). We found the best $g_\mathrm{pol}$ value to be 0.54 (reduced $\chi^{2}$=1.6). However, as shown in Fig.\,\ref{fig:sameg}a, the best-fit model with this method failed to reproduce the north-south asymmetry observed in the data (Fig.\,\ref{fig:dpi_modeling}a), and produced excess intensity to the north as shown in the residual (Fig.\,\ref{fig:sameg}b). 

Then, we implemented the modified density function in GRaTeR and created 144000 synthetic models. The values of the radius and the inclination remained fixed as before. The rest of the five free parameters ($g_\mathrm{pol}$, $\varphi_0$, $w$, $\alpha_\mathrm{in}$ and $\alpha_\mathrm{out}$) were explored with a coarse grid-spacing. The second column of Table\,\ref{tab:bestParam_DPI} shows the range of the inner and outer bounds of the free parameters, total number of values explored, and the best-fit values and error bars obtained. The data, the best-fit model from the grid-search approach, and the residuals are shown in Fig.\,\ref{fig:dpi_modeling}a,\,b, and c, respectively.

The best-fit model favors a high value of $g_\mathrm{pol}$, which is indicative of a strong forward scattering, and a density peaking at $\varphi_0=219\pm\,18^{\circ}$, which is the southwest direction in the image. The width of the density function is rather large expanding on $145\pm\,43^{\circ}$ (errors are computed according to \citet{Bhowmik2019}). Counter-intuitively, this first analysis suggests that the peak of density is on the southwest part of the disk, opposite to the observed intensity peak with respect to the major axis. Therefore, the large $g_\mathrm{pol}$ value (very forward scattering dust) and an azimuthally dependent density profile are necessary to reproduce the intensity peak positioned on the southeast as observed in the polarimetric image.

The best values for the two parameters describing the slope of the radial density function, $\alpha_\mathrm{in}$ and $\alpha_\mathrm{out}$, correspond to a sharp ring inward and a shallower profile outward. As a comparison, \citet{Perrot2016} found that the ring was very steep both inwards and outward when observed in total intensity. However, \cite{Perrot2016} did not systematically explore  several values of $\alpha_\mathrm{in}$ and $\alpha_\mathrm{out}$, but instead tested only three configurations with  $\alpha_\mathrm{in}=-\alpha_\mathrm{out} =[5,10,20]$. In addition, the differences with \cite{Perrot2016} can also arise from the self-subtraction as long as the ADI method acts as a high-pass filter \citep{Milli2012}, with a tendency to remove the faint signal produced by the diffuse dust population extending outside of the main ring. In this particular case, DPI allows us to better constrain the actual radial width of this dust ring. 

\begin{figure*}
    \includegraphics[width=\textwidth]{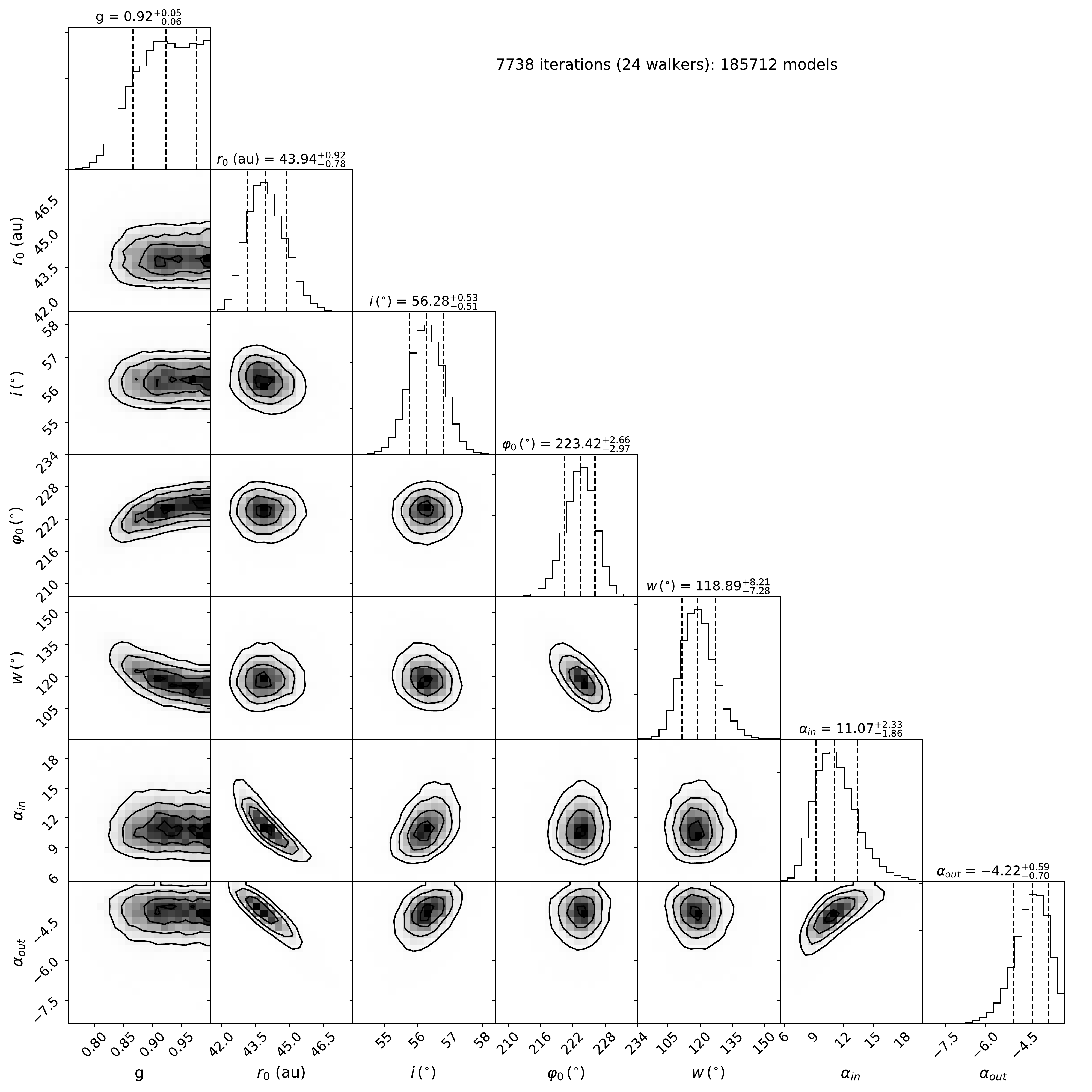}
    \caption{Posterior probability distribution of the free parameters used to constrain the polarized intensity. Vertical lines represent the 16$^{th}$ ($-1\sigma$), 50$^{th}$ (median value), and 84$^{th}$ ($+1\sigma$) percentiles of the samples in the distributions which are used to derive uncertainties.}
   \label{fig:DPIcornerPlot}
\end{figure*}

Given the outcome of the grid search in terms of the presence of a density enhancement at the opposite of the polarized intensity peak, we embarked on a more detailed exploration of the parameter space with MCMC to check the robustness of the solution. In this case, we considered seven free parameters with 24 walkers. We chose priors (shown in Table\,\ref{tab:bestParam_DPI}) that were exactly the same as the range of parameters in the grid search. The initial guess values of the free parameters were chosen to be close to the best-fit parameters provided by grid-search. Even though the length of the burn-in phase was 500 iterations, we kept running the MCMC with a total of 7738 iterations to roughly compute  the equal number of models as generated in the grid search. Similar results were obtained with a mean acceptance fraction of 0.45, indicating a converging solution \citep{Gelman1992}. Figure\,\ref{fig:DPIcornerPlot} shows the posterior probability distributions of all the free parameters, which were computed using Python's corner module \citep{corner2016}. We derived uncertainties based on the 16$^{th}$ ($-1\sigma$), 50$^{th}$ (median value), and 84$^{th}$ (+1$\sigma$) percentiles of the samples in the distributions, which are plotted in vertical lines in Figure\,\ref{fig:DPIcornerPlot}. The median and the error bars for each parameter are shown in the third column of Table\,\ref{tab:bestParam_DPI}, which are found to be consistent with the grid-based approach, particularly with regard to the fact that the density is peaking at the southwest ($\varphi_{0}=223.4^{\circ+2.7^\circ}_{-3.0^\circ}$) and that the anisotropic scattering factor is very high ($g_\mathrm{pol}=0.92^{+0.05}_{-0.06}$). Figures\,\ref{fig:dpi_modeling}d and \ref{fig:dpi_modeling}e show the model and the corresponding residual, which also look similar to the solution suggested by grid-search (Figs.\,\ref{fig:dpi_modeling}b, and\,c). To depict the dust distribution in the ring inferred by the model, we provide a deprojected view of the ring density variation in Fig.\,\ref{fig:dpi_modeling}f, which is convolved with the PSF.

We compared these geometric parameters with those obtained with GPI observations in \citet{Bruzzone2020}. They found an inclination of $60^\circ\,\pm\,10^\circ$, which is compatible with our findings. Radial extension is also consistent with a radius $r_0=44^{+8}_{-12}$\,au and a low density slope outside the main ring $\alpha_\mathrm{out}$ to be $-1.0^{+0.5}_{-1.0}$. This is much shallower than the value found in our analysis ($\alpha_\mathrm{out}=-4$). However, the outer slope relies on the sensitivity of a specific observation and therefore depends on the observing conditions. Moreover, we also restrained the fit to the location of R3 to avoid mixing R2 and R3, which inevitably would result in a shallower slope. 

We also performed a sanity check by running another MCMC with different initial conditions of the parameter space. The goal was to evaluate the possibility of a solution around the first intuitive interpretation, where both the intensity and the density peak lie in the southeast of the polarized intensity image. We started an MCMC chain of 70 walkers by considering the initial value of free parameters $g_\mathrm{pol}$, $r_0$, $i$, $\varphi_0$, $w$, $\alpha_\mathrm{in}$, and $\alpha_\mathrm{out}$ to be 0.3, 48\,au, 58$^\circ$, 100$^\circ$, 18$^\circ$, 15, and\,-4, respectively. We definitely observed a local minima around this position. The model corresponding to this local minima (reduced $\chi^{2}$=1.1) and its residual are shown in Fig.\,\ref{fig:localMin}. This model does not visually show an obvious difference with the one previously obtained (shown in Fig.\,\ref{fig:dpi_modeling}d). However, the MCMC  eventually clearly disfavors this local minima and converges toward the former best solution, where the density peak is in the southwest (same MCMC results were obtained as shown in last column of Table\,\ref{tab:bestParam_DPI}).

%
%
%
%
\begin{table*}[th!]
\caption{Free parameters constrained with the grid-search and MCMC methods when modeling the polarimetric data.} 
\begin{center}
\ra{1.5}
\begin{tabular}{l|ccc|ccc}\hline
\hline
&  \multicolumn{3}{c|}{grid-search} & \multicolumn{3}{c}{MCMC, 24 walkers}  \\
&  \multicolumn{3}{c}{(144000 models)} & \multicolumn{3}{|c}{(185712 models)}  \\
Parameters & Range & Number & Result & Prior & Initial value & Result  \\ 
 & &of values & & & &      \\ \hline 
 $g_\mathrm{pol}$ & [0, 0.99] & 10 & 0.90$\pm$0.09 & [0, 0.99] & 0.7 & 0.92$_{-0.06}^{+0.05}$ \\ 
$r_0$ (au) & 45 (fixed) & 1 & 45 & [40, 55] & 48 & 43.9$_{-0.8}^{+0.9}$ \\
 $i\,(^\circ)$ & 57 (fixed) & 1 & 57 & [50, 60] & 58 & 56.3$_{-0.5}^{+0.5}$ \\
$\varphi_{0}\,(^\circ)$ & [90, 300] & 15 & 219$\pm18$ & [90, 300] & 245 & 223.4$_{-3.0}^{+2.7}$ \\
%
$w\,(^\circ)$ & [10, 200] & 15 &145$\pm$43 & [10, 200] & 60 & 118.9$_{-7.3}^{+8.2}$  \\
$\alpha_\mathrm{in}$ &  [3, 20] & 8 & 14.0$\pm$3.9 & [3, 20] & 5 & 11.1$_{-1.9}^{+2.3}$   \\
$\alpha_\mathrm{out}$ &  [-3, -20] & 8 & -4.0$\pm$1.2 & [-3, -20] & -10 & -4.2$_{-0.7}^{+0.6}$ \\ 

\hline%

$\chi^{2} $ &  &  & 0.57 & & & 0.53\\
\hline%
\end{tabular}
\end{center}
\label{tab:bestParam_DPI}
\end{table*}

%
%
%
%

%
%
%
%

%
%
%
%
\subsection{Modeling of the total intensity image}
\label{s:unpol_model}

\begin{figure*}
  \centering
\includegraphics[width=0.24\textwidth]{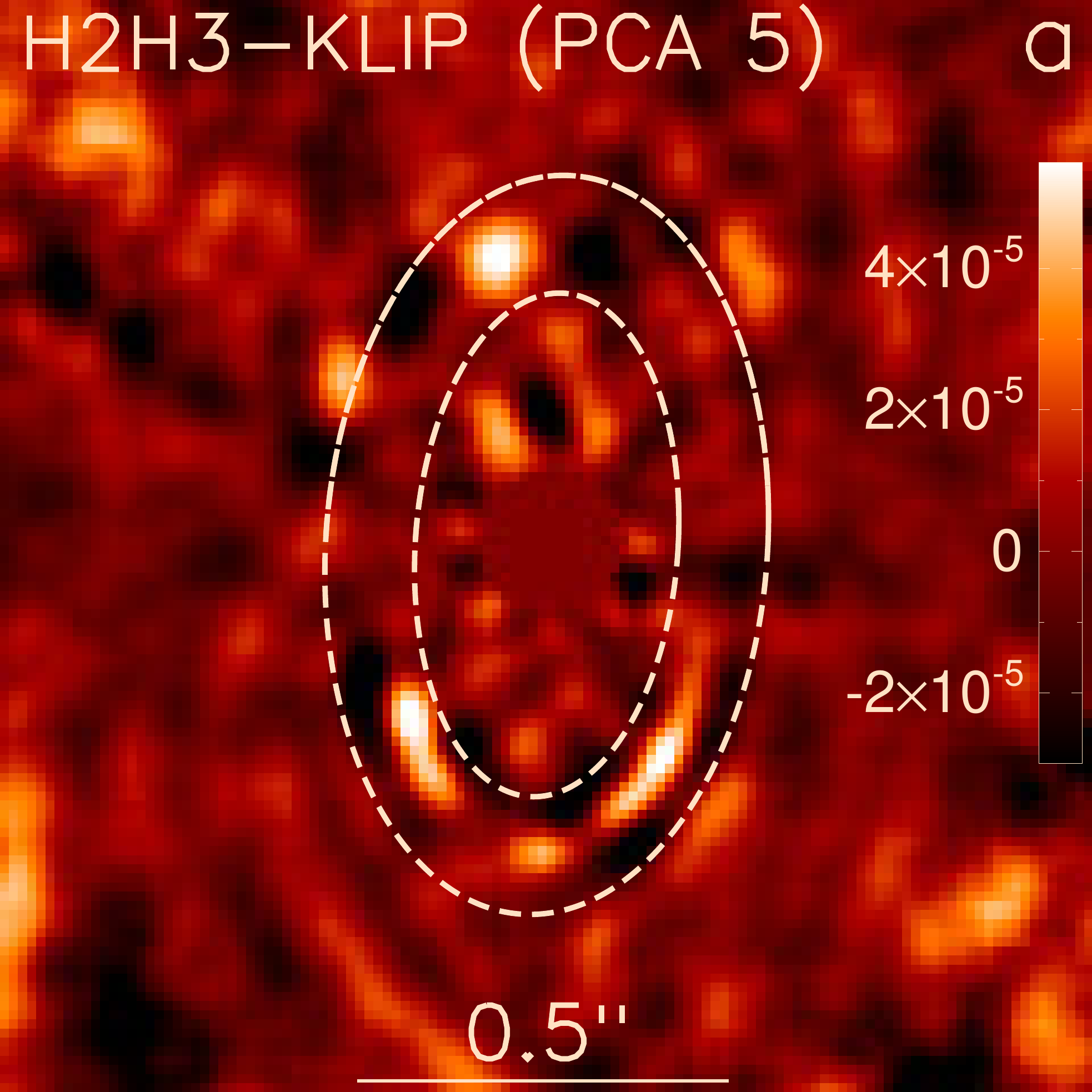} 
\includegraphics[width=0.24\textwidth]{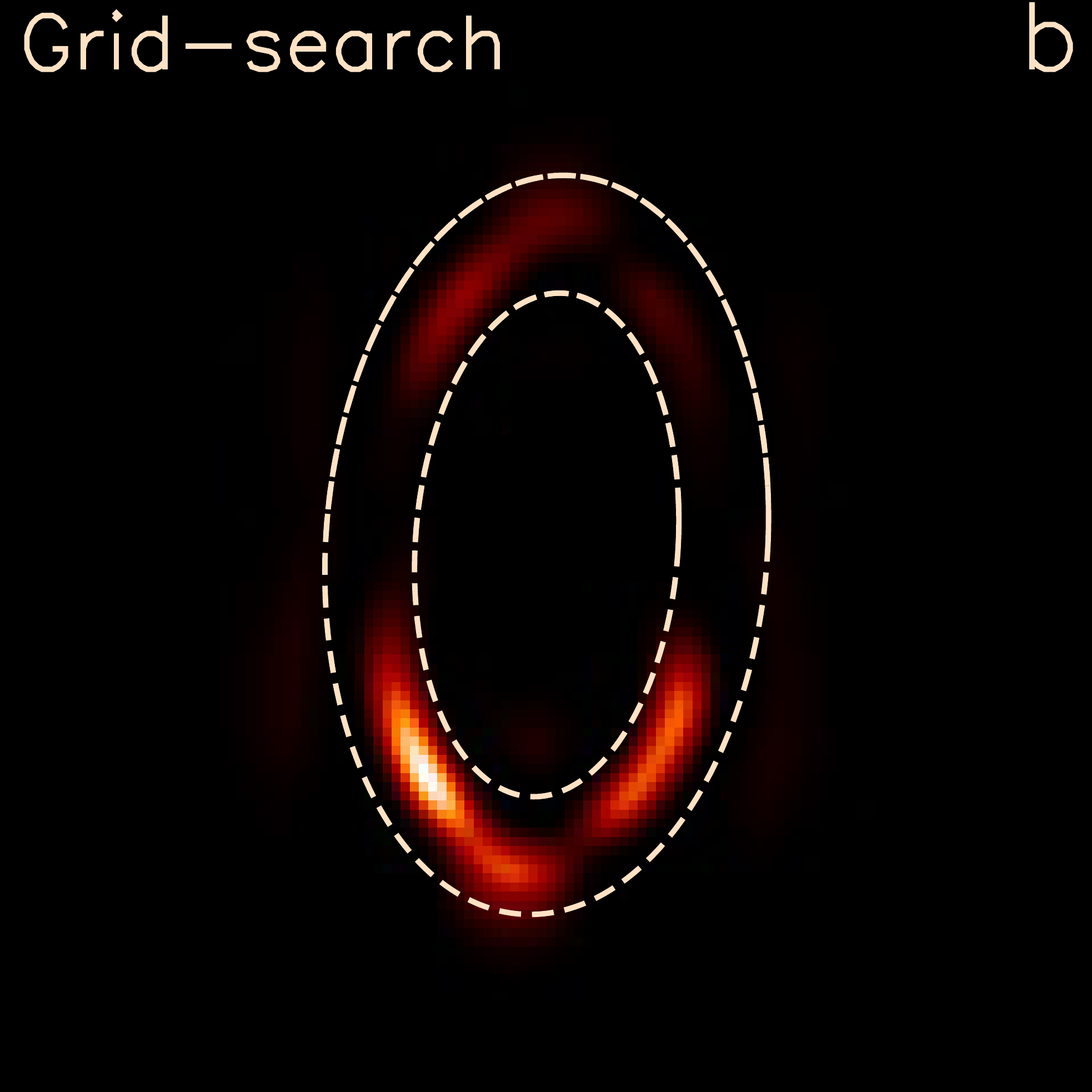}
\includegraphics[width=0.24\textwidth]{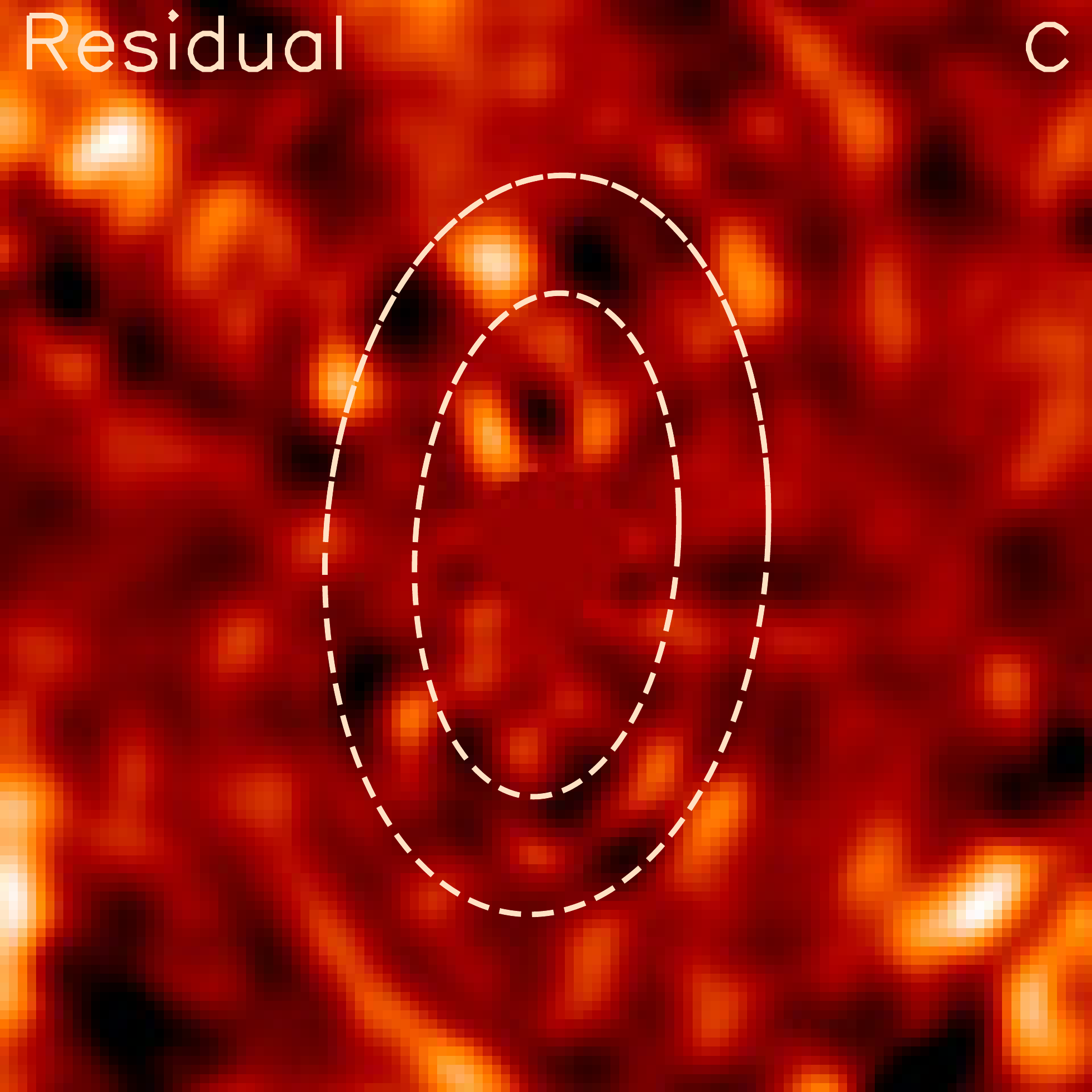}
  \caption{ADI-reduced intensity image (left) compared to the best-fit model based on grid-search (middle) and residuals (right). }
  \label{fig:dbi_grid_model}
\end{figure*}

In this section, we use a similar modeling analysis on the total intensity data reduced using a KLIP-ADI algorithm with KLIP truncated at five modes. Given the lower quality of the data, which is further heavily self-subtracted by the post-processing algorithms, we chose only three free parameters ($g_\mathrm{total\_int}$, $\varphi_{0}$, $w$) out of the total seven parameters mentioned in Sect.\,\ref{s:density}, with an objective to check whether the total intensity data were compatible with a non-uniform azimuthal density. We started our first analysis by fixing the values of the rest of the parameters based on \citet{Perrot2016}, such that $r_{0}$=45\,au, $i$=57$^\circ$, $\alpha_\mathrm{in}$=20 and $\alpha_\mathrm{out}$=-20. We generated 206400 synthetic forward models following the ADI processing \citep[e.g.,][]{Bhowmik2019}, which took the self-subtraction bias into account similarly to the total intensity data (Fig.\,\ref{fig:hd141_fig}d). A grid of models with fine sampling was generated with GRaTeR, while the reduced $\chi^{2}$ minimization provided the best-fit values in Table\,\ref{tab:bestParam_ADI}. Figures\,\ref{fig:dbi_grid_model}a,\,b, and\,c show the total intensity  data, the best-fit model, and the corresponding residual. We also used an MCMC analysis by choosing the best-fit values from the grid-search as the initial guess while maintaining the same prior as in DPI modeling in Table\,\ref{tab:bestParam_DPI}. Both the grid search and MCMC simulations (burn-in phase of 100 iterations) reached a similar conclusion: the density peak $\varphi_0$ is roughly located at around 230$^\circ$, which is consistent with the DPI modeling results. On the contrary, the width of the density function,  $w\,\sim\,40^\circ$ is much smaller than in DPI. This could be attributed to the strong self-subtraction effects because there is less flux along the minor axis.

We performed another test with MCMC and considered the values of fixed parameters similar to the DPI modeling results, namely,\,$r_{0}$=44\,au, $i$=56.3$^\circ$, $\alpha_\mathrm{in}$=11.1 and $\alpha_\mathrm{out}$=-4.2. The last column of Table\,\ref{tab:bestParam_ADI} shows the median results and 1$\sigma$ uncertainty, which although marginally consistent with former tests, still confirms the peak of density in the southwest in both DPI and ADI data. We also double checked our outcome by running MCMC with less aggressive ADI parameters (three modes of the KLIP-PCA instead of five). The resulting analysis is of similar quality than that involving five modes ($\chi^{2}=3.1$ instead of 3.7), but it is still marginally compatible to the values in the last column in Table\,\ref{tab:bestParam_ADI} ($g=0.14\pm0.02$, $\phi_0=228.5\pm3.9$, $w= 69.8 \pm6.0$). As a result, the modeling of the total intensity image seems quite sensitive to the input data, although the location of the density peak is still very consistent within the error bars.

Because of the  high impact of the self-subtraction on the ADI dataset, we can assume that the former MCMC results on total intensity data are more prone to biases than in DPI in what concern the geometrical parameters. For this reason, we performed one more MCMC run where the only free parameter is the $g_\mathrm{total\_int}$ value, which we can expect to vary between DPI and ADI data sets as opposed to the others. The six geometrical parameters were kept fixed according to the DPI modeling: \,$r_{0}$=44\,au, $i$=56.3$^\circ$, $\varphi_0$=224$^\circ$, $w$=117$^\circ$, $\alpha_\mathrm{in}$=11.1, and $\alpha_\mathrm{out}$=-4.2. After 1000 iterations (24 walkers), we obtained the best value of $g_\mathrm{total\_int}$ as $0.17_{-0.02}^{+0.02}$ (reduced $\chi^{2}$=3.9). This value is consistent at 3 $\sigma$ with the value ($0.23_{-0.02}^{+0.02}$) obtained in the previous test, although still in the lower range of $g_\mathrm{total\_int}$. Given the poor quality of the data (the northern side of the disk is not detected), we did not develop further this analysis.

%
%
%
%

%
%
%
%
\begin{figure}
   \includegraphics[width=8cm]{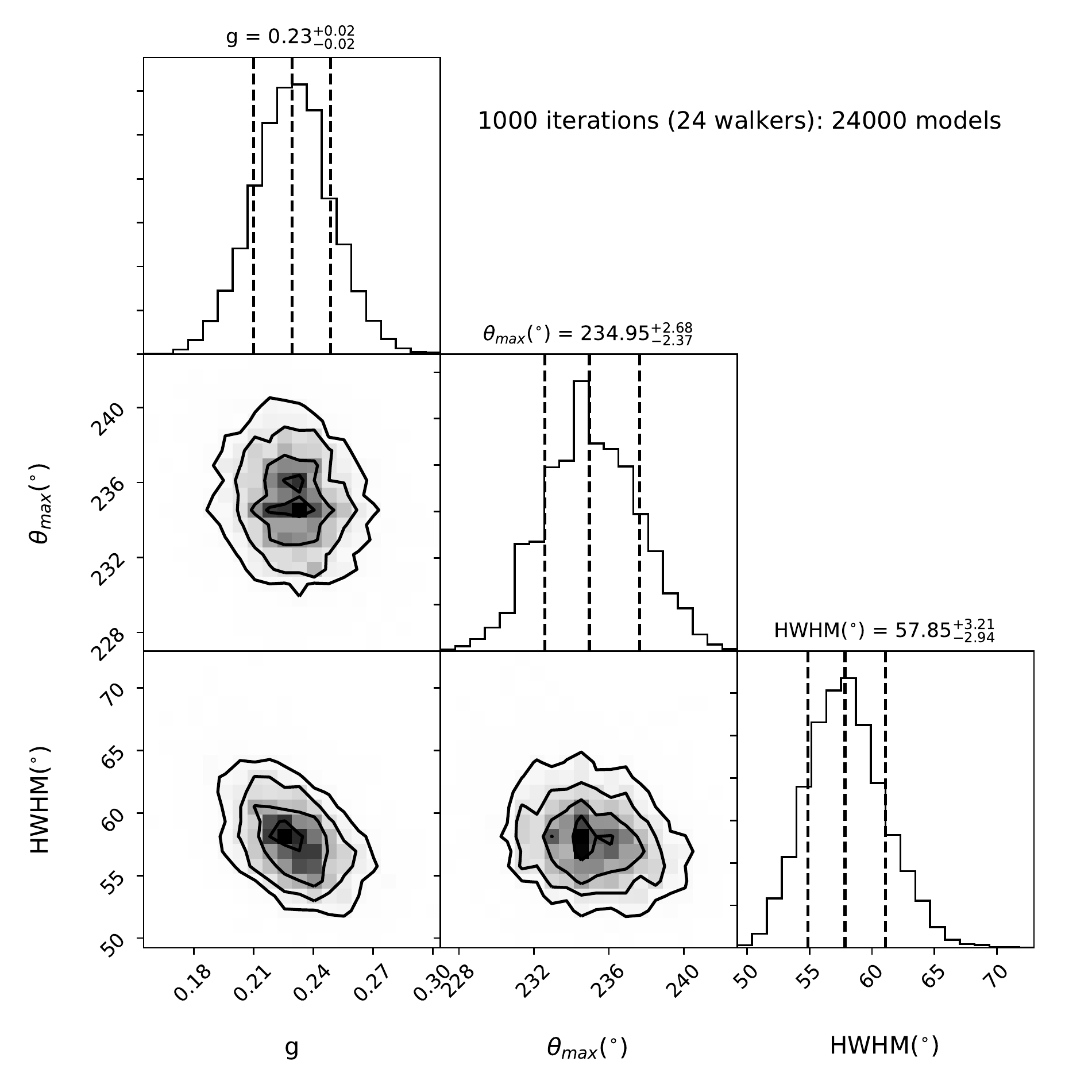}
      \caption{Posterior probability distribution of the free parameters used to constrain the total intensity data. Vertical lines represent the 16$^{th}$ ($-1\sigma$), 50$^{th}$ (median value), and 84$^{th}$ ($+1\sigma$) percentiles of the samples in the distributions which are used to derive uncertainties.}
         \label{fig:dbi_corner}
   \end{figure}

%
%
%
%
\begin{table*}[th!]
\caption{Free parameters constrained with the grid-search and MCMC methods while modeling the total intensity data. 
The second column assumes fixed parameters from \citep[][]{Perrot2016}: $r_{0}$=45\,au, $i$=57$^\circ$, $\alpha_\mathrm{in}$=20, $\alpha_\mathrm{out}$=-20, and the third column assumes the values obtained for the MCMC best-fit DPI model (Table\,\ref{tab:bestParam_DPI}): $r_{0}$=44\,au, $i$=56.3$^\circ$, $\alpha_\mathrm{in}$=11.1, $\alpha_\mathrm{out}$=-4.2.
} 
\begin{center}
\ra{1.5}
\begin{tabular}{l|ccc|ccc|cc}\hline
\hline
&  \multicolumn{6}{c|}{Fixed parameters from \citet{Perrot2016}} & \multicolumn{2}{c}{Fixed parameters from DPI model}  \\
\cmidrule{2-9}
&  \multicolumn{3}{c|}{grid-search (206400 models)} & \multicolumn{3}{c|}{MCMC (102144 models)} & \multicolumn{2}{c}{MCMC (24000 models)} \\
Parameters & Range & Number & Result  & Prior & Initial value & Result  & Initial value & Result \\ 
 & & of values &  & &  & & &     \\ \hline  
%
$g_\mathrm{total\_int}$ & [0, 0.99] & 100 &  0.46$\pm$0.14 & [0, 0.99] & 0.48 & 0.42$_{-0.04}^{+0.04}$ & 0.3 & 0.23$_{-0.02}^{+0.02}$\\ 
%
$\varphi_{0}\,(^\circ)$  & [90, 300] & 43 & 230.2$\pm$8.3  & [90, 300] & 230 & 234.0$_{-2.6}^{+2.6}$ & 240 & 235.0$_{-2.4}^{+2.7}$\\
$w\,(^\circ)$ &  [10, 200]& 48 & 41.5$\pm$11.8 & [10, 200] & 38 & 36.5$_{-2.4}^{+2.6}$ & 150  & 57.9$_{-2.9}^{+3.2}$ \\

\hline%

 $\chi^{2} $ &  &  & 1.9 & &  & 1.9 & & 3.7\\
\hline
\end{tabular}
\end{center}
\label{tab:bestParam_ADI}
\end{table*}

%
%
%
\section{Polarimetric phase function}
\label{s:SPF}

In the model presented in Sect.\,\ref{s:model}, we assume that the azimuthal intensity distribution in the polarimetric image depends on the product of two terms, the polarized scattering phase function (being itself a product of the Henyey-Greenstein function and Rayleigh scattering), and the density function, as a consequence of the disk being optically thin. Given that in the previous sections we limit ourselves to analyzing the grain distribution of the disk, here we aim to interpret relevant information about the grain properties from the extracted polarized scattering phase function (pSPF) and a spectrum in the NIR wavelengths.

\begin{figure}
\centering
\includegraphics[width=0.49\textwidth]{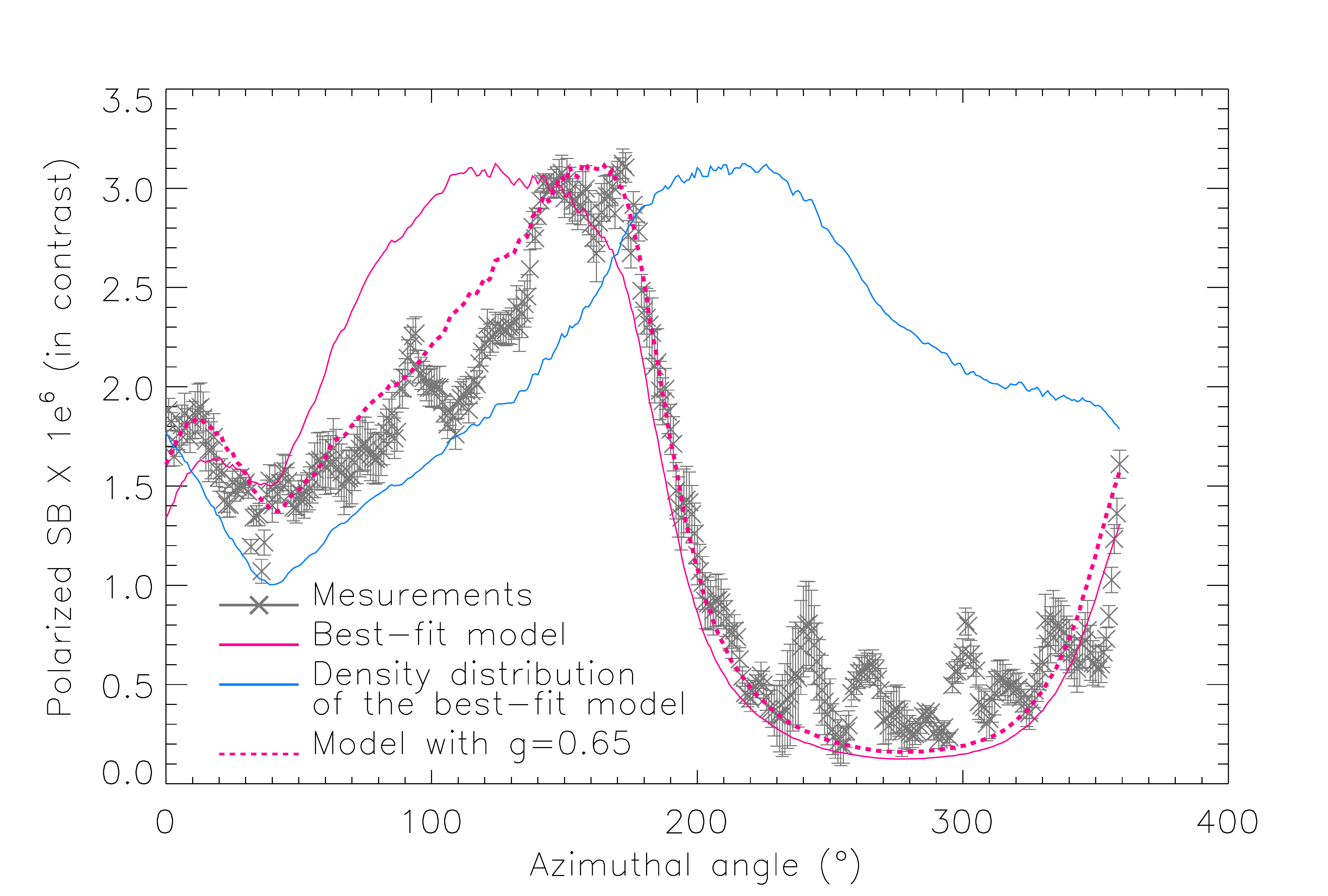}
\caption{Surface brightness of the ring R3 (in gray crosses) and of the best-fit model (in solid pink) shown as a function of the position angle. The dotted pink line represents the surface brightness of a model with g\,=\,0.65. The plot in solid blue shows the density distribution of the disk. }
\label{fig:intensity}
\end{figure}

In this section, we provide a measurement of the pSPF assuming the density term is determined from the model fitting performed in the previous section. For that we need to demodulate the density term from the surface brightness (SB). We first extracted the SB of the ring R3 with an azimuthal sampling of 1$\degb$ by fitting a Gaussian function radially from the star within the same aperture as shown in Fig.\,\ref{fig:dpi_fit}. The SB as a function of azimuthal angle is shown in Fig.\,\ref{fig:intensity}. The error bars are the standard deviations measured in the $U_{\mathrm{\phi}}$ image within the same aperture at every azimuthal angle. The SB of the best-fit model is also over-plotted to compare with the observed data. The SB of the ring peaks at 172$\degb$ and there is a steep decrease beyond this PA in accordance with the absence of flux in the western part of the ring. We notice in the same figure that the model does not exactly match the data and it deviates significantly from an azimuthal angle of $\approx$ 40$\degb$ to 140$\degb$. The largest deviation of 71\% is reached at an azimuthal angle of 109$\degb$. 

\begin{figure}[ht]
\centering
\includegraphics[width=0.49\textwidth]{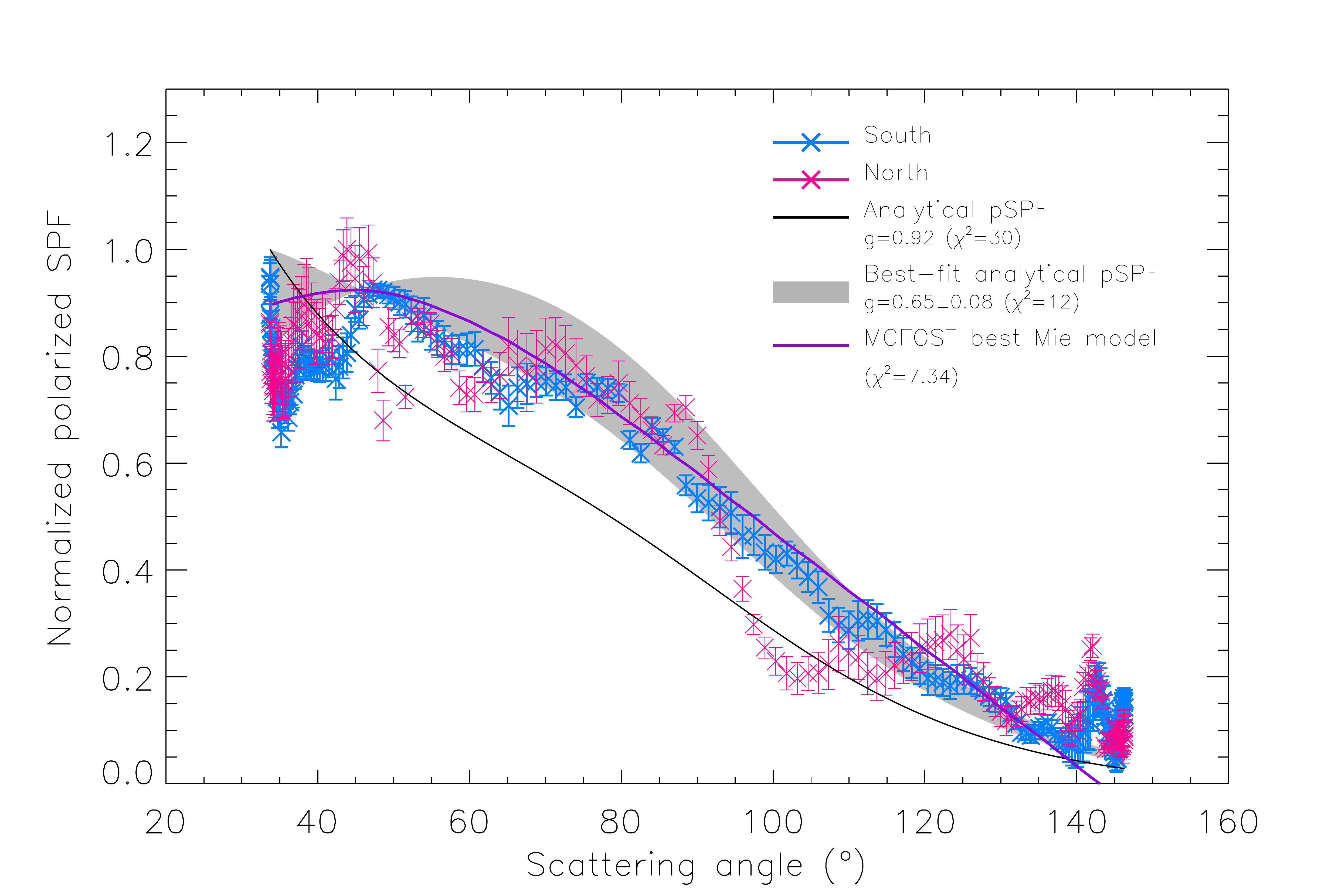}
\caption{Polarized scattering phase function of the northern (pink) and southern (blue) part of the ring as a function of scattering angles. The solid black line shows the analytical SPF with $g$\,=\,0.92 and the shaded area represents the analytical best-fit SPF with $g$\,=\,0.65$\pm{0.08}$. The best fit to the spectrum with Mie grains is shown as a  violet line. All plots are normalized to the maximum value of pSPF at the scattering angle of 43.8$\degb$.}
\label{fig:pf_dpi}
\end{figure}

To estimate the contribution of the density term to the surface brightness, we built a synthetic total intensity image in which the photometric effect of the scattering is artificially removed, with an anisotropic scattering parameter $g_\mathrm{pol}$=0 (isotropic scattering). The other model parameters were kept the same as the ones found with the MCMC as given in Table\,\ref{tab:bestParam_DPI}. This GRaTeR model is convolved to the observational PSF, and scaled to the $Q_{\mathrm{\phi}}$ image. Hereafter,  we refer to the density model as long as it is based solely on the density function. In the optically thin regime, the SB of the density model provides the dust density distribution of the ring. Therefore, in this framework, the pSPF is obtained by taking the ratio of the SB of the ring by that of the density model. The scattering phase angle is calculated as in \citet[][Eq. 2]{Milli2017}, which here corresponds to values ranging from 33$\degb$ to 147$\degb$ in the ring R3 {(which is $90 \degb \pm i$)}. The resulting pSPF  is plotted  in Fig. \,\ref{fig:pf_dpi}, separately for the northern and southern sides of the ring, and normalized to its maximum value which is reached at a scattering angle of 43.8$\degb$. 

While the process of measuring the pSPF from the data is dependent on the model due to the hypothesis regarding the density function, it should be noted that this is in disagreement with the numerical relation based on the Henyey-Greenstein and Rayleigh functions for $g_\mathrm{pol}=0.92$ ($\chi^2\,\sim\,$30), as determined by the best-fit model. Therefore, we tried to estimate the value of $g_\mathrm{pol}$ that generates an analytical SPF  conforming to our measured pSPF. We found that a better matched pSPF can be obtained for $g_\mathrm{pol}=0.65\pm{0.08,}$ although it is not fully satisfactory with regard to $\chi^2\,\sim\,$12. The model and its corresponding residual is shown in Fig.\,\ref{fig:spf_test}. This value happens to be in agreement, within 3$\sigma$ error bars, with $g_\mathrm{pol}\,=\,0.90\pm{0.09,}$ as obtained by the grid-search (Table\,\ref{tab:bestParam_DPI}) method.

Several debris disks have been modeled with two anisotropic scattering factors \citep{Milli2017, Bhowmik2019}, essentially one for the forward ($g_\mathrm{1}$) and the other for the backward ($g_\mathrm{2}$) direction. We also explored this scenario and found a higher reduced $\chi^2$ value ($\chi^2\,\sim\,$30) compared to that of the best-fit model with a single anisotropic scattering factor with  $g_\mathrm{pol}=0.65\pm{0.08}$. Using the latter instead of the MCMC value from Sect.\,\ref{s:model} ($g_\mathrm{pol}=0.95$), we generated the surface brightness curve while keeping the same density function. This resulted in a much better match of the actual SB, particularly between 40$\degb$ and 140$\degb$, as can be seen in Fig. \,\ref{fig:intensity} (pink dotted line). 

%
%
%
%

\section{Spectroscopy}
\label{s:spectro}

We performed a spectral analysis of R3 using the IFS and IRDIS data obtained in March 2016, as these cover the largest spectral range (YJH-K1K2) available and the S/N is higher in the IFS than for the May 2015 IFS data. Due to the large background noise in the IRDIS K2 band image, we retained the K1 band data only. In this section, we extract a total intensity spectrum of the southern part of the disk (Fig. \ref{fig:dbi_grid_model}a) that we correct for the self-subtraction. We then constrain the grain properties by fitting the spectrum with synthetic models.

To account for the self-subtraction biases in the ADI reduced data, we used the best-fit GRaTeR model obtained from the MCMC analysis (Table\,\ref{tab:bestParam_ADI}), following the procedure detailed in \citet{Bhowmik2019}. 
The model is convolved with a cropped PSF (0.4$''$ radius for IRDIS and 0.3$''$ radius for IFS) such that 99.99\% of PSF's total flux is retained. In our analysis, we considered only the southern part of the disk inside the elliptical contour shown in Fig.\,\ref{fig:dpi_fit} and normalized the best-fit KLIP model to the data.

\begin{figure}
\centering
\includegraphics[width=0.49\textwidth]{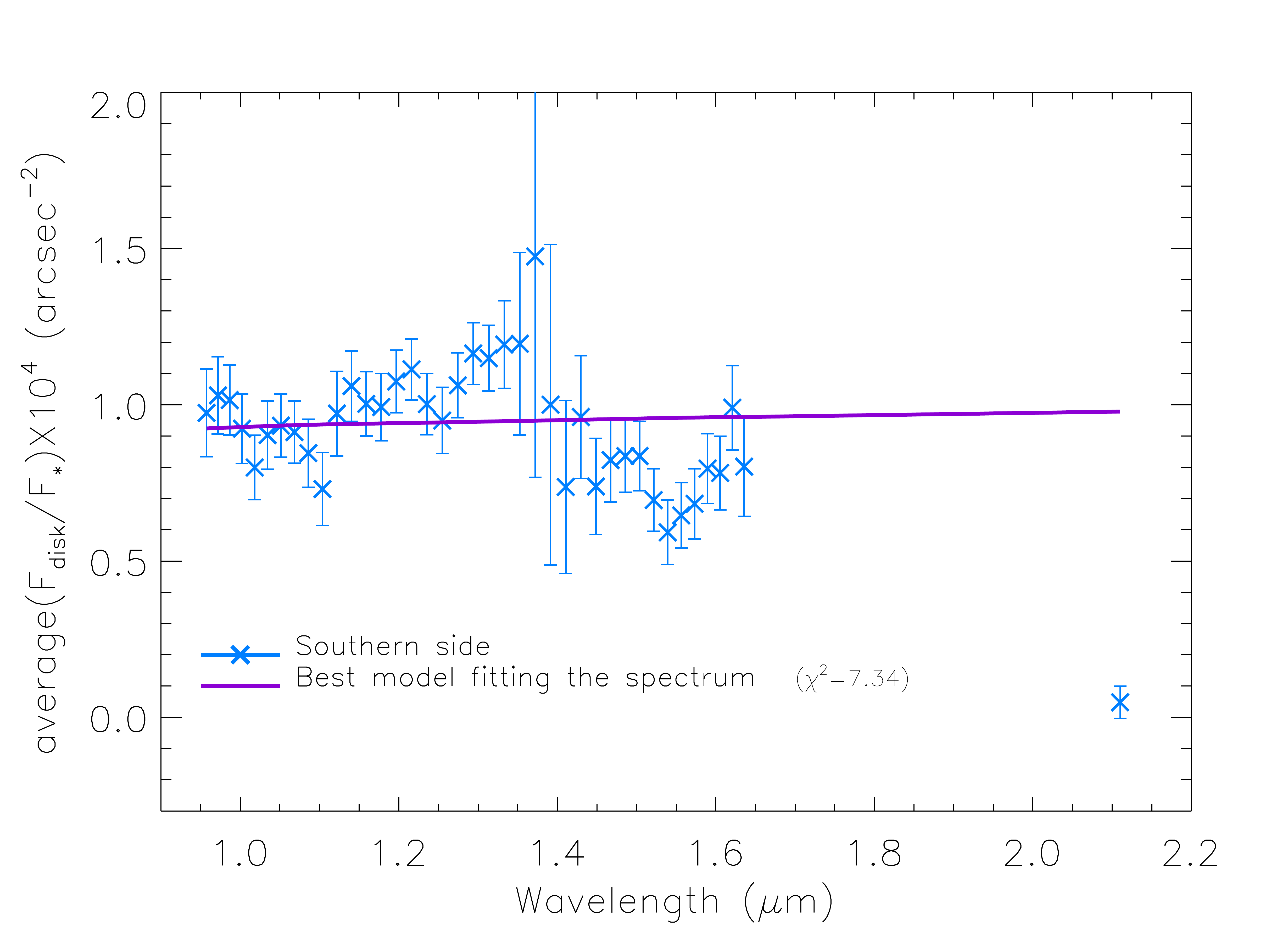}
\caption{Reflectance spectrum of the ring R3. The violet line represent the spectra corresponding to MCFOST's best Mie model that is normalized to the reflectance spectrum.}
\label{fig:spectrumGM}
\end{figure}

We measured the average azimuthal intensity of R3 at each wavelength on deprojected images
assuming $i=56.3\degb$ and $PA=356\degb$, then converted it into contrast per arcsec$^{-2}$ by normalizing it to the PSF flux, along with using the respective pixel scales  \citep[12.25$\pm${0.01}\,mas/pix for IRDIS and 7.46$\pm{0.02}$\,mas/pix for IFS][]{Maire2016}.
The outcome is the average spectral reflectance as a function of wavelength displayed in Fig.\,\ref{fig:spectrumGM}. 

Globally, the spectrum has a mild negative slope (blue color) of $(-7.97\pm{0.53})\times10^{-5}$\,arcsec$^{-2}$\,$\muup$m$^{-1}$ across the $YJHK1$ spectral range. The error bars shown in Fig.\,\ref{fig:spectrumGM} are the dispersion in the measurements of the de-projected residual images (similar to Fig.\,\ref{fig:dbi_grid_model}c) at each wavelength within the same southern-elliptical aperture used to extract the photometric measurements. The larger error bars at $1.4\pm0.05$\,$\muup$m are because of the telluric water absorption feature as a result of which the disk is very faint. At around 1.5\,$\muup$m, the spectral reflectance features a broad absorption pattern which might be attributed to OH bonding resonance in materials. However, the reliability of this feature is questionable given the error bars, calling for better S/N images in the future.

%
%
%
%

\section{Modeling the grain properties }
\label{s:grain_properties}

The SPFs of debris disks such as HR\,4796 \citep{Milli2019, Olofsson2020} and HD\,15115 \citep{Engler2019} were compared with theoretical models of SPF generated using Mie theory
in order to put constraints on grain properties. Following this approach, we considered models simulated
using the radiative transfer code MCFOST \citep{Pinte2006}. 
The grains in these models are porous in nature and composed of a mixture of astro-silicates \citep{Draine1984}, carbonaceous grains \citep{Li1997}, and water-ices \citep{Li1998} that can partially occupy the porous fraction of the mixture. We created 780 synthetic pSPF models
using the following parameters: the minimum dust grain size, $s_\mathrm{min}$, varying between 0.1-100\,$\muup$m, the fraction of vacuum removed by the ice, $p_\mathrm{H_20}$, ranging between 1\% and 29\%, the porosity without ice, $P$, between 0\%\,-\,80\% \citep{Augereau1999b}, and the silicates to carbonaceous grains 
volume fraction, $q_\mathrm{sior}$, which can take values of 1 (silicates only), 2 (same amount of silicates and carbonaceous grains), or $\infty$ (carbonaceous only).
Beside, we fixed the slope of the power size distribution $\kappa$ to -\,3.5 as predicted by \cite{Dohnanyi1969} and maximum grain size $s_\mathrm{max}\,=\,1\,$mm.
The total porosity of the grains, $p_\mathrm{wH_20}$, is a derived quantity defined as $p_\mathrm{wH_2O}\,=\,P\,(1\,-\,p_\mathrm{H_2O}/100)$ \citep{Augereau1999b}. All the models are normalized to the maximum value of the pSPF before performing the reduced $\chi^{2}$ analysis. Similarly, we used the 780 MCFOST models to fit the reflectance spectrum, by integrating over the range of accessible scattering angles (33$\degb$ to 147$\degb$), sampled and normalized on the 40 spectral channels corresponding to the IFS and IRDIS data, followed by the reduced $\chi^2$ analysis. 

\begin{figure}
  \centering
    \includegraphics[width=0.38\textwidth, trim={0.58cm 0.1cm 1cm 0.8cm},clip]{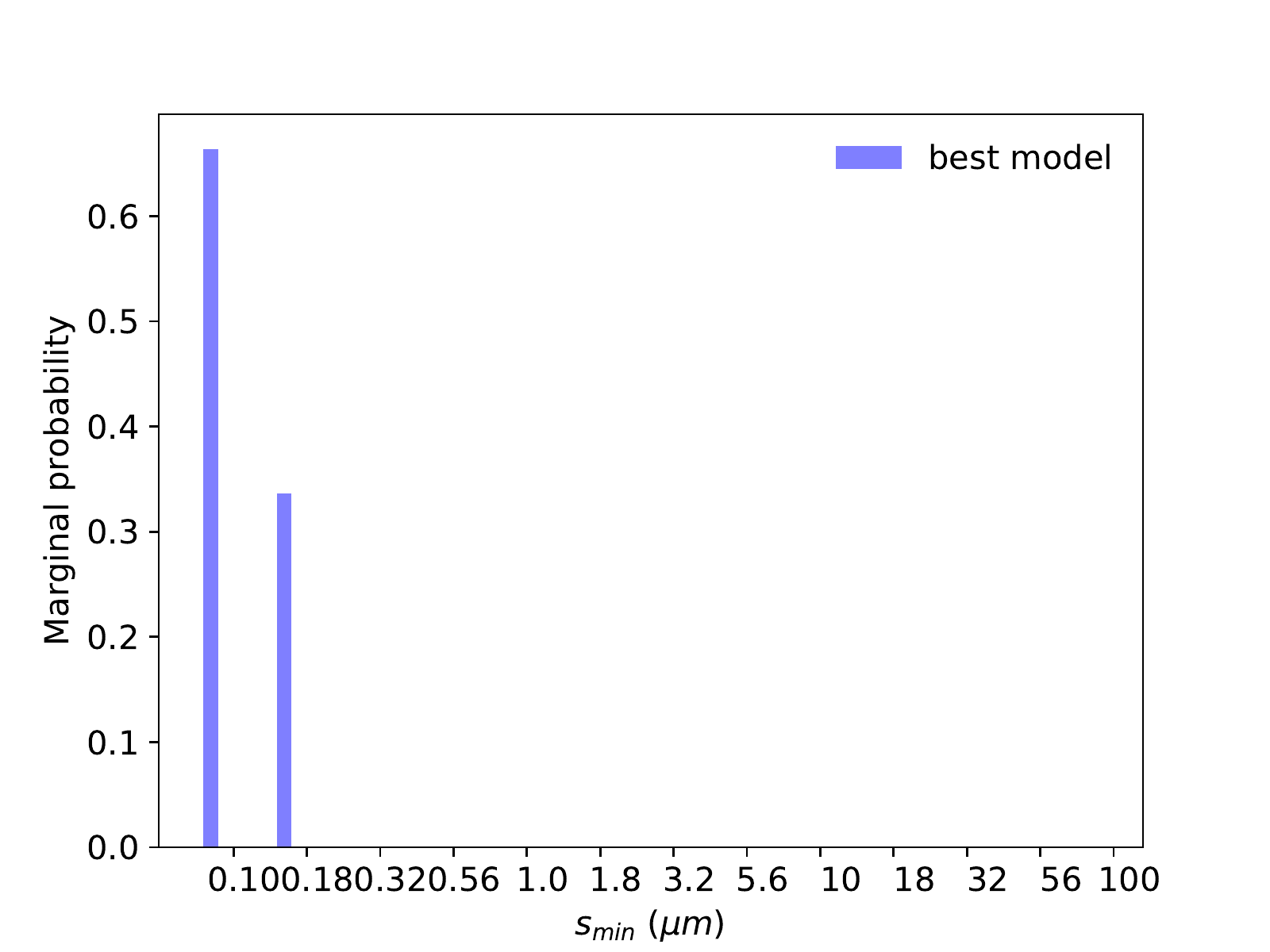}
  \includegraphics[width=0.38\textwidth, trim={0.32cm 0.1cm 1cm 0.8cm},clip]{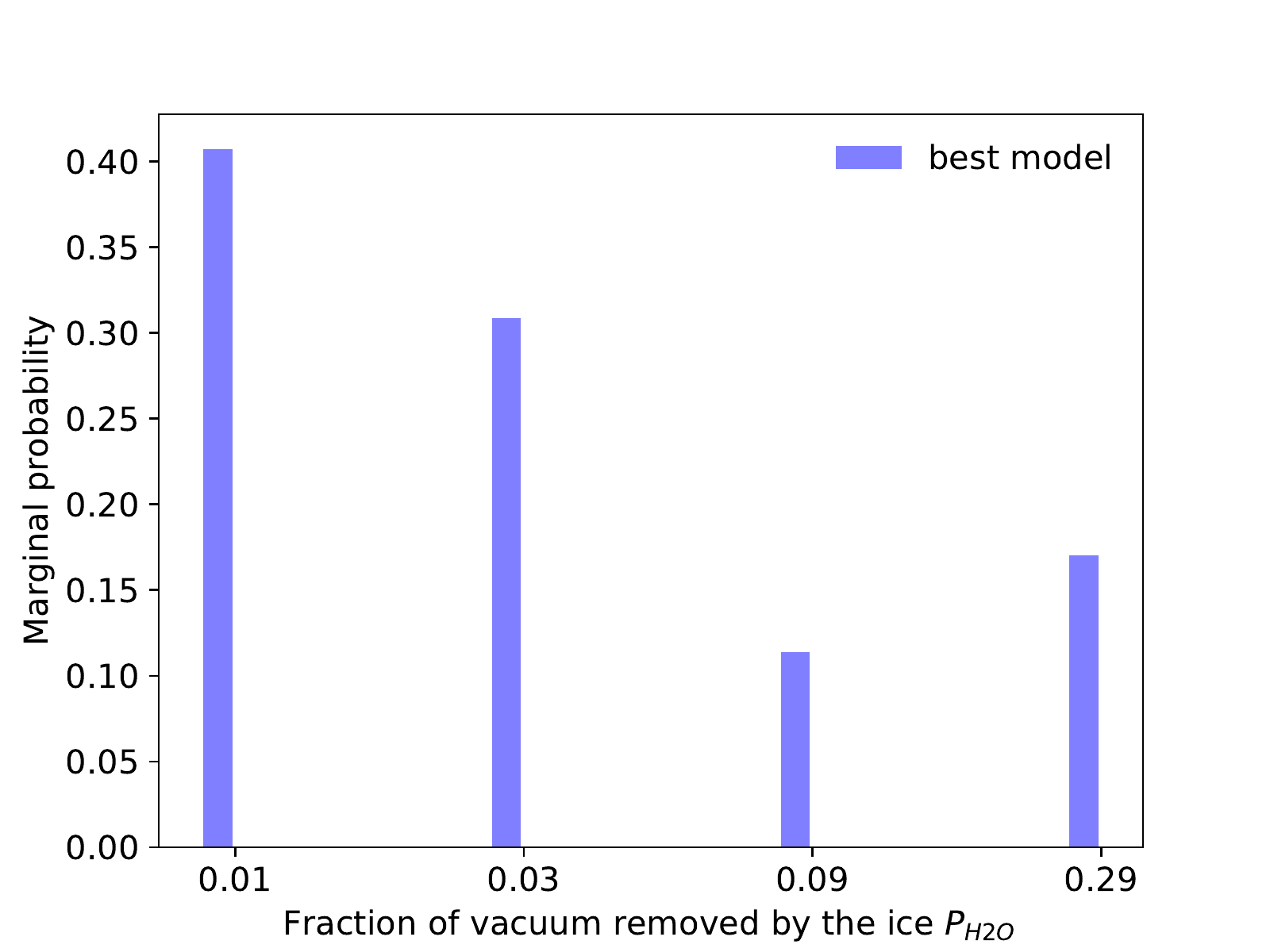}
        \includegraphics[width=0.38\textwidth, trim={0.58cm 0.1cm 1cm 0.8cm},clip]{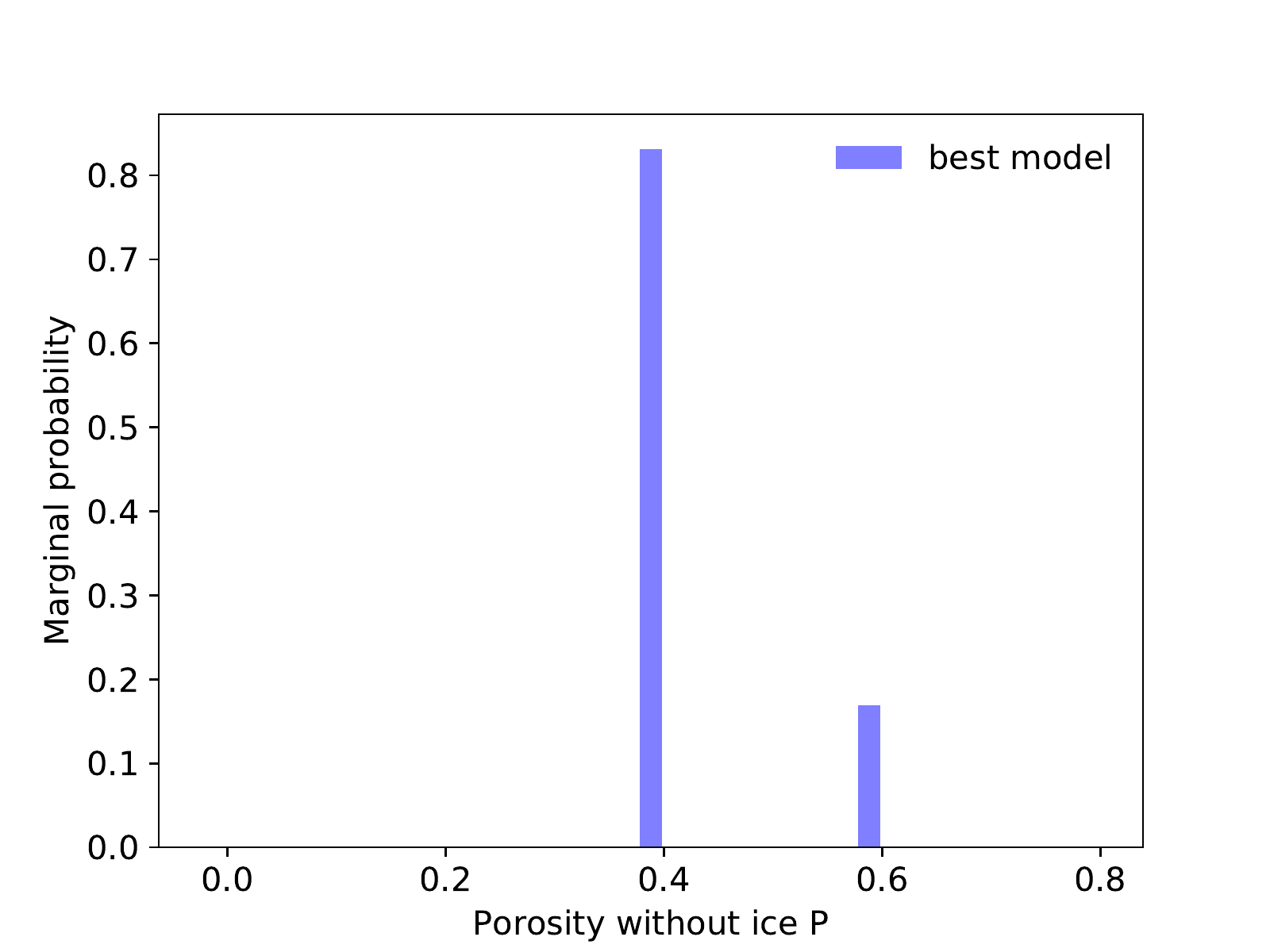}
    \includegraphics[width=0.38\textwidth, trim={0.58cm 0.1cm 1cm 0.8cm},clip]{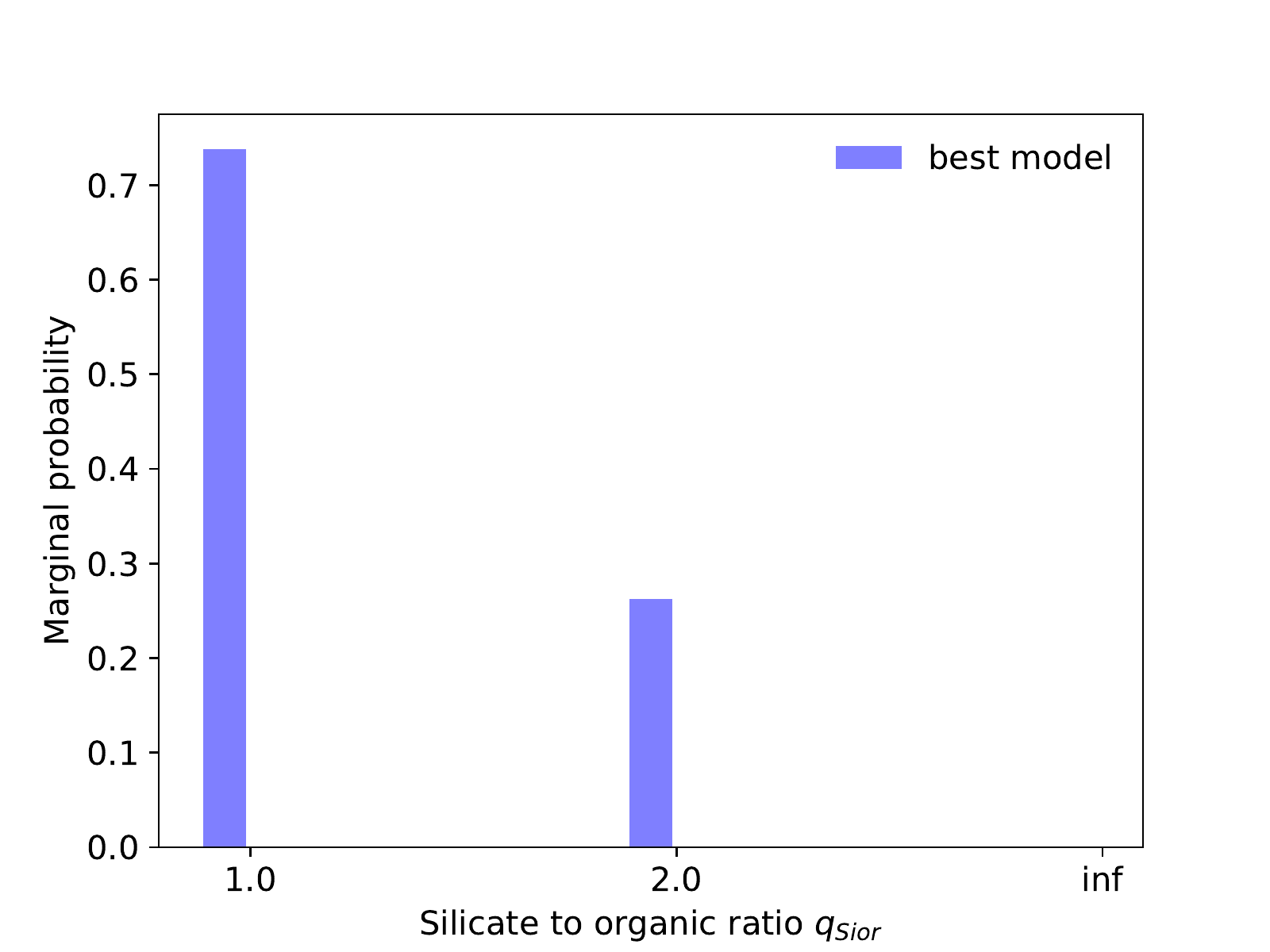}
  \caption{Marginal probability distributions of parameterized grain properties fitting both the pSPF and spectral reflectance. These distributions are inferred from fitting the data with 780 MCFOST models created using Mie theory.}
  \label{fig:pdf_spf}
 \end{figure}

Motivated by the goal to derive global grain properties, we aimed to find a single model fitting both the pSPF and the reflectance spectrum, based on the average $\chi^2$. The best fit displayed in Fig.\,\ref{fig:pf_dpi} and Fig.\,\ref{fig:spectrumGM} (violet line, reduced average $\chi^{2}\,$=\,7.34) corresponds to $s_\mathrm{min}=0.1\,\muup$m, $p_\mathrm{H_20}=1\%$, $P=40\%$, and $q_\mathrm{sior}=1$ (Table\,\ref{tab:spf}). The probability distribution (Fig.\,\ref{fig:pdf_spf}) shows that the minimum grain size is ranging between 0.1\,-\,0.2\,$\muup$m. This is apparently inconsistent with the larger value of $4\,\muup$m derived by \cite{Bruzzone2020}. However, we recall here that \cite{Bruzzone2020} directly fits MCFOST synthetic images considering an azimuthally uniform ring and limited the composition to a spherical astro-silicates type of grains, so these results cannot be directly compared.
In addition, the analysis favors $P=0.4$ and $p_\mathrm{H_2O}=0.01$, which translates to a total porosity of 40\%, so again it is discrepant with respect to a value of 0\% by \cite{Bruzzone2020}. Finally, while the maximum probability is reached for $q_\mathrm{sior}=1$, there is a 25\% probability that the ring has twice the amount of astro-silicates than carbonaceous grains. 

Also, as seen in Fig.\,\ref{fig:spectrumGM}, the best model fails to fit the data point corresponding to the K1 filter at 2.11$\muup$m. If we ignore the K1 data point, the average $\chi^2$ values tend to reduce noticeably to about 4. Making the spectrum bluer would have required a steeper grain size distribution which is unusual, hence, it is not considered here. Moreover, fitting  the pSPF and the reflectance spectrum separately provides a similar solution for the grain properties for the former, but a much larger value of $s_\mathrm{min}=3.2\,\muup$m for the latter. Therefore, the reflectance spectrum seems less relevant to derive the grain properties, in this particular case. Overall, the exercise is showing the limits in deriving grain properties with both a limited range in scattering phase angles and in wavelengths, together with a strong assumption on the grains sphericity. A more detailed analysis taking into account all the available observables is beyond the scope of this work but this will be key in setting meaningful constraints on the grain properties in future studies.

%
%
%
%
\begin{table}[th!]
\caption{Best-fit parameters obtained by fitting both the pSPF and spectral reflectance with a total of 780 simulated models following Mie theory.
} 
\begin{center}
\ra{1.5}
\begin{tabular}{c|l|c|c}
\hline \hline
Parameters &  Range & Number of values & {Best-fit model}   \\
 &   & in grid & Mie\\ \hline 
$s_\mathrm{min}$ ($\muup$m) & [0.1, 100] & 13 & 0.1  \\
$p_\mathrm{H_20} (\%)$ & [1, 30] & 4 & 1.0 \\
$P (\%)$ & [0, 80] & 5 &40 \\
$q_\mathrm{sior}$ & [1, 2, $\infty$] & 3 & 1 \\
\hline%
 $\chi^{2}$ & & & {7.34} \\
\hline
\end{tabular}
\end{center}
\label{tab:spf}
\end{table}

%
%
%
%

%
%
%
%
\section{Discussion}
\label{s:discuss}

\subsection{Grain scattering and composition}

The estimation of the scattering asymmetry factor $g_\mathrm{pol}$ shows a strong dispersion. Fitting synthetic images on the polarimetric data yields $g_\mathrm{pol}=0.92^{+0.05}_{-0.06}$ or $g_\mathrm{pol}=0.90\pm0.09$. Even though the residuals of the fit are acceptable, these are extremely high values as far as grain properties are concerned. This would imply very large grains with potentially negative signal in Q$_{\mathrm{\phi}}$ and non-null signal in U$_{\mathrm{\phi}}$, at some scattering phase angles, which is incompatible with the data. We note that the error bars on the $g_\mathrm{pol}$ parameter are particularly small and potentially unrealistic with the MCMC approach, although the mean values are consistent. 

When relying on the measured SB of the same polarimetric data, we find a relatively smaller estimate for the $g_\mathrm{pol}$ parameter ($g_\mathrm{pol}=0.65$), which appears more consistent with the other constraints and typical values already measured in such systems \citep{Engler2019, Milli2019}. The SB is indeed inconsistent with the $g_\mathrm{pol}=0.92$ model at scattering phase angles between $40$ and $140\degb$. However, in this region of the ring, the SB is essentially determined by the azimuthal density function, as can be seen in Fig. \ref{fig:intensity}. In terms of image residuals, a model with $g_\mathrm{pol}=0.65$ provides a larger reduced $\chi^2$ (0.90 instead of 0.53) but still acceptable visually. We also note that the proximity of the rings R2 and R3 in the eastern part can cause biases in the image fitting procedure, while the SB extraction should be less sensitive to that. From this analysis we conclude that degeneracies between the density and the scattering parameters prevent us from deriving an accurate value of $g_\mathrm{pol}$, thus it is more reliable to adopt a lower limit $g_\mathrm{pol}>0.65$.

MCMC model fitting on total intensity images provides an even lower value of $g_\mathrm{total\_int}=0.14 - 0.23$. While the scattering asymmetry factor can differ from polarized to total intensity data as a result of grain properties, and as already witnessed in other debris disks, the present case appears rather extreme. However, even if we generated forward models, we cannot exclude that the self-subtraction inherent to ADI, by suppressing the disk flux along the minor axis (which is oriented along the forward-backward direction, where the observation should be the most sensitive to scattering effects) would lead to an underestimation of $g_\mathrm{total\_int}$, as it was the case for HR\,4796 \citep{Milli2017}. Reconciling the derived values of the anisotropic scattering factor would certainly require higher S/N polarimetric images and less biased total intensity images, which is beyond the scope of this paper.

\subsection{Considering planetesimal disruption as a possible origin for the ring}
\label{s:breakup}

Large planetesimal disruptions in the recent past are a natural way of producing density asymmetries along the ring. Such disruptions may happen through direct collisions between planetesimals \citep{jackson2014, lawler2015}, or, if a massive planet is present in the vicinity, planetesimals can also be tidally disrupted if they pass within the tidal disruption radius of the planet \citep{cata18}. Collisions tend to dominate disruption rates in cases where the parent planetesimals are small, while tidal disruption dominates in cases where the planetesimals are large \citep{jans20}.
\citet{jackson2014} have shown that the collisional breakup of large planetesimals "naturally" produces strongly asymmetric rings, which are made of the small particles produced by the collisional cascade created in the initial breakup's aftermath. This disk is asymmetric because all fragments have eccentric orbits imparted by the velocity kick they get in the aftermath of the breakup, with a pericenter that is necessarily located at the location of the breakup. As a result, this creates a ring that is narrow and bright on the side of the breakup and more extended, diffuse, and faint on the opposite side. The survival time of these asymmetries is set by the progressive spread due to collisions among the fragments \citep{kral2015} or by orbital precession due to other planetary bodies \citep{jackson2014} and can typically be as large as a few million years at 50\,au.

In this collisional-origin scenario, one crucial parameter is the initial velocity kick imparted to the collisional fragments, which constrains the eccentricities of the fragment orbits and thus determines the ring's extent and shape. This velocity kick can be parameterized as a departure, $\sigma_{k}$, with respect to the local circular Keplerian velocity ($\varv_{k}$). To a first order, $\sigma_{k}$ is set by the mass $M_0$ of the shattered large body and the location of the breakup.
For example, bodies with masses of Ceres and Pluto colliding at 50\,au around a Sun-like star would result in $\sigma_{k}/\varv_{k}=0.05$ and 0.1, respectively. The azimuthal extension of the high-density region around the breakup location is set by the value of $\sigma_{k}/\varv_{k}$: the higher this ratio, the smaller the width, $w,$ of the azimuthal Lorentzian distribution becomes and the more asymmetric the disk (see Fig.\,9a of \citet{jackson2014}).
While \citet{jackson2014} give no explicit analytical relation between $w$ and $M_0$, we can use their Fig.9a to derive an approximate relation between $w$ and $\sigma_k/v_k$, for which we get:
\begin{equation}
w \sim 2^{\rm{o}}\times(\sigma_v/v_k)^{-1}.
\label{width}
\end{equation}

We can now use the fact that $\sigma_{\mathrm{k}}/v_{\mathrm{k}} \propto M_{\mathrm{{0}}}^{1/3}$ to estimate the mass $M_0$ of the object, whose breakup produces a Lorentzian density distribution having a width, $w, $ that is compatible with the one obtained in our best-fit. We remain careful and consider the value obtained in our intensity data-fit, $w=57^{\rm{o}}$, and the value obtained in our polarimetric fit, $w=119^{\rm{o}}$, as bracketing values for $w$. In this case, we get from Eq.\ref{width} that $0.015\leq (\sigma_{\mathrm{v}}/v_{\mathrm{k}}) \leq 0.03$. For a typical asteroidal density of $\sim3g/cm^3$, this corresponds to an object of mass $M_{\mathrm{0}}$ with $7.5\times10^{-6}M_{\oplus} \leq M_{\mathrm{0}} \leq 6\times10^{-5}M_{\oplus}$ and a radius comprised between $\sim\,$150 and $\sim$300\,km. 

This size domain is well below the smallest object (of Ceres size) that was explored in the simulations of \citet{jackson2014}. As the brightness of the breakup-produced ring is directly linked to the mass of the initially shattered object, this raises the question of whether or not a 150-to-300\,km object can account for the brightness of the inner ring, which is around $10^{-2}$ that of the star. This value is obtained by integrating the flux of the ring within the aperture as per the procedure explained in Sect.\,\ref{s:spectro}. If we consider the most optimistic case (in terms of the total geometrical cross-section), where all the mass of the initial object is entirely transformed into a cloud of micron-sized grains, we obtain that this dusty cloud intercepts a fraction of stellar light (supposing a perfect albedo=1 case) comprised between $2\times10^{-6}$ (for a 150\,km parent body) and $1.5\times10^{-5}$ (for a 300\,km one). This is at least three orders of magnitude less than the observed luminosity contrast. This observed luminosity contrast of $\sim0.01$, which should be of the order of the ring's optical depth, would correspond to a lunar-sized progenitor of at least radius $\sim1500$\,km.

The collisional breakup scenario, as proposed in \citet{jackson2014}, thus seems highly unlikely, essentially because there is an inconsistency between the high luminosity of the disk, which would be the signature of a very large collisional progenitor, and the wide azimuthal extension of the density profile, which would on the contrary be the signature of a smaller progenitor. 

We note that the alternative breakup scenario by tidal disruption in the vicinity of a giant planet encounters the same problem as the collisional breakup hypothesis: the velocity dispersion of the produced fragments has to be at least comparable to the escape velocity of the shattered object. This means that a large object (which is needed to generate the bright disc of debris that is observed) would also necessarily produce debris on high-$e$ orbits, leading to a ring that is too peaked in azimuth when compared to observations.

A way to circumvent this problem would be to assume that the ring is the result of a succession of more than a thousand collisional breakups (or tidal disruptions) of $\sim150-300\,$km progenitors happening in close succession at exactly the same spatial location. While this scenario cannot be ruled out beforehand, it seems, however, highly contingent and non-generic.

\subsection{Effect of the gas on the ring}
\label{s:gas}

\subsubsection{Gas to the rescue of the disruption scenario}
\label{s:gascoll}

The presence of gas in HD\,141569 has always been considered an important factor for explaining its particular morphology. The most recent measurements obtained with ALMA indicate large amounts of CO gas distributed at the same  distances as the ringlets detected with SPHERE \citep{Difolco2020}. Such a large quantity of gas could have a dynamical effect on the smallest grains in the system.
The fact that the CO gas distribution is almost azimuthally homogeneous means that gas should naturally induce a more axisymmetric dust distribution than what is predicted in \citet{jackson2014} in a gas-free case after the breakup of a massive object. Therefore, the azimuthally peaked distribution of dust observed with SPHERE could have been produced by a more massive object than what is estimated in Sect.\,\ref{s:breakup}, and gas would broaden the extent of the smallest dust grains azimuthally. 
This scenario could help solving the main inconsistency in the disruption scenario in which the progenitor is too small to account for the actual luminosity of the ring. 
Here, we assess the validity of this hypothesis, based on orders-of-magnitude estimates of the gas mass in the disk and of its dynamical effect on the dust.

Based on ALMA data, \citet{Difolco2020} estimated the mass of CO gas, respectively of dust, to be $M_{\rm CO} \sim 0.1 M_\oplus$, and $M_{\rm dust}\,\sim\,0.03-0.52\,M_\oplus$, leading to a CO-to-dust mass ratio in the range 0.19-3.3.
Under the assumption that the gas is of secondary origin \citep[and not primordial as posited in][]{Difolco2020}, the hydrogen content can be much lower than in a protoplanetary disk \citep{kral2016}, but there may be a substantial amount of carbon and oxygen produced by the photodissociation of CO \citep{kral2019,marino2020}.
The CO mass in the system is high enough, so that it is self-shielding \citep{Visser2009} and carbon may also start shielding CO from photodissociating. Then, CO can viscously spread, which would explain why it is so broad (extending over 200 au) in a secondary origin context \citep{kral2019}. The total carbon mass then may be lower or greater than the CO mass depending on unconstrained parameters, such as the viscosity $\alpha$ of the gas disk. However, assuming a high $\alpha$ value considering that the magnetorotational instability may be very active in these disks (see \citealt{kral2016b} and the first observational confirmations in \citealp{marino2020}), the carbon and oxygen masses should supersede the CO mass \citep{kral2019,marino2020}, leading to a gas-to-dust ratio larger than 1. Therefore, the gas-to-dust ratio lower limit is 0.19, but it is likely to be greater than 1.

Given this rather high gas-to-dust ratio for a potentially secondary gas disk \citep{kral2017,matra2017}, we go on to explore the effect of gas on bound and unbound dust grains in this system. The gas surface density in CO at 44 au in HD\,141569 is $\Sigma_{\rm CO} \sim 1.5 \times 10^{-4}$ g/cm$^2$ \citep{Difolco2020}, or $8 \times 10^{-5}$ g/cm$^2$ using the more optically thin $^{13}$CO line (and assuming a typical isotopic ratio of 77 between $^{12}$CO and $^{13}$CO) giving a lower limit to the total gas surface density. Now, we can compute the time for a bound grain to couple to gas. Using \citet{richert2018} and \citet{burns1979}, the dimensionless stopping time (in orbital units) reads as:

\begin{equation}
T_s=1.6\left( \frac{s}{1 \, \mu{\rm m}} \right) \left( \frac{\rho}{1 \, {\rm g/cm}^3} \right) \left( \frac{\Sigma_g}{10^{-4} \, {\rm g/cm}^2} \right)^{-1},
\end{equation}

\noindent where $s$, $\rho$ and $\Sigma_g$ are the grain size, grain density, and gas surface density, respectively\footnote{We note that if primordial the total gas mass and surface density could be several orders of magnitude higher than for a secondary gas production.}. Therefore, grains that are $\lesssim 50$ microns should see their orbits strongly affected by gas in fewer than 100 orbits \citep{takeuchi2001}, which is shorter than the typical collisional time for a disk of optical depth $\sim0.01$. 
This means that any azimuthal clustering of dust grains should broaden on a timescale of only a dozen orbits for grains close to the blow-out size (which is quite high in this system with $s_{\rm blow} \sim 8$ microns\footnote{We assumed a luminosity of 27.0 L$_\odot$ from \citealt{merin2004} and M$_\star=2$ M$_\odot$ from \citealt{Difolco2020} and $\rho = 2000$ kg/m$^3$, but we note this is uncertain as it depends on an unknown grain composition.}).

Given the SPHERE images are representative of light scattered by dust, we also need to investigate a similar quantity for unbound grains which can be smaller than the blow-out size. These potentially unbound micron to sub-micron grains would be affected by gas drag and following the same procedure as in \citet{Bhowmik2019} (assuming T=35 K from \citealt{Difolco2020} and $M_\star=2 \, M_\odot$ from \citealt{merin2004}), we find:

\begin{equation}
T_{\rm un}=0.22\left( \frac{s}{1 \, \mu{\rm m}} \right) \left( \frac{\rho}{1 \, {\rm g/cm}^3} \right) \left( \frac{\Sigma_g}{10^{-4} \, {\rm g/cm}^2} \right)^{-1} \left( \frac{\Delta R}{10 \, {\rm au}} \right)^{-1},
\end{equation}

\noindent where $\Delta R$ is the width over which a grain feels the gas. This stopping time for unbound grains can be even shorter than $T_s$. Grains below the blow-out size could strongly feel the gas and may circularize rapidly (and become bound), thus accumulating instead of leaving the system. This conclusion is similar to what was found for HD\,32297 in \citet{Bhowmik2019}.

These results show that small grains produced in a disruptive-like scenario, such as the one described in the previous section, may quickly move onto different orbits, which would broaden the expected density peak, both azimuthally and radially. Therefore, we can conclude that the disrupted mass we found in the previous section is, in fact, a lower limit. 

When the gas mass is taken into account, the disruptive scenario cannot be ruled out any longer because a larger disrupted body could indeed both explain the disk brightness and its azimuthal extent. It also means that fewer disrupting events may actually be needed to explain the current disk brightness if several disrupted bodies are to be involved. However, dedicated simulations with, for example, an improved version of LIDT-DD \citep{kral2013,kral2015,thebault2018} that includes gas drag would be needed to explore this avenue further and give a stronger conclusion.

\subsubsection{Potential of photoelectric instability to explain the observations}
\label{s:gasinstab}

The photoelectric instability \citep{lyra2013} has been identified as a possible mechanism that could potentially alter disk structures and create asymmetric arc-like features, such as those observed in this study \citep{richert2018}. The amount of CO gas derived by \citet{Difolco2020} corresponds to the regime of ``high-level of gas'' described in \citet{richert2018}, for which the instability is the strongest and shows no anti-correlation between gas and dust. Interestingly, \citet{Difolco2020} demonstrated that the CO is not entirely axisymmetric, but is instead enhanced  between 220$^{\circ}$ and 290$^{\circ}$, thus peaking in the west-southwest part of the ring; hence, this finding is in qualitative agreement with the dust distribution inferred from scattered light observations. Both gas and dust seem slightly correlated. 

The photoelectric linear instability may be stabilized for gas-to-dust ratio $<$ 1. As explained previously, we expect the gas-to-dust ratio to exceed 1 based on only CO and its daughter species carbon and oxygen. This instability may then explain the presence of the three rings (located within 100 au) observed with SPHERE as the gas observed with ALMA extends up to more than 200 au \citep{white2016,miley2018,Difolco2020}. This is one tempting possibility, although it goes beyond the scope of this paper to explore it further by running dedicated simulations of this instability for HD\,141569.

\subsubsection{Explaining the slight asymmetry observed in the gas}
\label{s:gasasym}

Finally, a relevant question considers whether the asymmetric structure of the gas revealed with ALMA at separations of $\sim$50\,au (see previous subsection) can produce a pressure trap that could be responsible for the dust distribution. First, we note that the radial location of the gas asymmetry \citep{Difolco2020} seems to be at a slightly larger value than the ring studied in this paper, but there is some overlap and a higher resolution would be needed to pinpoint the gas peak location. Also, the excess of gas emission (CO clump) is 6 $\sigma$ above the noise in the data by \citet{Difolco2020}, but given the line is optically thick, it does not give any information on the surface density compared to the rest of the gas disk, but we do note that an increased temperature at the ring location would indeed create more emission in CO. We cannot rule out a pressure trap (a more optically thin line would be needed) but we can suggest another possibility.

Due to the high quantity of dust in the ringlets and due to the photoelectric effect, there could be a lot of electrons at these specific locations that would heat the CO gas and increase its brightness, hence creating the illusion of a CO clump in the ALMA data. In this case, the significant dust asymmetry observed could thus be responsible for the CO clump.

\section{Conclusions}
\label{s:conclusion}

We obtained SPHERE polarimetric data of the gas rich debris disk around HD\,141569, as well as new total intensity data complementing the study presented in \citet{Perrot2016}. We definitively confirmed the presence of several concentric rings around the star at few tens of au. 

In this paper, we focus on constraining the dust distribution in the innermost detected ring located at $\sim$45\,au, based on the combined information from polarized and total intensity images. We posited that the dust density must be non-uniform azimuthally to account for the observed intensity distributions. Assuming that the ring azimuthal density can be described with a Lorentzian function, we found that the dust is peaking in the southwest, while the intensity peak is located in the southeast, as a result of the light being preferentially scattered forward. This main finding is robust whatever the approach taken to model the polarimetric and total intensity images. 

The values derived from total intensity and polarimetry are consistent providing overall a position angle of $229.1\degb\pm9.0\degb$, as determined with MCMC. Interestingly, this location is consistent with the enhancement of CO measured with ALMA by \citet{Difolco2020}, possibly pointing to a link between the distributions of gas and dust. On the contrary, the azimuthal width of the density function is rather broad but differs by a factor of 2 between polarimetric ($118.9^{\circ+8.2^\circ}_{-7.3^\circ}$) and total intensity data ($57.9^{\circ+3.2^\circ}_{-2.9^\circ}$). In addition, the polarimetric image fitting procedure converges to a very high value of the anisotropic scattering factor ($g_\mathrm{pol}=0.92$), while we obtained a moderate value for total intensity ($g_\mathrm{total\_int}=0.14 - 0.23$).

Based on the modeling of the images we extracted the polarized scattering phase function (pSPF) by demodulating the azimuthal density variation from the surface brightness. We concluded that a lower value of the anisotropic scattering factor ($g=0.65$) provides a better match to the pSPF as compared to the image fitting. However, we note that the degeneracies between scattering and density can stand as an issue impeding a thorough measurement of this parameter.  We also derived the spectral reflectance of the ring in the wavelength range of $0.98-2.1\,\muup$m. Modeling the pSPF and spectral reflectance with MCFOST and assuming Mie scattering provides a composition of porous sub-micron sized mixture of carbonaceous and astro-silicate grains. We argue that broader wavelength coverage and higher S/N data would be needed to derive more meaningful conclusions about the grain properties. 

Finally, we discuss the implications of the dust distribution in terms of the dynamical history of the system. While massive collisions of planetesimals can produce a ring-like pattern with dust enhancement at the place of the collision, this scenario fails to account for the azimuthal dust distribution in the ring, since it suggests a small size for the progenitor, incompatible with the total luminosity of the ring. However, the gas mass is high enough ($>0.1$ M$_\oplus$) that gas can broaden any azimuthal clustering on a few dynamical timescales as grains move onto different orbits very quickly before being destroyed through collision. In this paper, we find that the disruption scenario can no longer be ruled out when gas dynamics is taken into account because a larger disrupting body can, in this case, account for both the high brightness of the disk and its azimuthal extent. We note that dedicated simulations including gas drag are required to further explore this scenario and reinforce our conclusion.

\begin{acknowledgements}
%
SPHERE is an instrument designed and built by a consortium consisting of IPAG (Grenoble, France), MPIA (Heidelberg, Germany), LAM (Marseille, France), LESIA (Paris, France), Laboratoire Lagrange (Nice, France), INAF–Osservatorio di Padova (Italy), Observatoire de Genève (Switzerland), ETH Zurich (Switzerland), NOVA (Netherlands), ONERA (France) and ASTRON (Netherlands) in collaboration with ESO. SPHERE was funded by ESO, with additional contributions from CNRS (France), MPIA (Germany), INAF (Italy), FINES (Switzerland) and NOVA (Netherlands).  SPHERE also received funding from the European Commission Sixth and Seventh Framework Programmes as part of the Optical Infrared Coordination Network for Astronomy (OPTICON) under grant number RII3-Ct-2004-001566 for FP6 (2004–2008), grant number 226604 for FP7 (2009 – 2012) and grant number 312430 for FP7 (2013–2016). 
G.\,S. would like to acknowledge the funding received from the European Union’s Horizon 2020 research and innovation programme under the Marie Sk\l{}odowska-Curie grant agreement No 798909. G.\,S. would also like to thank Dr. Elodie Choquet for her suggestions in Sect.\,\ref{s:model}.  
French co-authors acknowledge financial support  from the Programme National de Plan\'etologie (PNP) and the Programme National de Physique Stellaire (PNPS) of CNRS-INSU in France.
J.\,O. acknowledges support by ANID, -- Millennium Science Initiative Program -- NCN19\_171, from the Universidad de Valpara\'iso, and from Fondecyt (grant 1180395).
C.\,P. acknowledge financial support from Fondecyt (grant 3190691) and financial support from the ICM (Iniciativa Cient\'ifica Milenio) via the N\'ucleo Milenio de Formaci\'on Planetaria grant, from the Universidad de Valpara\'iso.
A.\,V. acknowledges funding from the European Research Council (ERC) under the European Union's Horizon 2020 research and innovation programme (grant agreement No. 757561).
A.\,Z. acknowledges support from the FONDECYT Iniciaci\'on en investigaci\'on project number 11190837. 
Finally, this work has made use of the the SPHERE Data Centre, jointly operated by OSUG/IPAG (Grenoble), PYTHEAS/LAM/CESAM (Marseille), OCA/Lagrange (Nice) and Observatoire de Paris/LESIA (Paris). We thank P. Delorme and E. Lagadec (SPHERE Data Centre) for their efficient help during the data reduction process. 

\end{acknowledgements}

\bibliographystyle{aa} 
\bibliography{references}

\begin{appendix}
%
%
%
\section{S/N map of the IRDIS image}
\label{app:irdis_s/n}
The S/N maps are generated by taking the ratio of the reduced image and the corresponding azimuthal standard deviation.

\begin{figure}
  \centering
    \includegraphics[width=0.4\textwidth,trim={15cm 0cm 15cm 0cm}, clip]{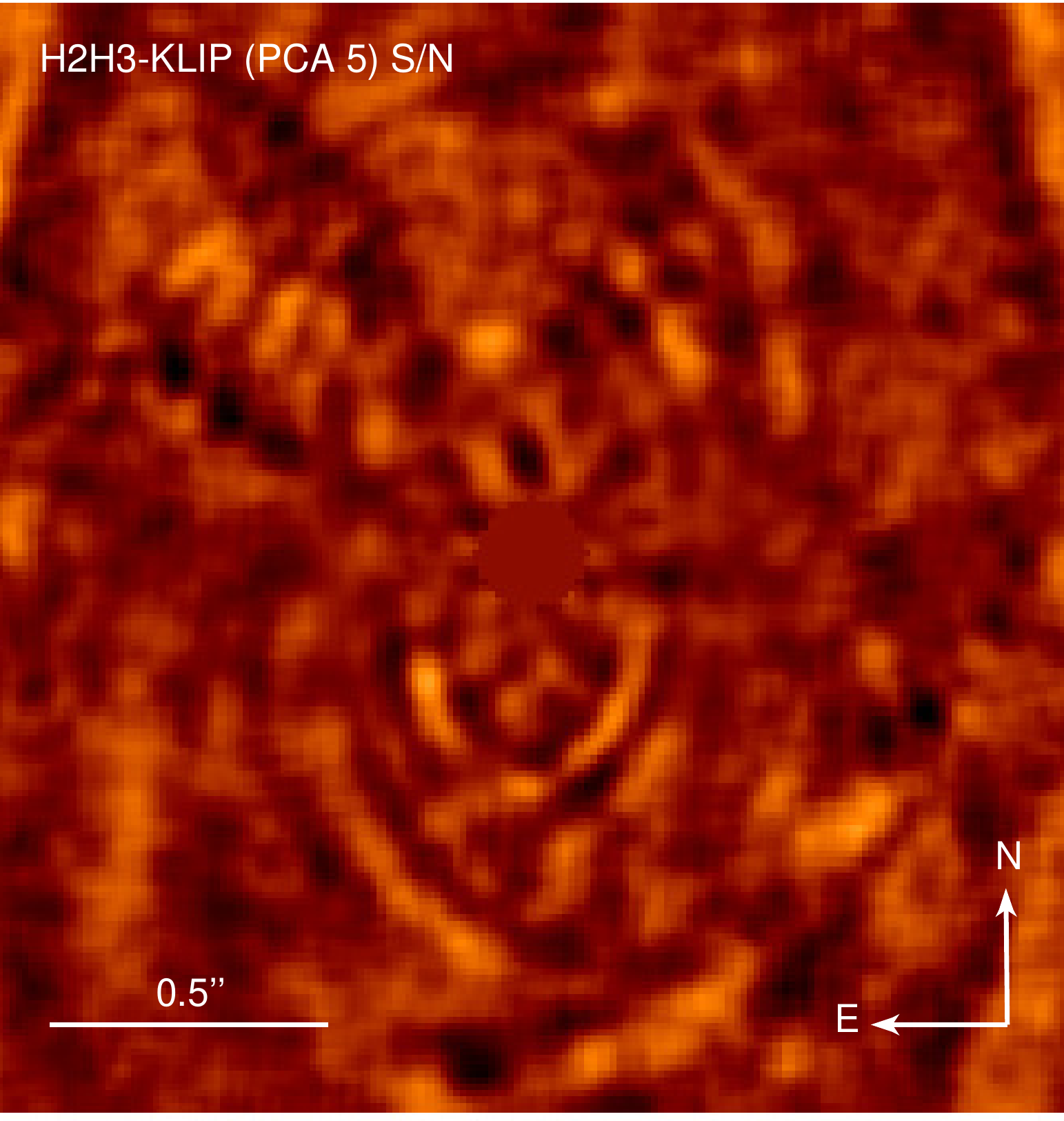}
    
    \includegraphics[width=0.4\textwidth,trim={0cm 0cm 0cm 13cm}, clip]{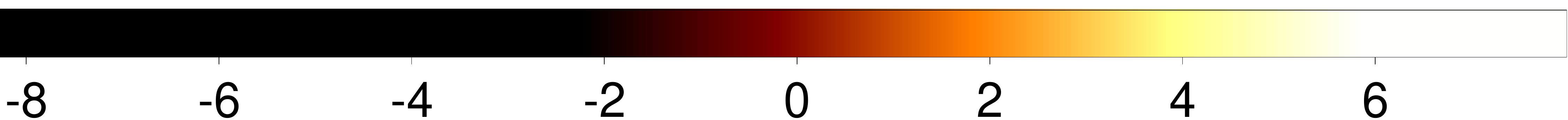}
  \caption{S/N map of the IRDIS image shown in Fig.~\ref{fig:hd141_fig}d.}
  \label{fig:hd141_fig_SNR}
\end{figure}

%
%
%
\section{Complementary IFS data reduction and S/N maps}
\label{app:ifs_s/n}

\begin{figure*}
  \centering
    \includegraphics[width=0.35\textwidth]{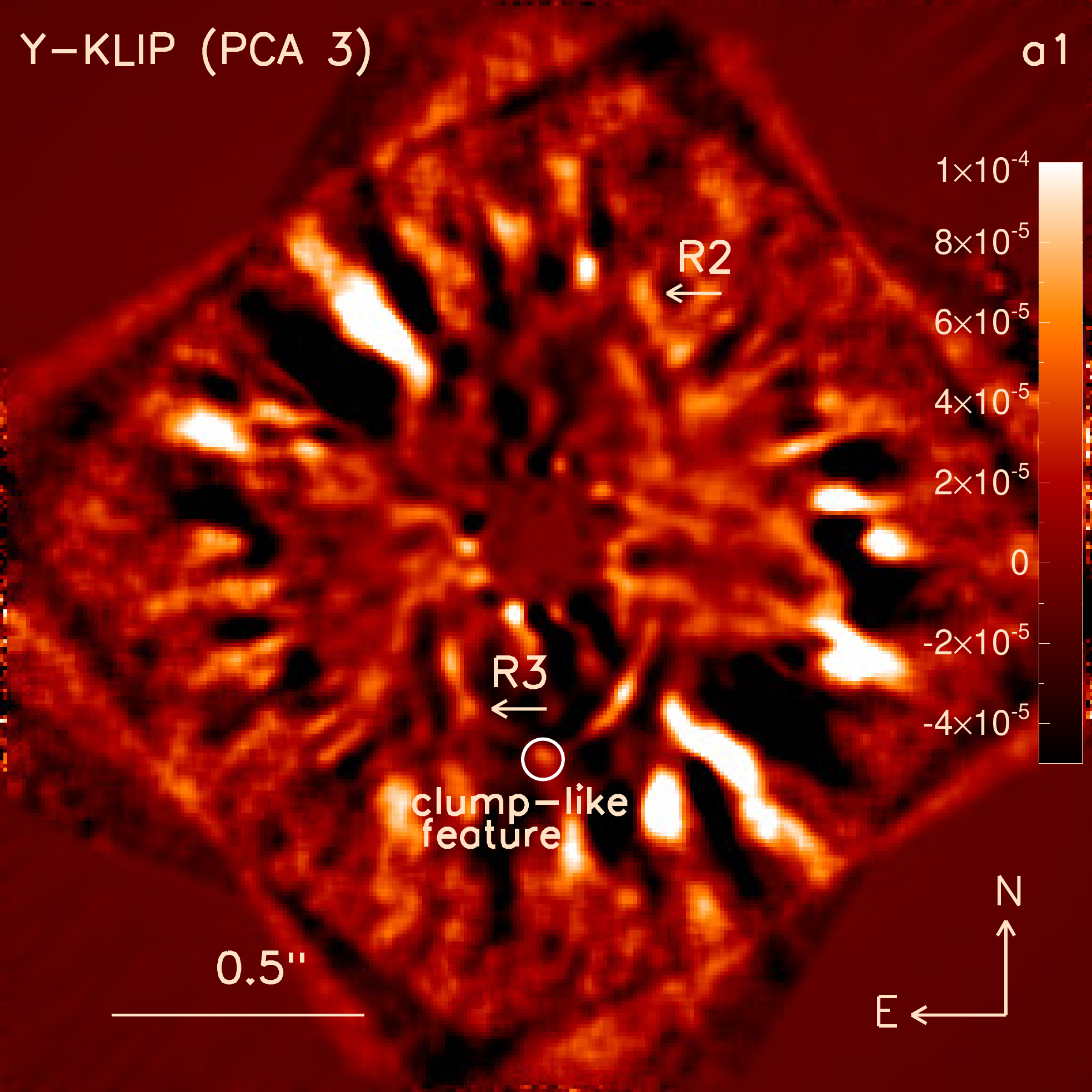}
        \includegraphics[width=0.35\textwidth]{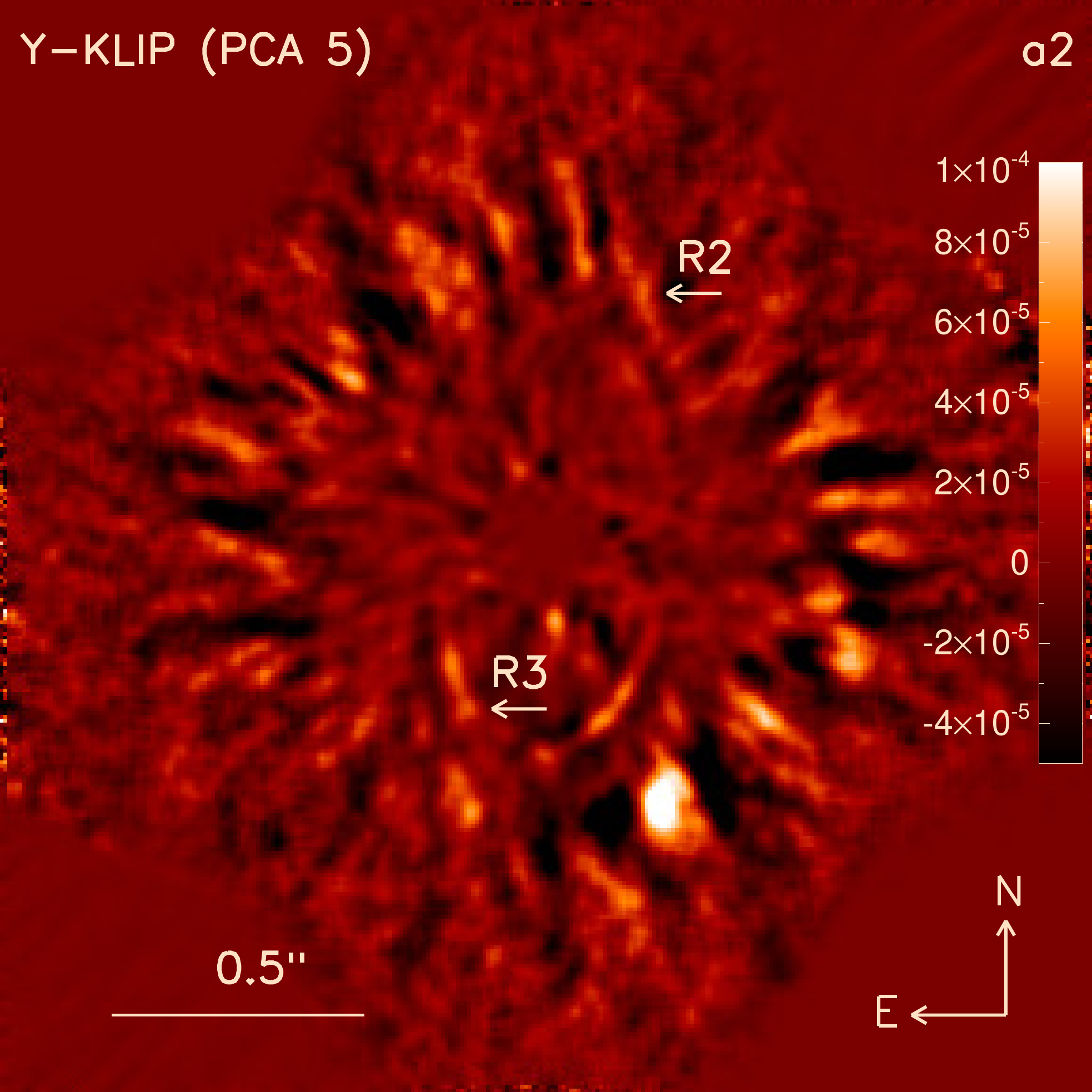} 
	
    \includegraphics[width=0.35\textwidth]{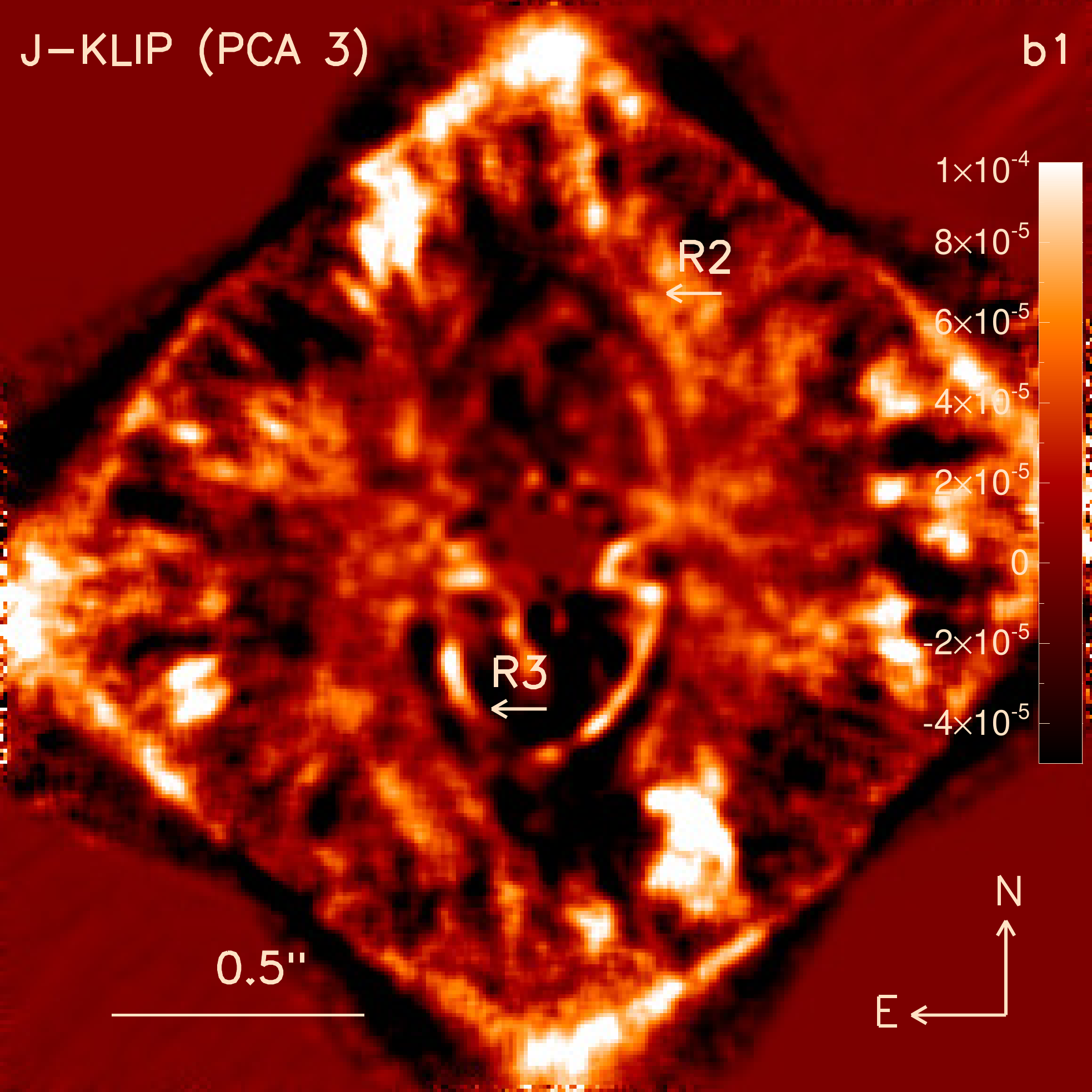}
        \includegraphics[width=0.35\textwidth]{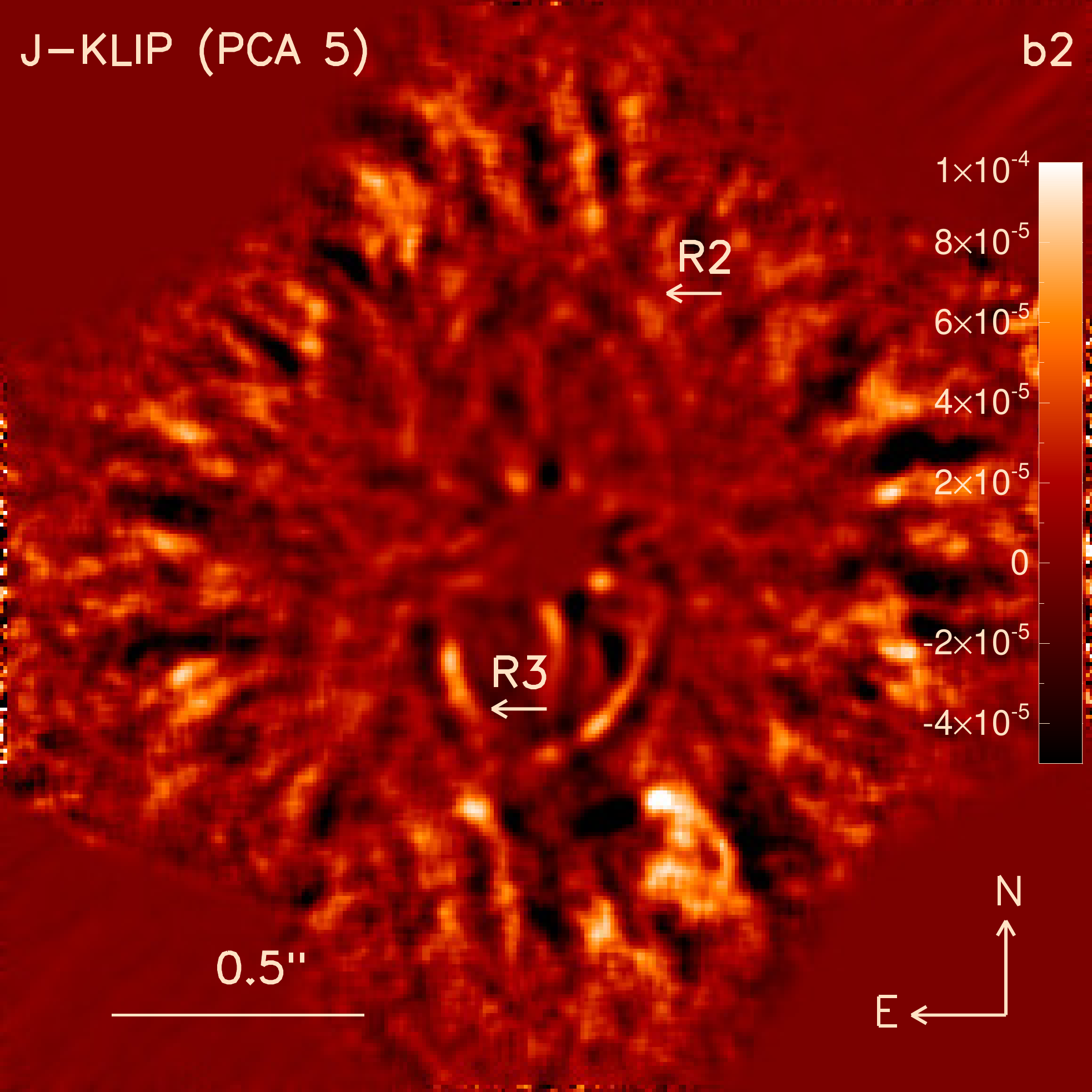}
         
    \includegraphics[width=0.35\textwidth]{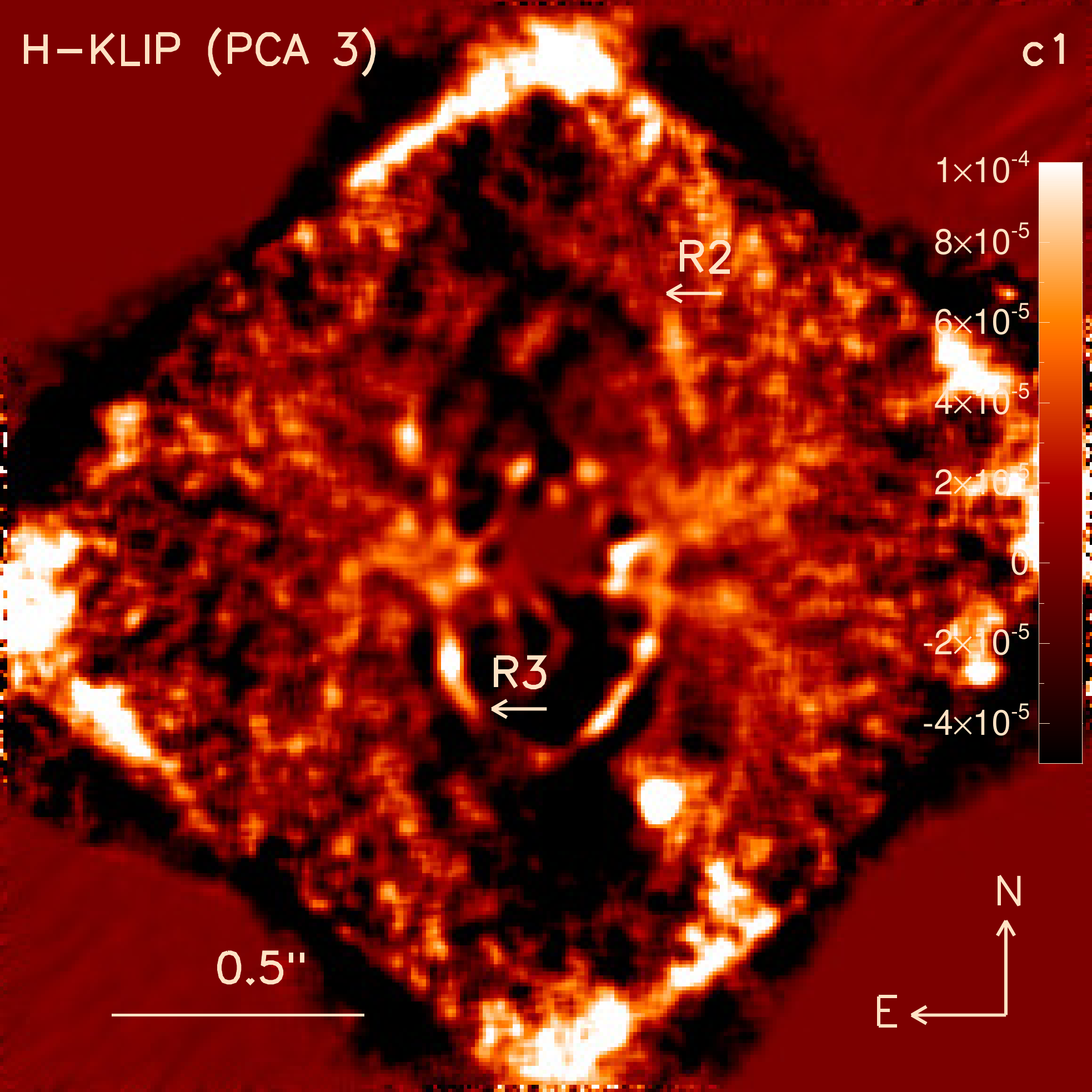}
        \includegraphics[width=0.35\textwidth]{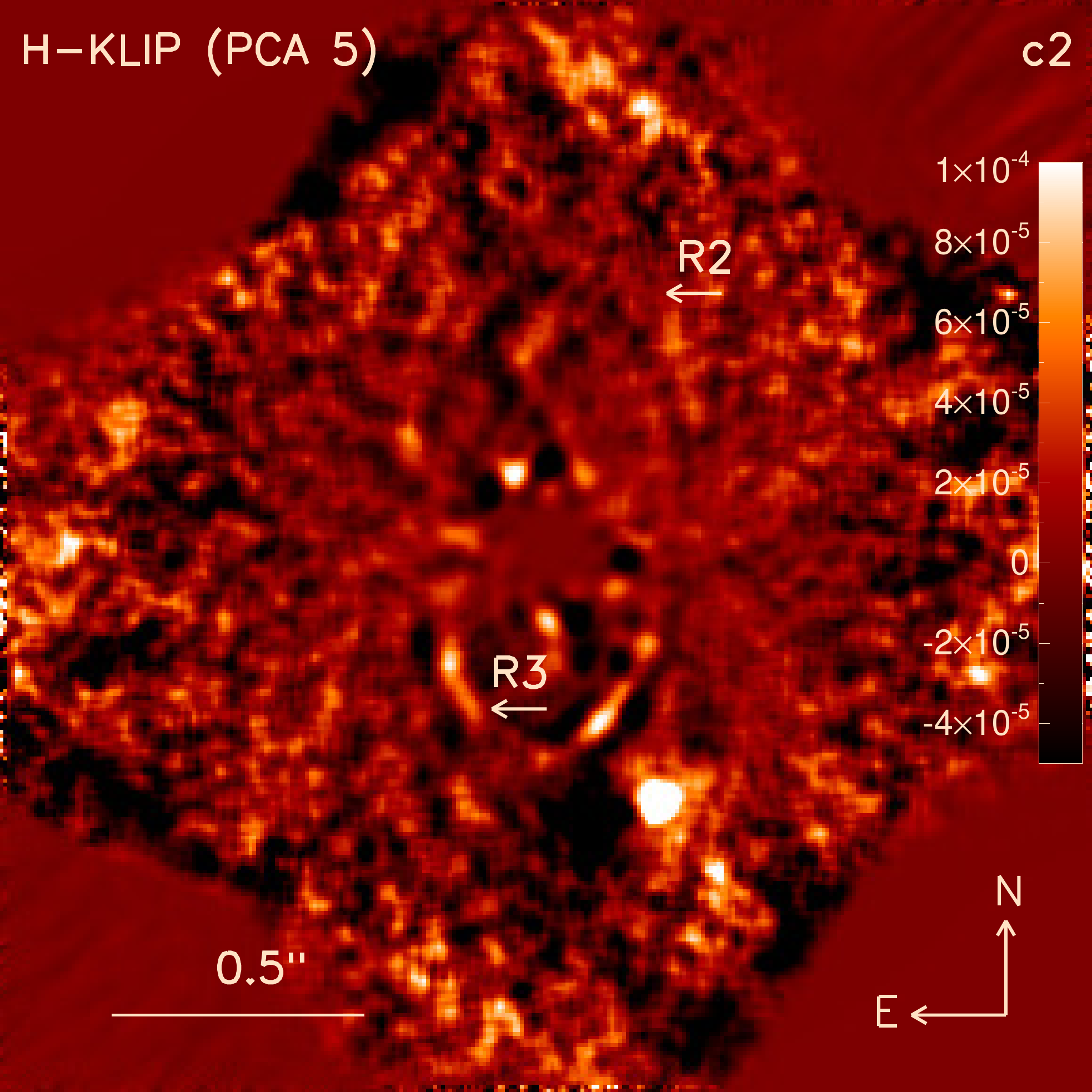}
  \caption{KLIP-reduced IFS images in the units of contrast of HD\,141569A observed with SPHERE in March 2016 in YJH narrow band. The images in column 1 and column 2 are processed with KLIP truncated at three and five modes respectively. From top to bottom, each row of images represent the median image of wavelength range of 0.95\,-\,1.14$\muup$m, 1.15\,-\,1.34$\muup$m and 1.45\,-\,1.65$\muup$m respectively. All the images have a field of view of 2.16$''\,\times\,$2.16$''$ and are smoothed with a five-pixel Gaussian kernel. }
  \label{fig:dbi_yjk}
\end{figure*}

\begin{figure*}
  \centering
    \includegraphics[width=0.35\textwidth, trim={15cm 0cm 15cm 0cm}, clip ]{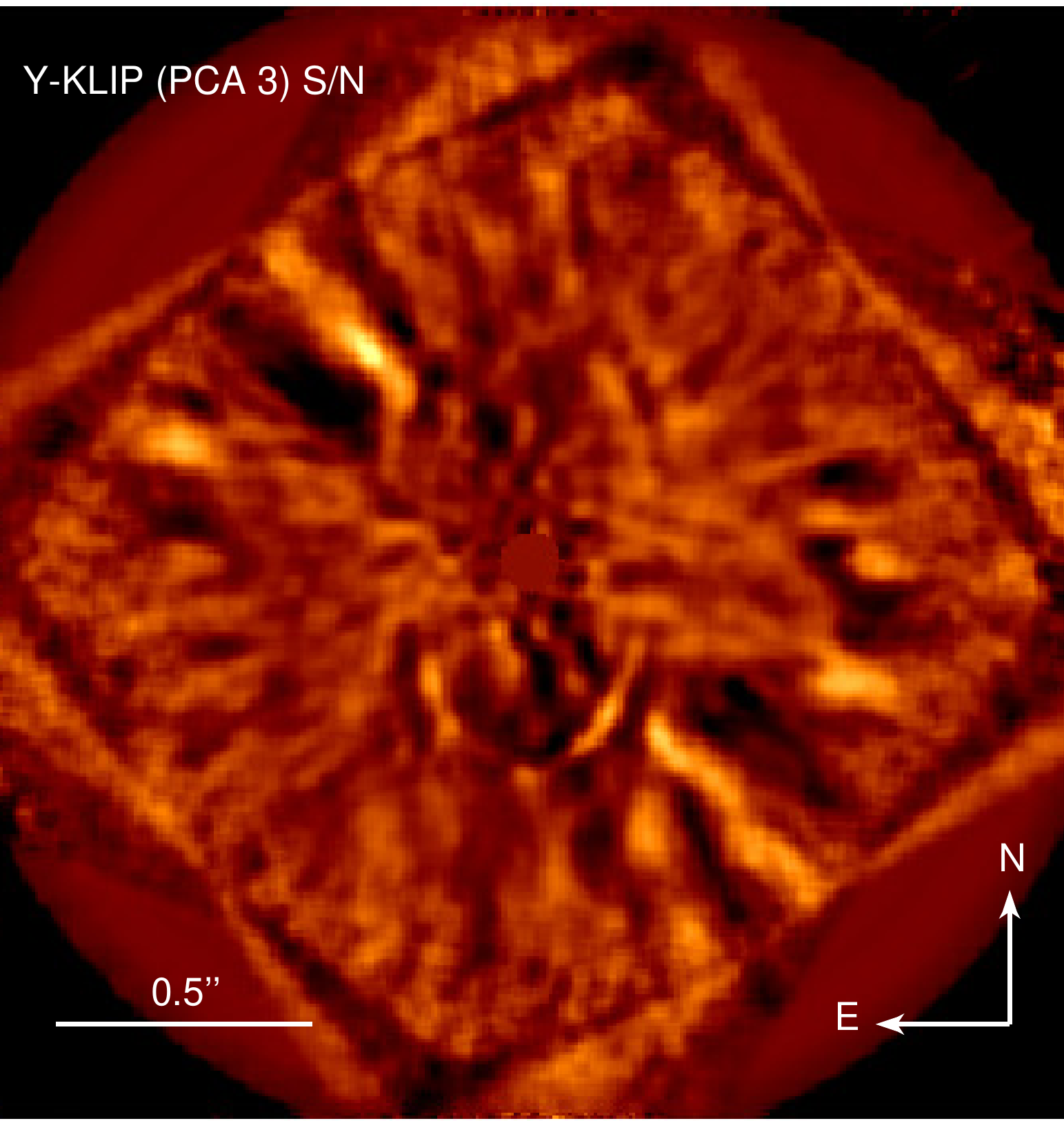}
        \includegraphics[width=0.35\textwidth,trim={15cm 0cm 15cm 0cm}, clip]{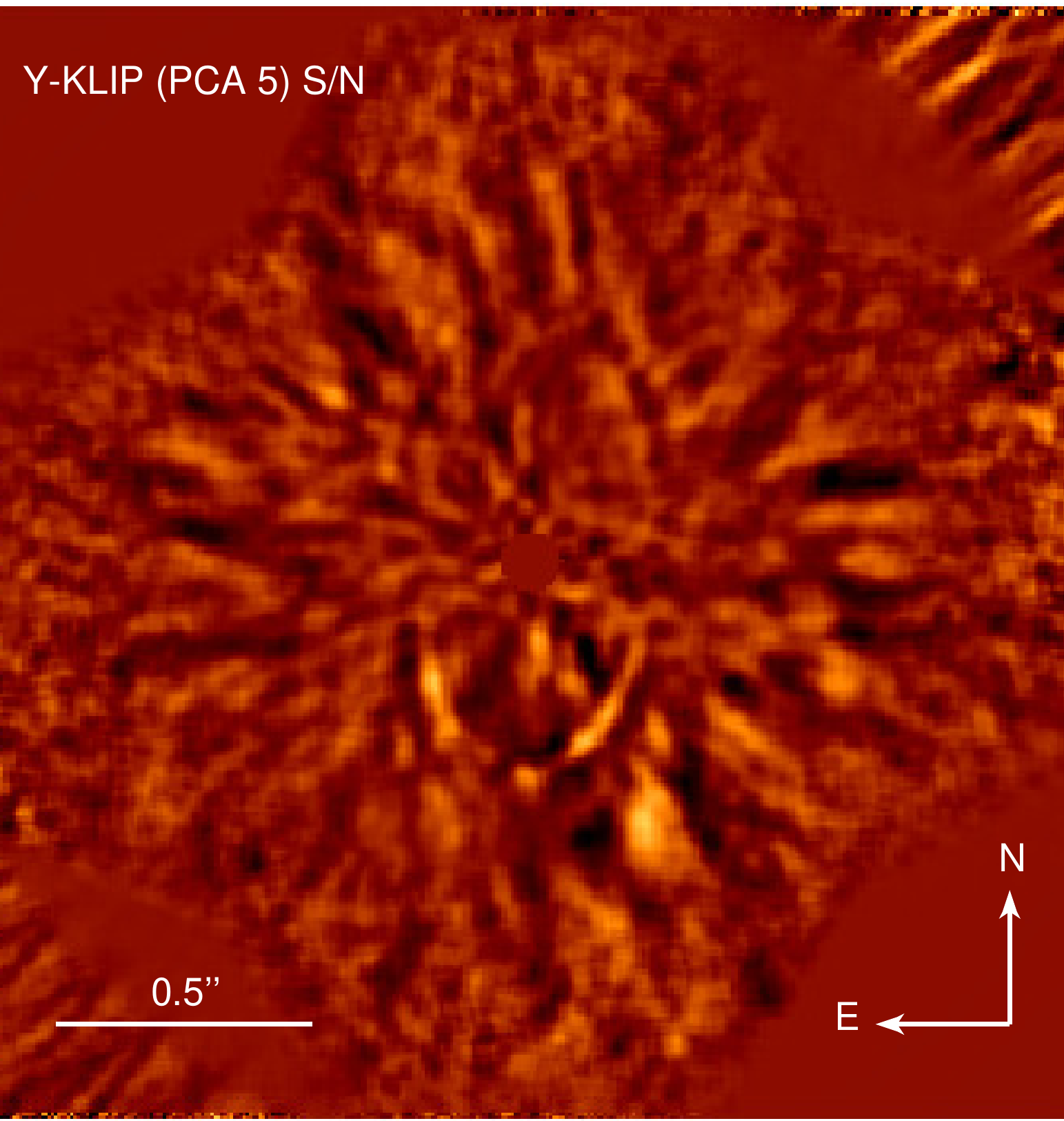} 
	
    \includegraphics[width=0.35\textwidth,trim={15cm 0cm 15cm 0cm}, clip]{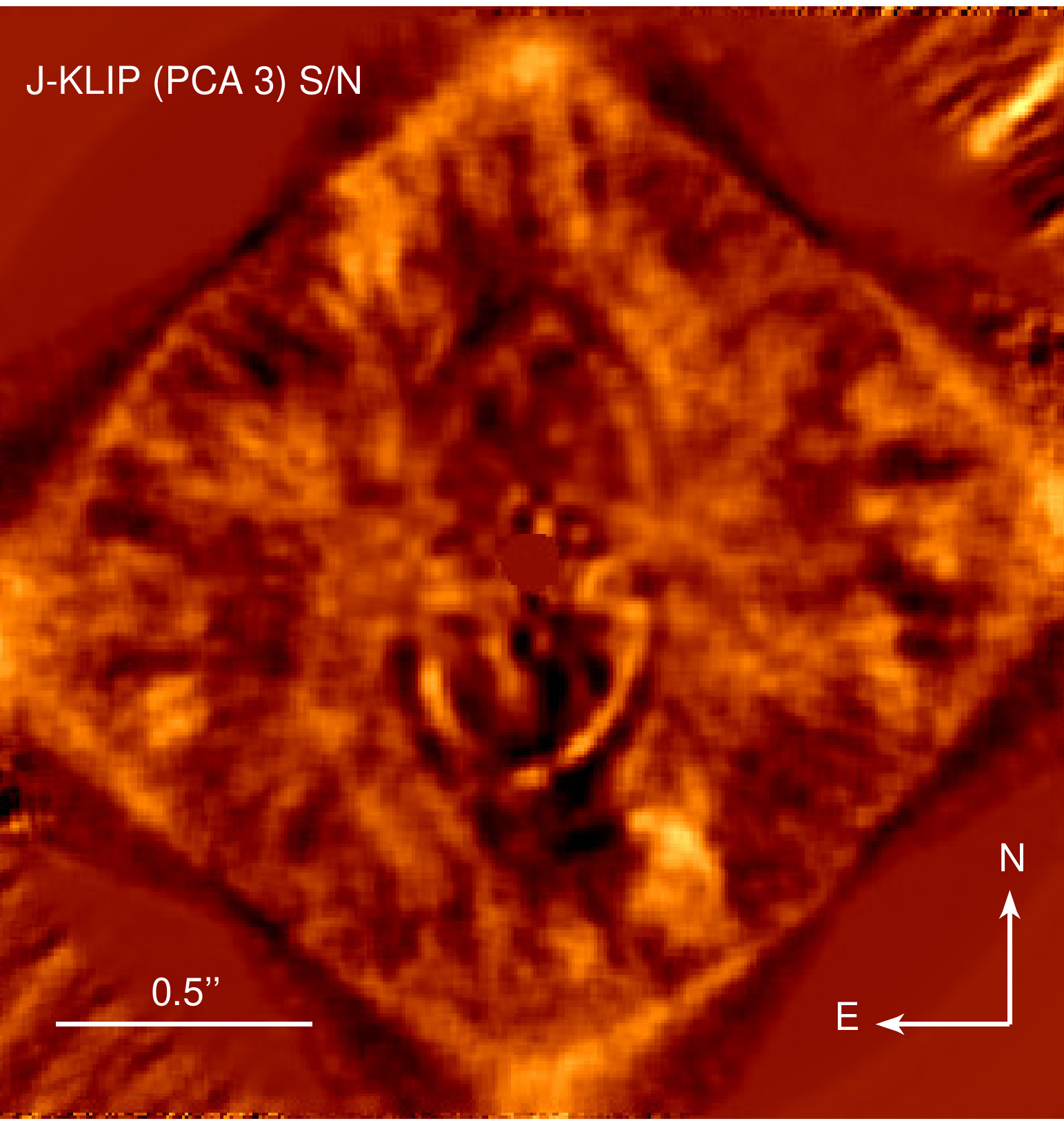}
        \includegraphics[width=0.35\textwidth,trim={15cm 0cm 15cm 0cm}, clip]{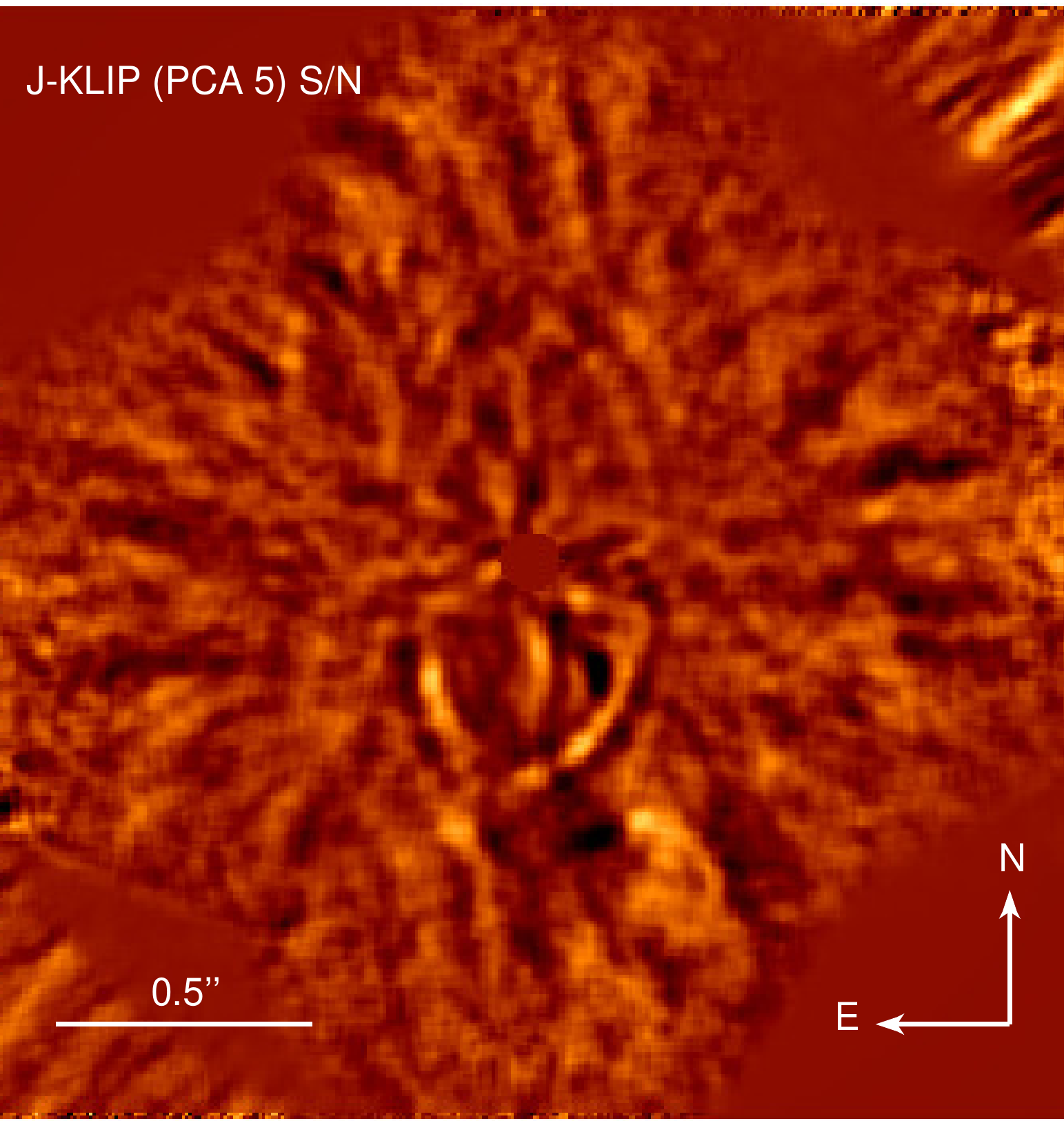}
         
    \includegraphics[width=0.35\textwidth,trim={15cm 0cm 15cm 0cm}, clip]{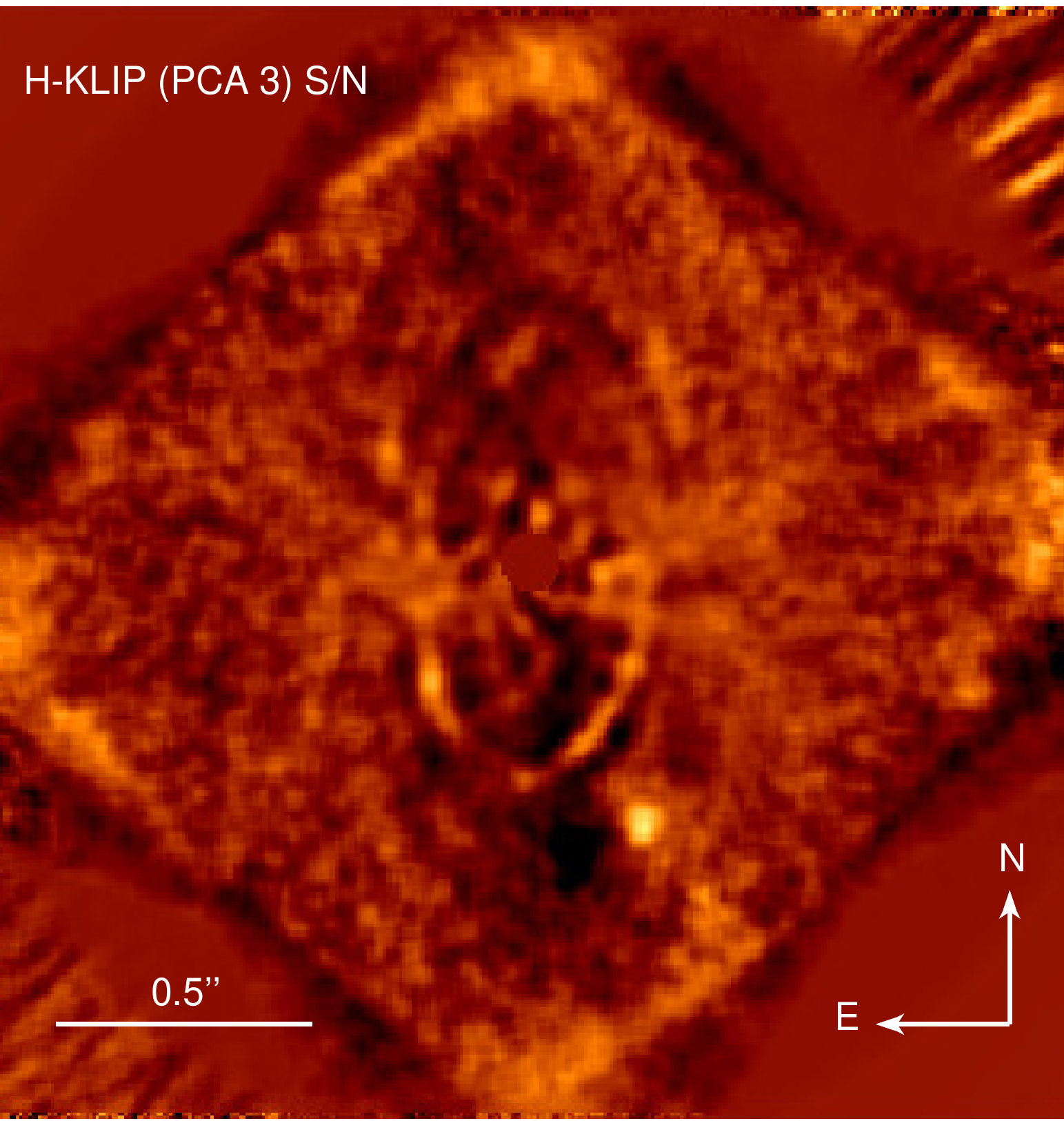}
        \includegraphics[width=0.35\textwidth,trim={15cm 0cm 15cm 0cm}, clip]{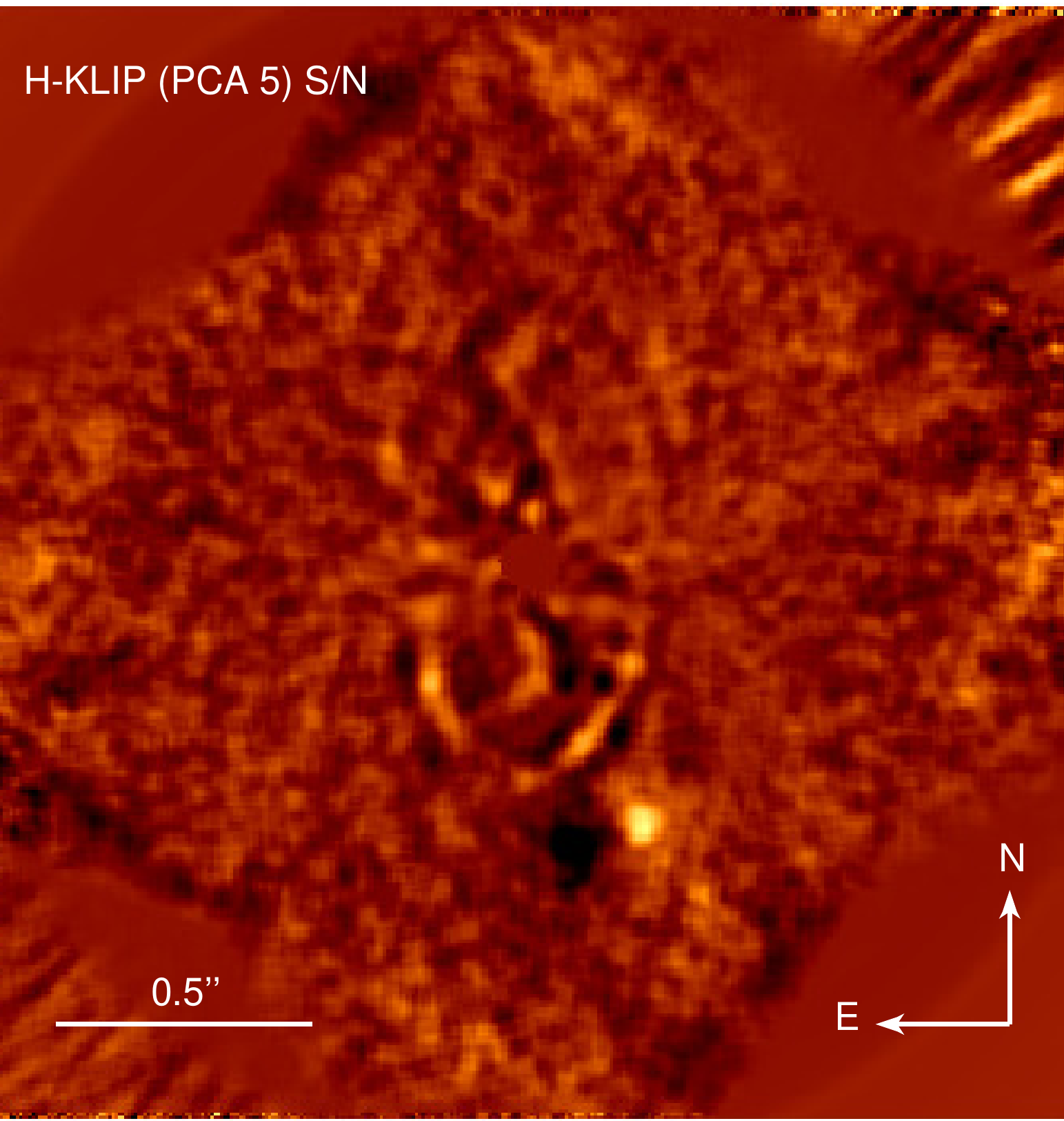}
	
    \includegraphics[width=0.65\textwidth,trim={0cm 0cm 0cm 13cm}, clip]{figures/colorbar}
  \caption{S/N map of the IFS images shown in Fig.~\ref{fig:dbi_yjk}.}
  \label{fig:dbi_yjk_SNR}
\end{figure*}

%
%
%
\section{Complementary ASDI reduced image}
\label{app:ifs_asdi}

\begin{figure}
\centering
\includegraphics[width=0.4\textwidth, trim={3cm 7cm 3cm 6cm}, clip ]{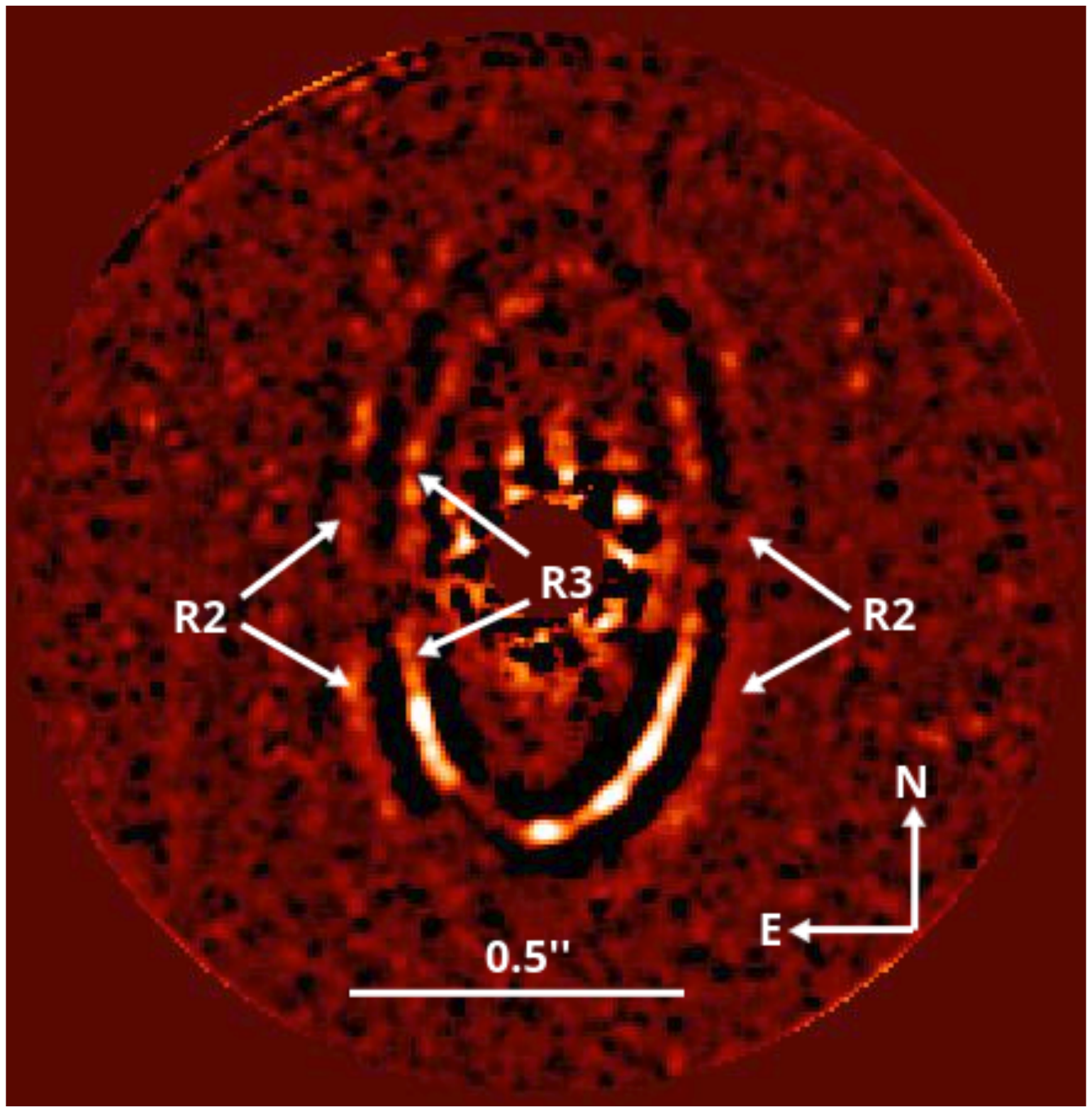}
\caption{IFS images from the three epochs processed using both angular and spectral differential imaging, confirming the detection of R2 and the north south asymmetry in R3.}
\label{fig:asdi}
\end{figure}

%
%
%
\section{Best-fit model assuming an azimuthally uniform dust density in the ring R3}
\label{app:uniform_den}
\begin{figure}
  \centering
\includegraphics[width=0.24\textwidth]{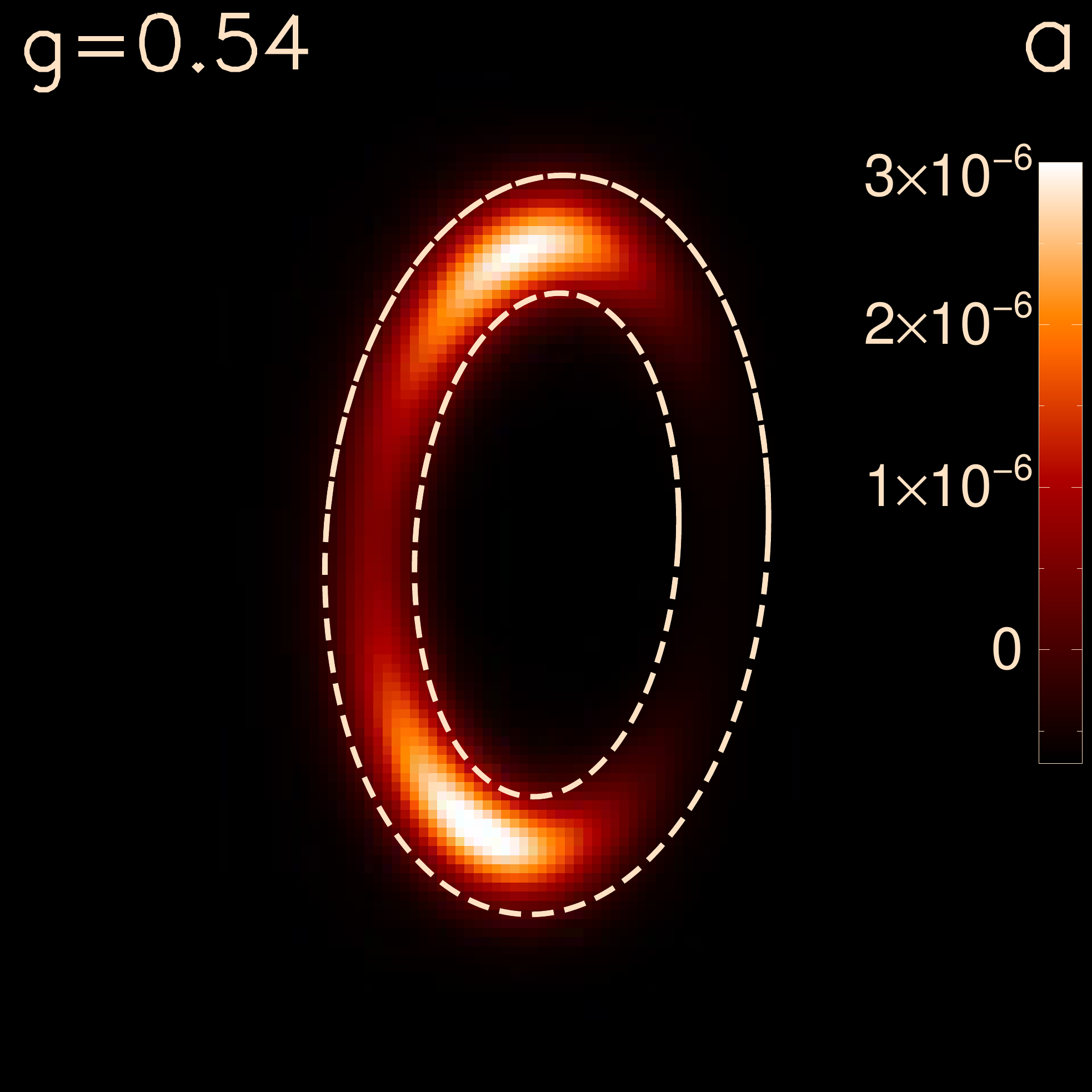}

\includegraphics[width=0.24\textwidth]{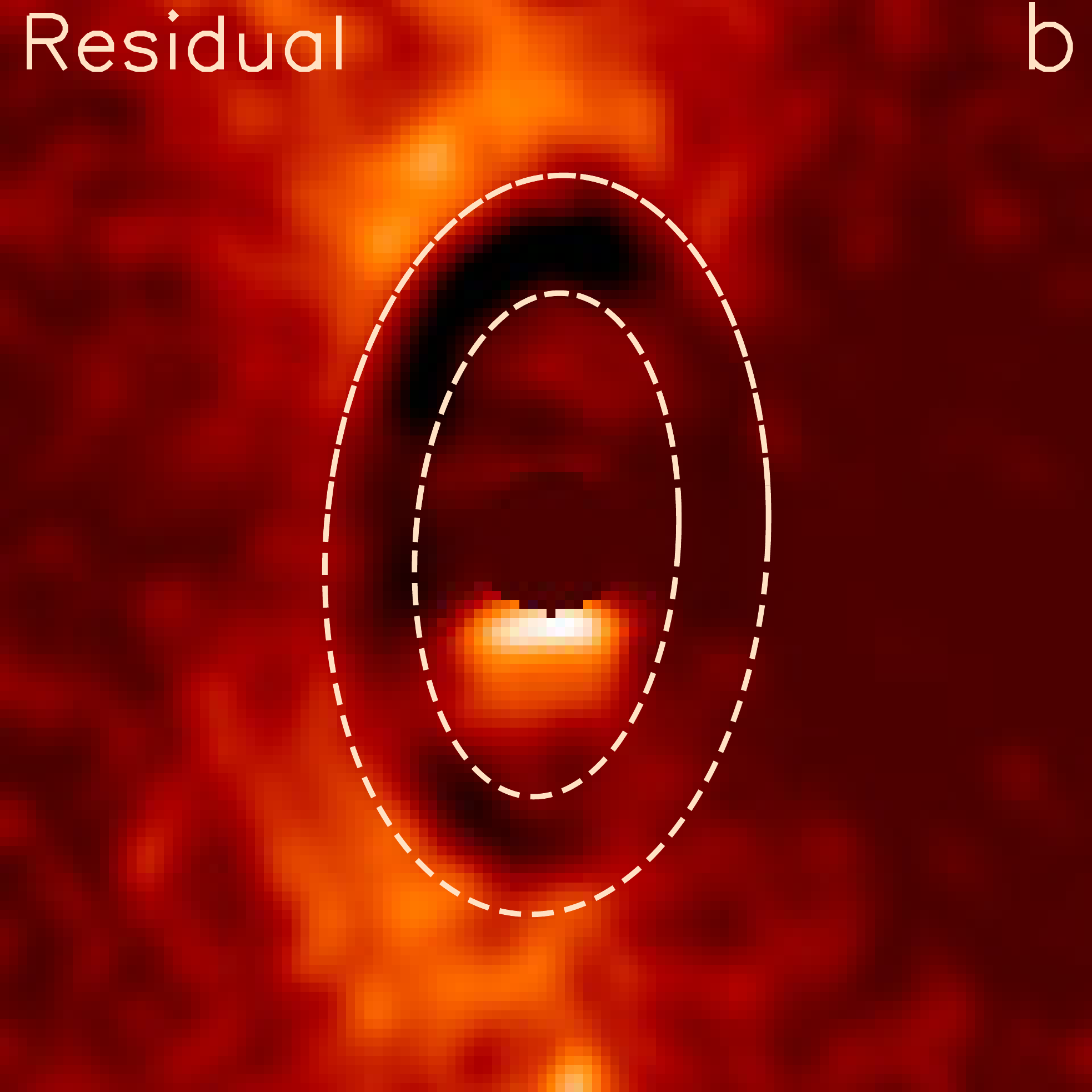}
  \caption{Best-fit model and the corresponding residual for polarimetric data considering homogeneous dust density distribution and $g_\mathrm{pol}$=0.54.}
  \label{fig:sameg}
\end{figure}

%
%
%
\section{Additional MCMC test}
\label{app:mcmctest}
\begin{figure}
  \centering
\includegraphics[width=0.24\textwidth]{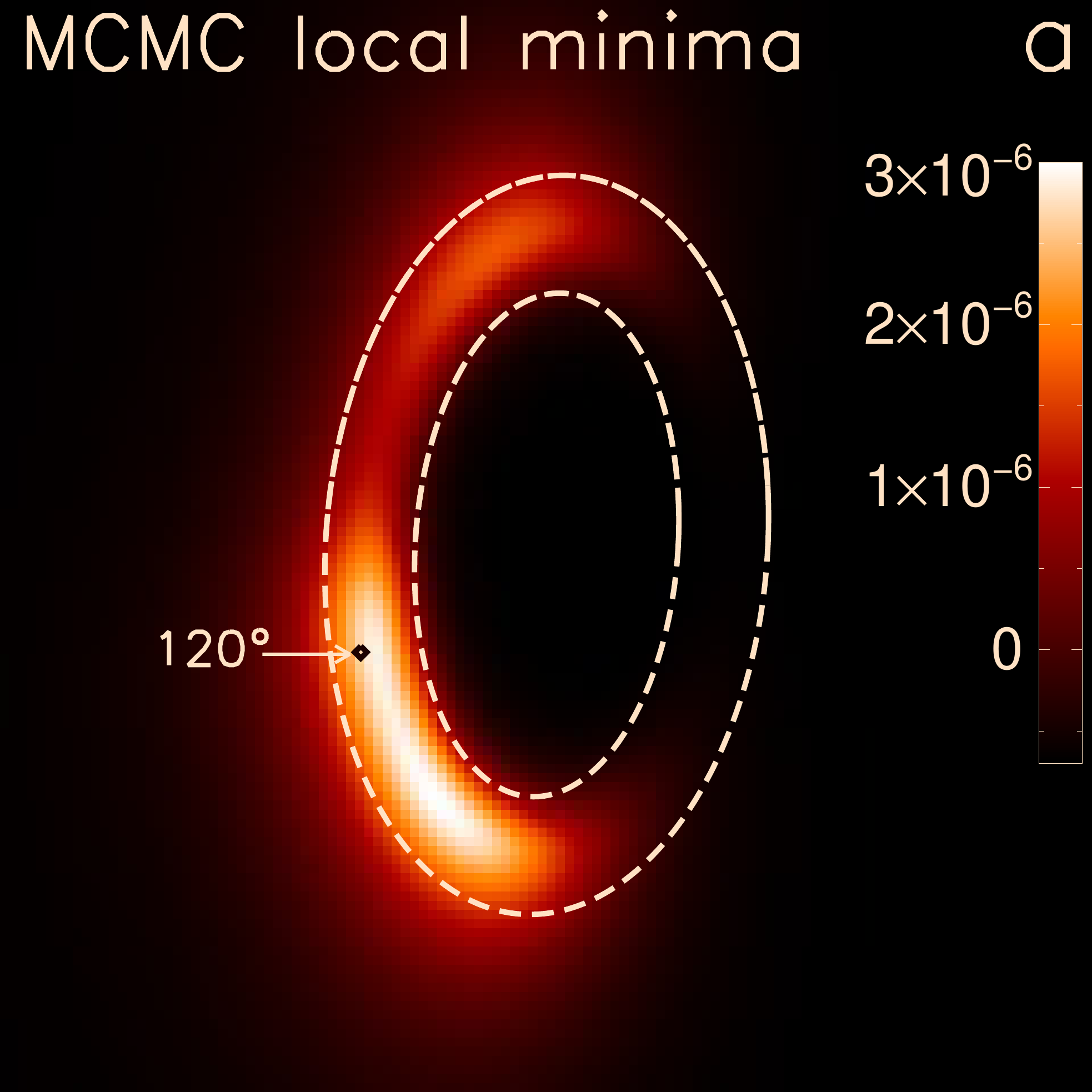} 

\includegraphics[width=0.24\textwidth]{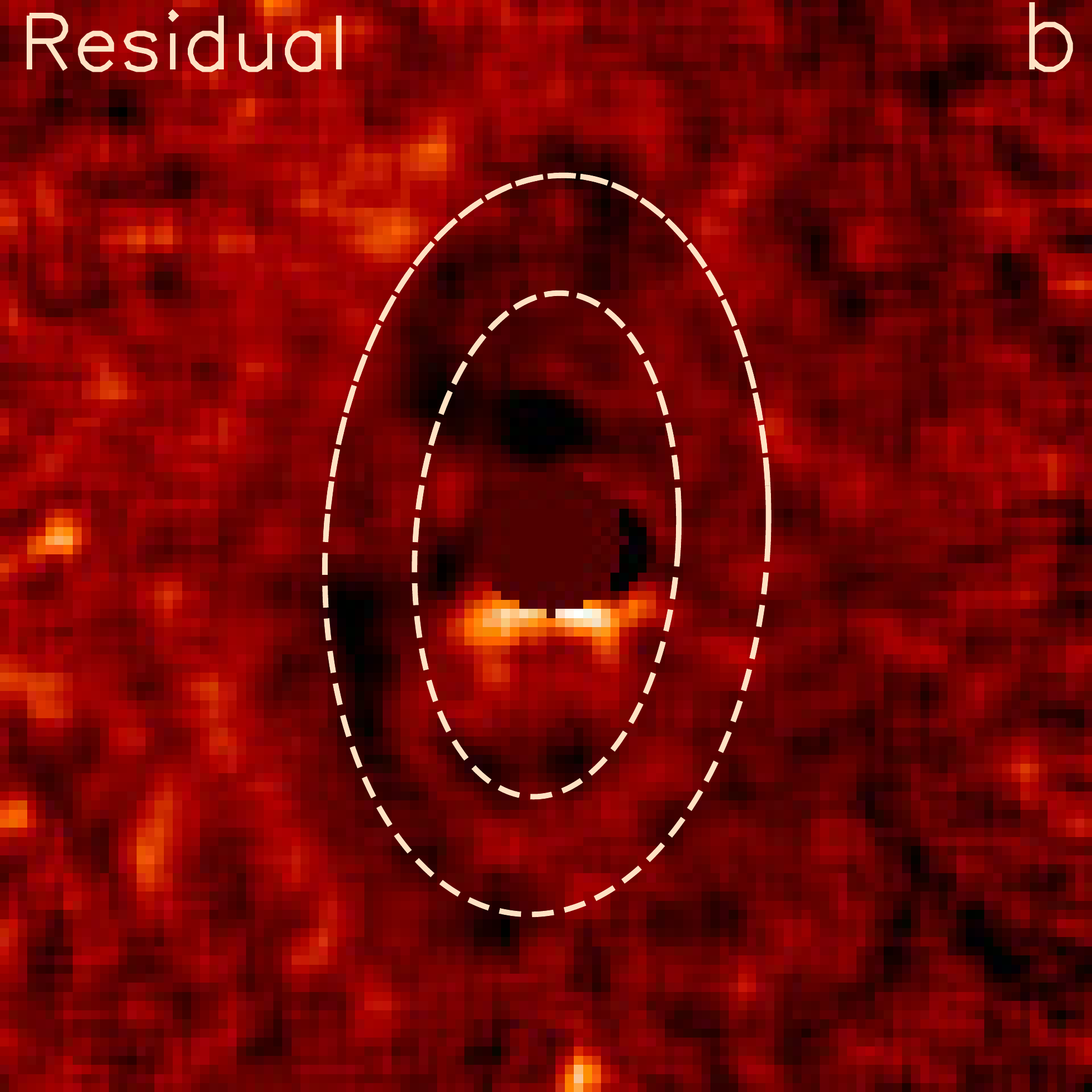}
  \caption{MCMC local-minima with the corresponding residual for the modeling of polarimetric data. }
  \label{fig:localMin}
\end{figure}

%
%
%
\section{Best-fit model proposed by pSPF analysis}
\label{app:pSPFanalysis}
\begin{figure}
  \centering
\includegraphics[width=0.24\textwidth]{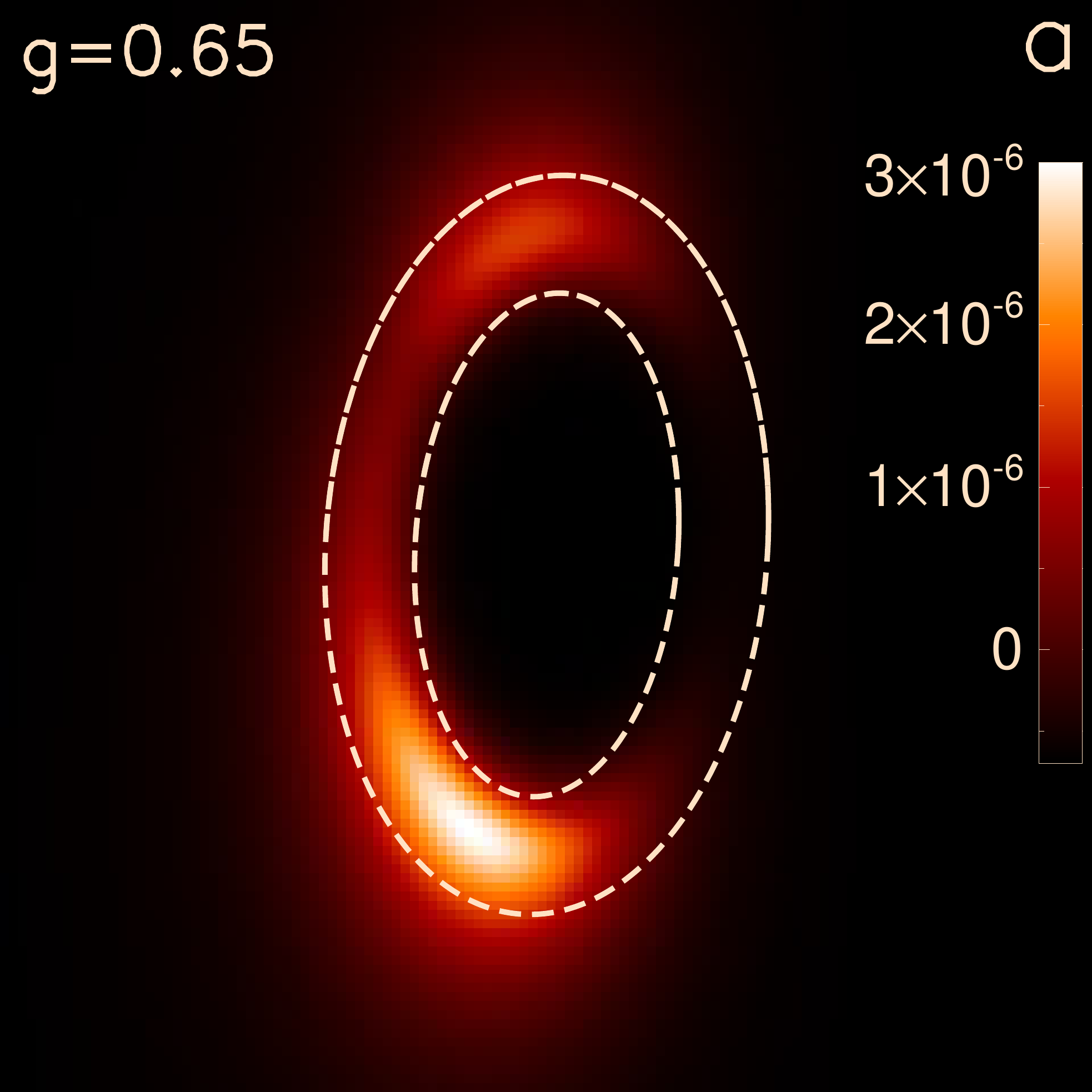}

\includegraphics[width=0.24\textwidth]{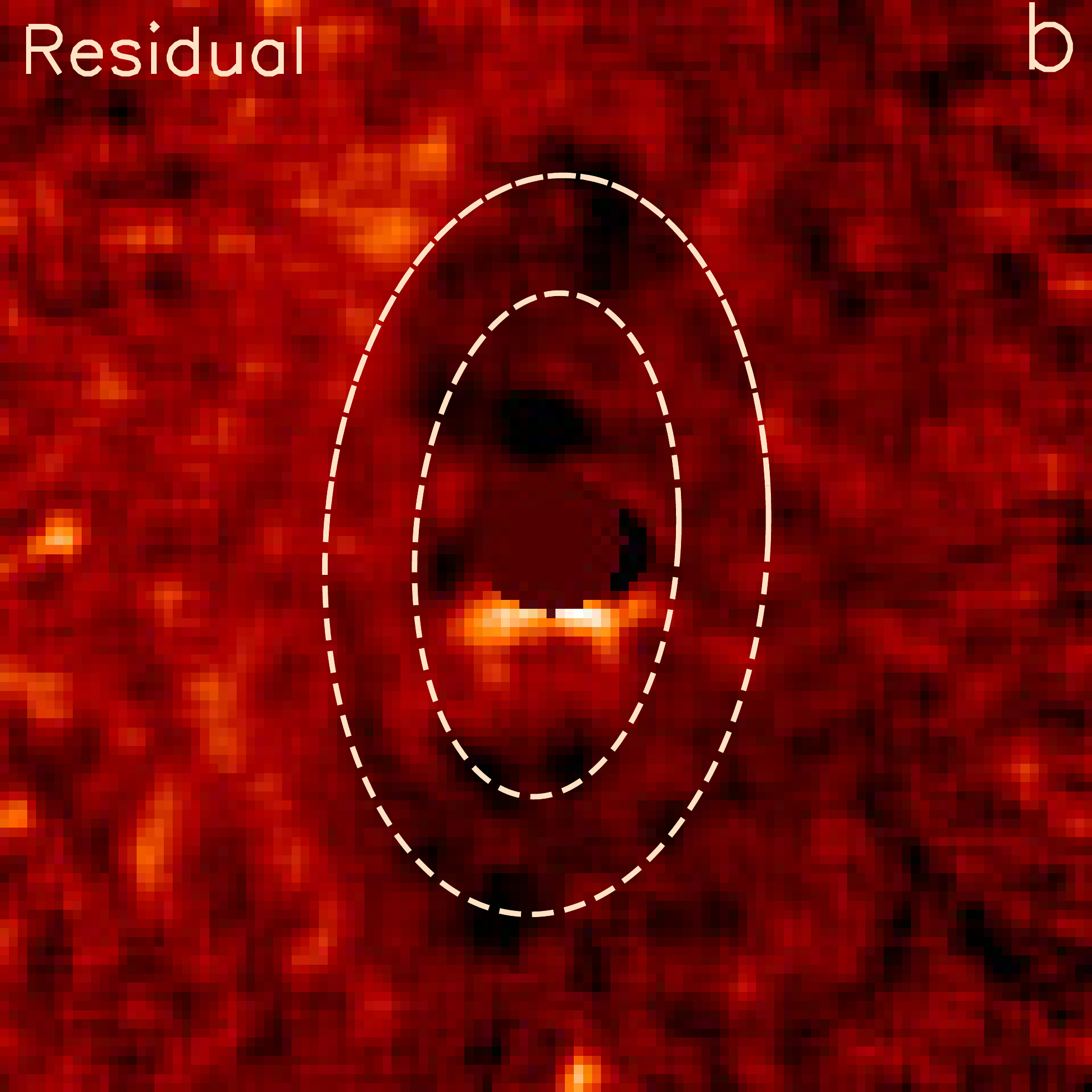}
  \caption{Best-fit model and the corresponding residual for the best-fit asymmetric scattering factor value ($g_\mathrm{pol}$=0.65) derived analytically from the pSPF analysis.}
  \label{fig:spf_test}
\end{figure}

\end{appendix}

\end{document}